\documentclass[twocolumn]{aastex62}
\usepackage{booktabs}
\usepackage{amsmath}
\usepackage{array}
\usepackage{float}
\usepackage{natbib}
\usepackage{booktabs}
\usepackage{amsmath}
\usepackage{array}
\usepackage{wasysym}
\newcommand{\msun}{M$_\odot$}

\newcommand{\msy}{M$_\odot$~yr$^{-1}$}
\newcommand{\mmsun}{\rm M_\odot}
\newcommand{\mlsun}{\rm L_\odot}

\newcommand{\Htwo}{H$_{2}$}
\newcommand{\HI}{\ion{H}{1}}
\newcommand{\HII}{\ion{H}{2}}
\newcommand{\HeI}{\ion{He}{1}}
\newcommand{\HeII}{\ion{He}{2}}

\newcommand{\Fe}{[\ion{Fe}{2}]}
\newcommand{\FeIII}{[\ion{Fe}{3}]}
\newcommand{\SII}{[\ion{S}{2}]}
\newcommand{\PII}{[\ion{P}{2}]}
\newcommand{\SIII}{[\ion{S}{3}]}
\newcommand{\OI}{[\ion{O}{1}]}
\newcommand{\OII}{[\ion{O}{2}]}
\newcommand{\OIII}{[\ion{O}{3}]}
\newcommand{\NII}{[\ion{N}{2}]}
\newcommand{\SiVI}{[\ion{Si}{6}]}
\newcommand{\CaVIII}{[\ion{Ca}{8}]}

\newcommand{\ArIII}{[\ion{Ar}{3}]}
\newcommand{\NI}{[\ion{N}{1}]}
\newcommand{\SIX}{[\ion{S}{9}]}

\newcommand{\Ha}{H\,$\alpha${}}
\newcommand{\Hb}{H\,$\beta${}}

\newcommand{\pab}{Pa\,$\beta${}}
\newcommand{\paa}{Pa\,$\alpha${}}
\newcommand{\pag}{Pa\,$\gamma${}}
\newcommand{\brg}{Br\,$\gamma${}}

\newcommand{\ecs}{erg~cm\pwr{-2}~s\pwr{-1}}
\newcommand{\es}{erg~s\pwr{-1}}

\newcommand{\kms}{km~s\pwr{-1}}
\newcommand{\um}{$\mu$m}
\newcommand{\Av}{$A_V$}

\newcommand{\hk}{$H-K$}

\newcommand{\squiggle}{$\mathtt{\sim}$}

\newcommand{\pwr}[1]{$^{#1}$}
\newcommand{\etal}{\mbox{\it{et al.}~}}
\makeatletter
\newcommand*{\rom}[1]{\expandafter\@slowromancap\romannumeral #1@}
\makeatother
\makeatletter
\def\blfootnote{\xdef\@thefnmark{}\@footnotetext}
\makeatother

\newcolumntype{R}[1]{>{\RaggedLeft\arraybackslash}p{#1}}
\newcolumntype{C}[1]{>{\centering\arraybackslash}p{#1}}

\shorttitle{NGC~5728 Ionization Cones : I - Excitation and Nuclear Structure}
\shortauthors{Durr\'{e} \& Mould}

\begin{document}

\title{The AGN Ionization Cones of NGC~5728 : I - Excitation and Nuclear Structure}
%
\author[0000-0002-2126-3905]{Mark Durr\'{e}}
\author[0000-0003-3820-1740]{Jeremy Mould}
\affil{Centre for Astrophysics and Supercomputing, Swinburne University of Technology, P.O. Box 218, Hawthorn, Victoria 3122, Australia}
\email{mdurre@swin.edu.au}
\begin{abstract}
	We explore the gas morphology and excitation mechanisms of the ionization cones of the Type II Seyfert galaxy NGC~5728. Near-IR and optical data from the SINFONI and MUSE IFUs on the VLT  \deleted{telescope} are combined with \textit{HST} optical images, \textit{Chandra} X-ray data and VLA radio observations. The complex nuclear structure has a star-forming (SF) ring with a diameter of 2 kpc. A radio jet impacts on the ISM at about 200 pc from the nucleus, with the SN remnants in the SF ring also present. Emission-line ratios of \Fe{} and \HII{} show heavy extinction towards the nucleus, moderate extinction in the SF ring and reduced extinction in the ionization cones. The AGN is hidden by a dust bar with up to 19 magnitudes of visual extinction; the dust temperature at the nuclear position is $\sim870$~K.  An X-ray jet is  aligned with the ionization cones, and associated with high-excitation emission lines of \SiVI{} in a coronal-line region extending 300 pc from the nucleus. 
	
	Molecular hydrogen is spatially independent of the cones, concentrated in a disk equatorial to the star-forming ring, but also showing entrainment along the sides of the bicone. Gas masses for warm and cold \Htwo{}, \HI{} and \HII{} are estimated, and the excitation mechanisms for ionized and molecular gas are elucidated, from both optical (which shows a clean SF-AGN mixing sequence) and infrared diagnostics (which show more complicated, multi-component excitation regimes).
\end{abstract}
\section{Introduction}
The Unified Model of active galactic nuclei (AGN) predicts that radiation from the accretion disk around the super-massive black hole (SMBH) is collimated by a torus of obscuring dust and gas, which is broadly symmetric around the accretion flow axis. This radiation impinges on the interstellar medium (ISM), exciting the gas by photo-ionization and  transferring mechanical energy to the gas by radiation pressure and disk winds; this appears in the form of cones extending from the nucleus. These cones are observed in local Seyfert galaxies, and are more frequently seen in type II (obscured) Seyferts than type I (un-obscured) Seyferts \citep{Schmitt1996}, consistent with unification. The kinematics and excitation of the gas in these cones is of interest to elucidate excitation mechanisms, to examine the interaction of the AGN with its host galaxy and to correlate optical and infrared observations with radio and X-ray.

\cite{Urry1995}, in the seminal review paper on AGN unification, cited NGC~5728 as the paradigm of a Type II AGN with these ionization cones. This makes it a prime target for investigation, particularly using observations in the near infrared, allowing penetration of obscuring dust.  In the present paper (`Paper I'), we will explore in detail the nuclear structure of this galaxy at multiple wavelengths and characterize the \added{narrow-line region} (NLR) gas morphology and excitation. In the companion paper (Durr\'{e} \etal, in submission, hereafter `Paper II'), we will use these data to explore the kinematics of the cones, showing that they are AGN-driven outflows, determining mass outflow rates and power and relating these to the AGN bolometric luminosity and the amount of gas that can be expelled during the AGN activity cycle; the super-massive black hole (SMBH) attributes are also derived. We will also determine the stellar kinematics and relate them to the gas kinematics. 

The basic parameters for NGC~5728 are given in Table \ref{tbl:ngc5728params}. Apart from the AGN activity, the nucleus has a highly complex structure. The Spitzer Survey of Stellar Structure in Galaxies (S4G) \citep{Buta2015} lists the morphology as (R$ _{1}$)SB(\underline{r}$'$l,bl,nr,nb)0/a, which indicates a barred spiral with a closed outer ring, an inner pseudo-ring/lens and a nuclear ring and bar/bar-lens. This morphology is important in the context of SF in the nuclear region.

\begin{table}[!htbp]
	\footnotesize
	\centering
	\caption{NGC~5728 Basic Parameters}
	\label{tbl:ngc5728params}
	\begin{tabular}{@{}lcl@{}}
		\toprule
		Parameter     &               Value               & Reference                \\ \midrule
		RA            &            14h42m23.9s            &                          \\
		Dec           &            -17d15m11s             &                          \\
		Distance      & 41.1$\pm$2.9 Mpc & \cite{Mould2000a}\tablenotemark{*}        \\
		Scale         &       200 pc arcsec\pwr{-1}       &                          \\
		z             &             0.009353              & \cite{Catinella2005}     \\
		Morphology    &             SAB(r)a?              & \cite{DeVaucouleurs1991} \\
		Activity      &          Sy 2 (Sy 1.9?)           & \cite{Veron-Cetty2006}   \\
		M$_H$ (2MASS) &              -23.85               & \cite{Skrutskie2006}     \\ \bottomrule
	\end{tabular}
\tablenotetext{*}{{\footnotesize Virgo+GA+Shapley flow-corrected distance, $H_0 = 73$ \kms Mpc\pwr{-1}}}
\end{table}

From the 2MASS \textit{H}-band absolute magnitude with a mass-to-light ratio of 1 at \textit{H} and a Solar absolute magnitude of M$_H$ = 3.3 \citep{Binney1998}, the estimated galaxy mass is $7.2 \times 10^{10}~\mmsun$.

\begin{figure}[htbp!]
	\centering
	\includegraphics[width=1\linewidth]{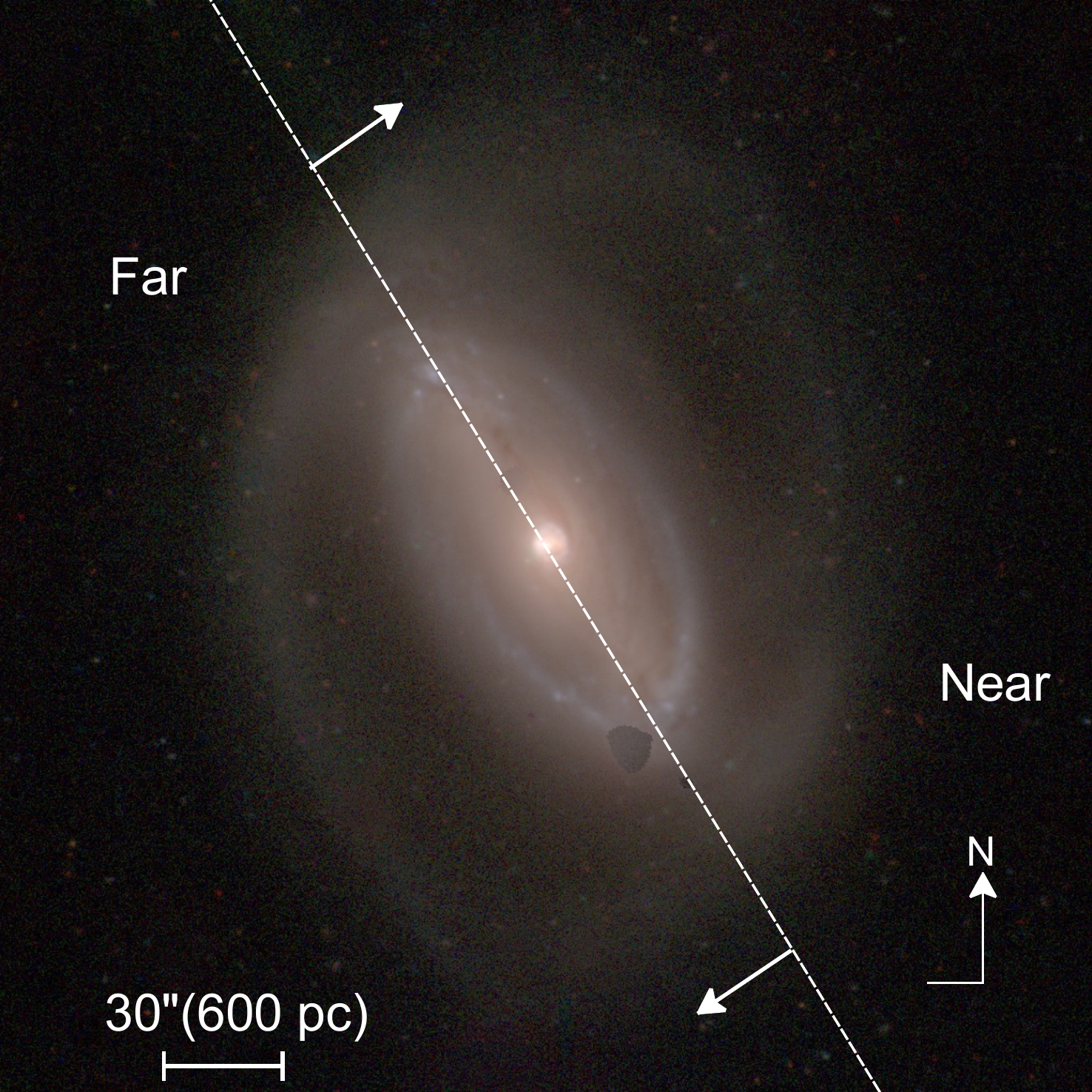}
	\caption{NGC~5728 star-cleaned color-composite image from CGS, created from the \textit{B}, \textit{V} and \textit{I} images and cleaned of stars, as described in \cite{Ho2011}. The image is oriented north up, east left. The sides of the image are 4\arcmin.6, equivalent to 55 kpc at the distance of NGC~5728. The features visible are the outer ring, faint arms trailing from SF peaks at the end of the major bar, a smooth disk with dust lanes, and the complex inner nuclear structure, composed of a ring with a bar. The dotted line indicates the morphological axis and kinematic line of nodes, and the arrows indicate the rotation direction. The near and far sides of the galaxy are labeled. }
	\label{fig:ngc5728cleancolor}
\end{figure}

The Carnegie-Irvine Galaxy Survey \citep{Ho2011} image (Fig. \ref{fig:ngc5728cleancolor}) clearly shows the outer ring with faint trailing spiral arms. The bar is weak, with the region between the nucleus and the outer ring being reasonably smooth with some faint dust rings. \cite{Rubin1980}, using \Ha{} emission line measurements, found that the rotation curve along the NE-SW axis was flat and that the NE axis was approaching; that study concluded that the near side was to the NW, based on the trailing faint spiral arms. She also deduced that the non-circular velocities could be modeled equally well by an additional axisymmetric expansion, or by a displaced inner disk or spheroidal/triaxial inner bulge.
NGC~5728 has ionization cones that are oriented across our line of sight. \cite{Gorkom1982} first observed asymmetrical optical and radio emissions from this galaxy; their VLA 6 and 20 cm observations showed a diffuse 10\arcsec{} region which was coextensive with the optical emission, with a compact nucleus and jet. These observations were followed up by \cite{Schommer1988}; they concluded that the gas was streaming inwards, however their data did not exclude outward motions, as suggested by the double-peaked \Ha{} emission line region near the nucleus. \cite{Arribas1993} noted that the kinematic center of the \Ha, \NII, and \SII{} emission lines and the radio flux did not coincide with the emission line flux maximum, suggesting that the nucleus was highly obscured. The \Ha+\NII{} and \OIII{} emission line images from the \textit{HST} observations of \cite{Wilson1993} revealed a spectacular bi-conical structure with an overall extent of 1.8 kpc. 

\textit{HST} UV imaging polarimetry of the nucleus \citep{Capetti1996}, using the Faint Object Camera, revealed a centro-symmetric pattern of scattered light originating in the hidden nucleus; the polarization upper limit was 1.9\%, in line with ground-based observations. This showed that the cones were much wider than that inferred from the emission lines (55--65\degr), implying that some of the torus is transparent to UV light while still blocking the ionizing radiation. The cone symmetry was used to locate the nucleus behind a dust lane. The activity type is classified as Seyfert 1.9 \citep{Veron-Cetty2006}, based on broad \Ha{} Balmer line visibility.

The kinematics and excitation mechanisms of \Htwo{} and \Fe{} lines in a sample of active galaxies were studied in \cite{Rodriguez-Ardila2004,Rodriguez-Ardila2005}, which included this galaxy, using long-slit observations from the 0.8--2.4 \um{} IRTF Spex spectrograph. They showed that, in general, the \Htwo{} was kinematically decoupled from the NLR, with this particular galaxy having the highest ratio of \Htwo{} to \brg{} flux in their sample of 22 AGN. They estimated a warm \Htwo{} mass of $\sim$900 \msun{} in the nucleus. \cite{Rodriguez-Ardila2011} also detected the coronal line of \SiVI{}, however they did not observe any coronal lines in this object with an ionization potential (IP) $>$ 167 eV, indicating a limit to the hard X-ray flux. From WiFeS optical IFU observations, \cite{Dopita2015a} reported that the NLR was much more extended than from \cite{Wilson1993}; 4.4 kpc rather than 1.8 kpc in extent. 
Using the spectra  for this object from \cite{Rodriguez-Ardila2004}, which were taken with a $ 0\arcsec.8\times15\arcsec $ slit, \cite{Riffel2009b} deduced a mixture of 70\% intermediate age (100 Myr to 2 Gyr) and 30\% old ($>$ 2 Gyr) stellar population contributions to the continuum, with corresponding mass fractions of 25\% and 75\%, with no black-body dust component. 

In this paper, we use the standard cosmology of H$_0 = 73$ \kms Mpc\pwr{-1}, $\Omega_{Matter}=0.27$ and $\Omega_{Vacuum}=0.73$. All images are oriented so that North is up, East is left.

\section{Observations, Data Reduction and Calibration}
\label{sec:ngc5728Observations}
We obtained near infrared (NIR) IFU data from SINFONI on VLT-U4 (Yepun), both from our own observations (\textit{J}-band filter) and from the MPE group (\textit{H+K}-band filter) who kindly let us use the data from their LLAMA (Luminous Local AGN with Matched Analogues) survey \citep{Davies2015}. Each dataset consists of two object frames, combined with a sky frame with the same exposure, in the observing mode `Object-Sky-Object'. For the \textit{J}-band SINFONI observations, the offset was 30\arcsec{} in Dec, plus a 0\arcsec.05 jittering procedure; the offset for the \textit{H+K} observations was was 60\arcsec{} in RA, plus a 0\arcsec.1 jitter. Standard stars are observed close to the observation at a similar airmass.

We also obtained archival data at optical wavelengths from the MUSE optical IFU instrument \citep{Bacon2010} on the ESO VLT for the TIMER (`Time Inference with MUSE in Extragalactic Rings') survey (P.I. Dimitri A. Gadotti , European Southern Observatory); this data was reduced before archival release and is used with the kind permission of Dr. Gadotti.

Our observations are summarized in Table \ref{tbl:ngc5728obslog}, which also includes the details of the MUSE TIMER program optical observations. 

\begin{table*}[!htbp]
\begin{center}
		\caption{NGC~5728 Observation Log}
\label{tbl:ngc5728obslog}
	\begin{tabular}{lccccc}
		\toprule
		Date                   &           Instrument/Filter            & Exp. Time\tablenotemark{a}/Frames & Airmass/ Seeing (\arcsec)\tablenotemark{b} &               R\tablenotemark{c}               &      FOV (pc)\tablenotemark{d}       \\
		Program ID/PI          & Plate Scale (\arcsec)\tablenotemark{e} &        Total Exp. Time (s)        &               Standard Star                & BC\tablenotemark{f}/$\Delta$V\tablenotemark{g} & Sp. Res.\tablenotemark{h} (pc/pixel) \\ \midrule
		2014 Apr 11            &               SINFONI  J               &               300/6               &                 1.085/1.3                  &                      2400                      &                 1500                 \\
		093.B-0461(A)/Mould    &                  0.25                  &               1800                &                 HIP091038                  &                    11.6/53                     &                  25                  \\ \midrule
		2015 Feb 23            &              SINFONI H+K               &               300/6               &                  1.011/1                   &                      1560                      &                 600                  \\
		093.B-0057(B)/Davies   &                  0.1                   &               1800                &                 HIP073266                  &                    28.3/85                     &                  10                  \\ \midrule
		2015 Mar 5             &              SINFONI H+K               &               300/6               &                 1.012/1.25                 &                      1560                      &                 600                  \\
		093.B-0057(B)/Davies   &                  0.1                   &               1800                &                 HIP071451                  &                    25.8/85                     &                  10                  \\ \midrule
		2015 Jun 14            &              SINFONI H+K               &               300/6               &                 1.016/1.25                 &                      1560                      &                 600                  \\
		093.B-0057(B)/Davies   &                  0.1                   &               1800                &                 HIP078968                  &                    -13.9/85                    &                  10                  \\ \midrule
		2015 Jun 25            &              SINFONI H+K               &               300/6               &                 1.203/0.75                 &                      1560                      &                 600                  \\
		093.B-0057(B)/Davies   &                  0.1                   &               1800                &                 HIP082670                  &                    -21.8/85                    &                  10                  \\ \midrule
		2016 Apr 3             &                MUSE  V                 &               480/1               &                 1.048/0.66                 &                      1800                      &                12800                 \\
		097.B-0640(A) /Gadotti &                  0.2                   &                480                &           \dots\tablenotemark{i}           &                    15.5/70                     &                  40                  \\ \bottomrule
	\end{tabular}
\end{center}
{
\tablenotetext{a}{Exposure time (sec) for each on-target frame.} 
\tablenotetext{b}{Seeing from the ESO Paranal Astronomical Site Monitoring (VLT) or the Mauna Kea Weather Center DIMM archive (Keck).}
\tablenotetext{c}{Measured spectral resolution.}
\tablenotetext{d}{Observed field of view}
\tablenotetext{e}{SINFONI plate scale is re-binned to half size in the final data cube.}
\tablenotetext{f}{Barycenter correction in \kms{} from `Barycentric Velocity Correction' website \url{http://astroutils.astronomy.ohio-state.edu/exofast/barycorr.html} \citep{Wright2014}}
\tablenotetext{g}{Spectral velocity resolution (\kms)}
\tablenotetext{h}{Spatial resolution}
\tablenotetext{i}{Data not available}}
\end{table*}
The SINFONI data reduction was performed using the recommendations from the ESO SINFONI data reduction cookbook and the \texttt{gasgano}\footnote{\url{http://www.eso.org/sci/software/gasgano.html}} \citep{ESO2012} software pipeline (version 2.4.8). Bad read lines are cleaned from the raw frames using the routine provided in the cookbook. Calibration frames are reduced to produce non-linearity bad pixel maps, dark and flat fields, distortion maps and wavelength calibrations. Sky frames are subtracted from object frames, corrected for flat field and dead/hot pixels, interpolated to linear wavelength and spatial scales and re-sampled to a wavelength calibrated cube, at the required spatial scale. The sequence of reduced object cubes are mosaicked and combined to produce a single data cube. Subsequent cube manipulation and computation is done using the the cube data viewer and analysis application \texttt{QFitsView} \citep{Ott2012}, which incorporates the \texttt{DPUSER} language.

Standard stars are used for both telluric correction and flux calibration; these had the spectral types of B2 to B9, which only have hydrogen and helium lines in their spectra. Flux calibration is done by extracting the stellar spectrum from the data cube within a aperture, chosen using logarithmic scaling on the collapsed `white-light' image. A segment of the spectrum is chosen that avoids telluric features and stellar absorption lines, on or near the effective wavelength for the filter. The total counts in a 10 nm window centered on the chosen wavelength is divided by the exposure time and the window wavelength width to get a value of counts s\pwr{-1} nm\pwr{-1}. This is equated to the stellar magnitude is found from the 2MASS catalog \citep[through the SIMBAD astronomical database\footnote{\url{http://simbad.u-strasbg.fr/simbad/}}]{Skrutskie2006}, converted to a flux density using the website `Conversion from magnitudes to flux, or vice-versa'\footnote{\url{http://www.gemini.edu/sciops/instruments/midir-resources/imaging-calibrations/fluxmagnitude-conversion}}. The telluric correction spectrum was created from the standard star spectrum by manually fitting and removing the hydrogen and helium absorption features, then dividing by a black-body spectrum of the correct temperature for the spectral type. This spectrum is normalized and is divided into each spectral element. This temperature is somewhat higher than given in standard references (e.g. 17000 K vs. 14300 K for B6 stars), as the hydrogen opacity is lower at infrared wavelengths; we observe a lower (and therefore hotter) layer in the stellar atmosphere \citep[see][]{Durre2017}. After telluric correction and flux calibration, the data cubes from the SINFONI \textit{H+K} observations for separate dates were combined into single cubes. These were centered on the brightest pixel (in the image created by collapsing the cube along the wavelength axis). The whole observed field for the \textit{H+K} filter is $ 3\arcsec\times3\arcsec $, which equates to $600\times600$ pc.

Initially, the \textit{J} SINFONI cube showed a strong telluric absorption feature at 1268.9 nm; normally this is not an issue, but for this object, the redshift moved the \Fe{} 1257 nm line to this wavelength and producing a pronounced `notch' in the emission line spectrum. This was corrected by using the \texttt{IRAF} \textit{telluric} procedure which shifts and scales the telluric spectrum to best divide out features from data spectra by minimizing the RMS error. The comparison between the telluric spectrum and a spectrum from the cube of the nuclear region showed a shift of 0.8 pixels, which, when applied to the data cube, reduced the feature considerably. 

The well-known `instrumental fingerprint' in the SINFONI \textit{J} data cube was characterized by PCA tomography \citep{Menezes2014,Menezes2015a}; this fingerprint is the result of detector array persistence \citep[see][]{George2016}. Over the characteristic broad horizontal stripe, the flux correction was substantial, in places up to $\pm$15\%. The correction was applied by interpolation over the fingerprint in the \textit{y} axis direction at each spectral pixel. For the final \textit{H+K} data cube, no fingerprint was visible in any of the tomograms, presumably because of the combination of cubes over 4 dates.

All the data cubes had some spikes and geometric artifacts, from the data-reduction package (DRP) or other noise which is not removed (e.g. cosmic rays). These were manually cleaned by interpolation over the offending pixels, especially over the range of emission lines wavelengths.

To check on the wavelength calibration, the OH skyline wavelengths were measured from the off-target sky data cubes. These lines are assumed to be very narrow and the wavelengths are published in \cite{Rousselot2000}. This showed a wavelength shift of $\sim0.44$ nm, equivalent to $\sim62$ \kms; this value is used in in the computation of the systemic velocity of the galaxy.

\section{Results}
\subsection{The Nucleus of NGC~5728}
\label{sec:ngc5728nucleus}
\begin{figure*}[!htbp]
	\centering
	\includegraphics[width=.7\linewidth]{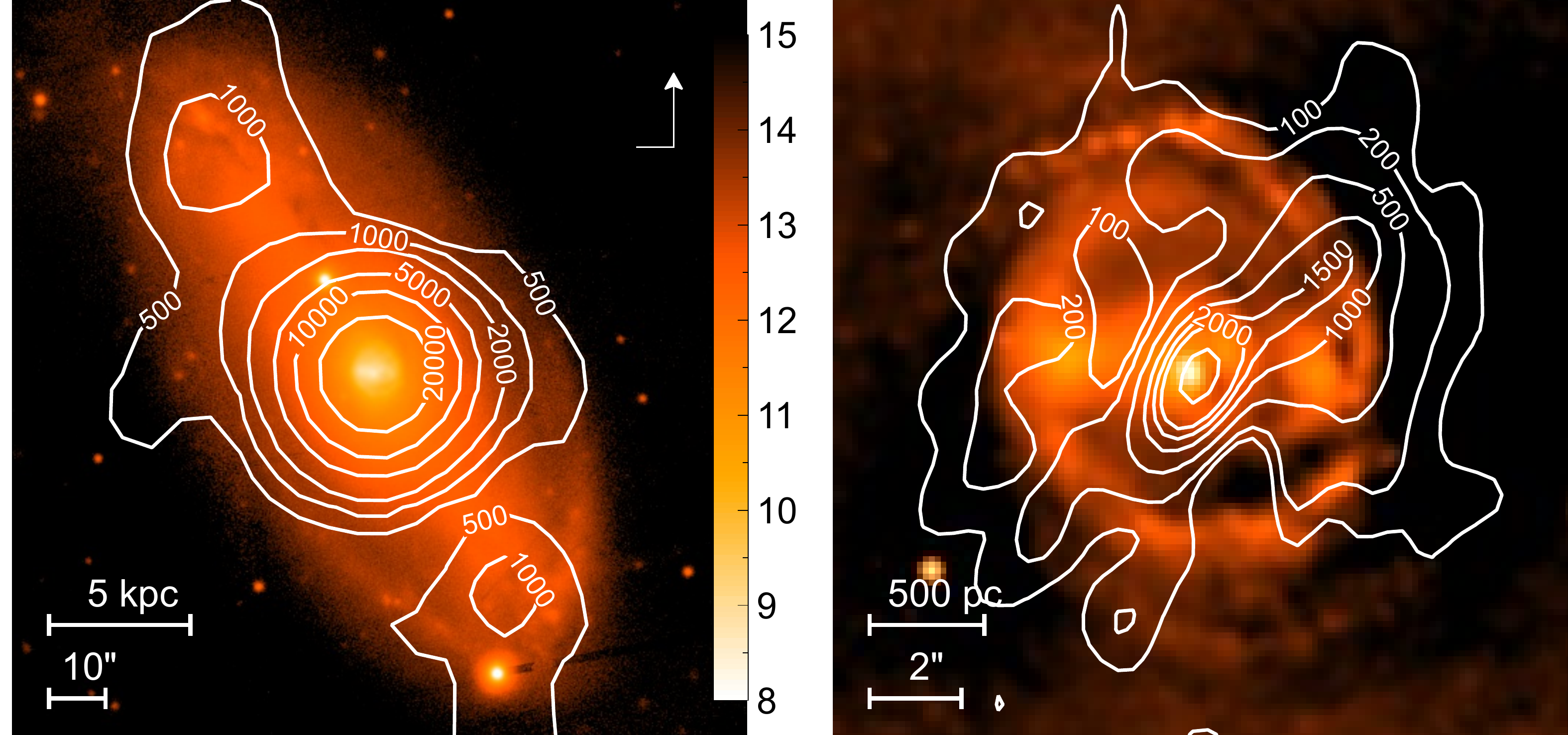}
	\caption{Left panel: Pan-STARRS \textit{i} image overlaid with VLA 20 cm large-scale flux contours with star-formation at the end of the main bar and nuclear AGN/SF emission. Color values in mag arcsec\pwr{-2}. Right panel: HST F160W structure map overlaid with VLA 6 cm flux contours showing the AGN jet and the SF ring. Contour values in $\mu$Jy. }
	\label{fig:ngc5728images11}
\end{figure*}

The nucleus of NGC~5728 is highly complex, as revealed by multi-wavelength images taken at high spatial resolution, with evidence of star formation, radio and X-ray jets and distorted kinematics. Fig. \ref{fig:ngc5728images11} shows the \textit{i} band image from the MAST Pan-STARRS image cutout facility\footnote{\url{https://archive.stsci.edu/}}, overlaid with the contours of the VLA 20 cm large-scale map. It also shows the \textit{HST} F160W structure map combined with the VLA 6 cm map. The VLA images \citep{Schommer1988} were acquired from the NRAO Science Data Archive\footnote{\url{https://archive.nrao.edu/archive/advquery.jsp}}, as set out in Table \ref{tbl:ngc5728VLAData}. There are 2 sets of 20 cm VLA data; large scale with 5\arcsec{} spaxels and small scale with 0\arcsec.3144 spaxels; the beam size given is the half-power beam width for the antenna configuration and frequency (from the NRAO resolution table\footnote{\url{https://science.nrao.edu/facilities/vla/docs/manuals/oss/referencemanual-all-pages}}). 

The \textit{HST} structure map \citep{Pogge2002} produces an image defined as:
\begin{equation}
S=\left(\frac{I}{I \otimes P}\right)\otimes P_t 
\end{equation}
where \textit{I} is the original image, \textit{P} is the HST point spread function (PSF) from the \textit{TinyTim} software\footnote{\url{http://www.stsci.edu/hst/observatory/focus/TinyTim}} \citep{Krist2011}, \textit{P$_t$} is the transpose of the PSF and $\otimes$ is the convolution operator.

The structure map enhances the nuclear bar and ring, with the whole central structure looking like a miniature barred spiral. The spiral aspect appears close to face-on, which is different to the rest of the galaxy. \cite{Prada1999} found that the core is counter-rotating with respect to the main galaxy, and considered that it was most probably caused by orbital instabilities associated with the secondary bar; however they could not rule out satellite or gas accretion with negative angular momentum. The filamentary dust lanes delineate the cold-phase fueling flows into the SMBH \citep[see e.g.][]{SimoesLopes2007, Mezcua2015}.

The 20 cm large-scale map (Fig. \ref{fig:ngc5728images11}) shows nuclear emission combined with two regions coincident with the star formation at the end of the galactic bar. At 6 cm, the higher spatial resolution nuclear emission shows an annulus structure, with the highest emission coincident with the structure map outer ring/arms. It also shows a jet in the NW-SE direction, aligned with the ionized gas emission. 

\begin{figure*}[!htbp]
	\centering
	\includegraphics[width=.7\linewidth]{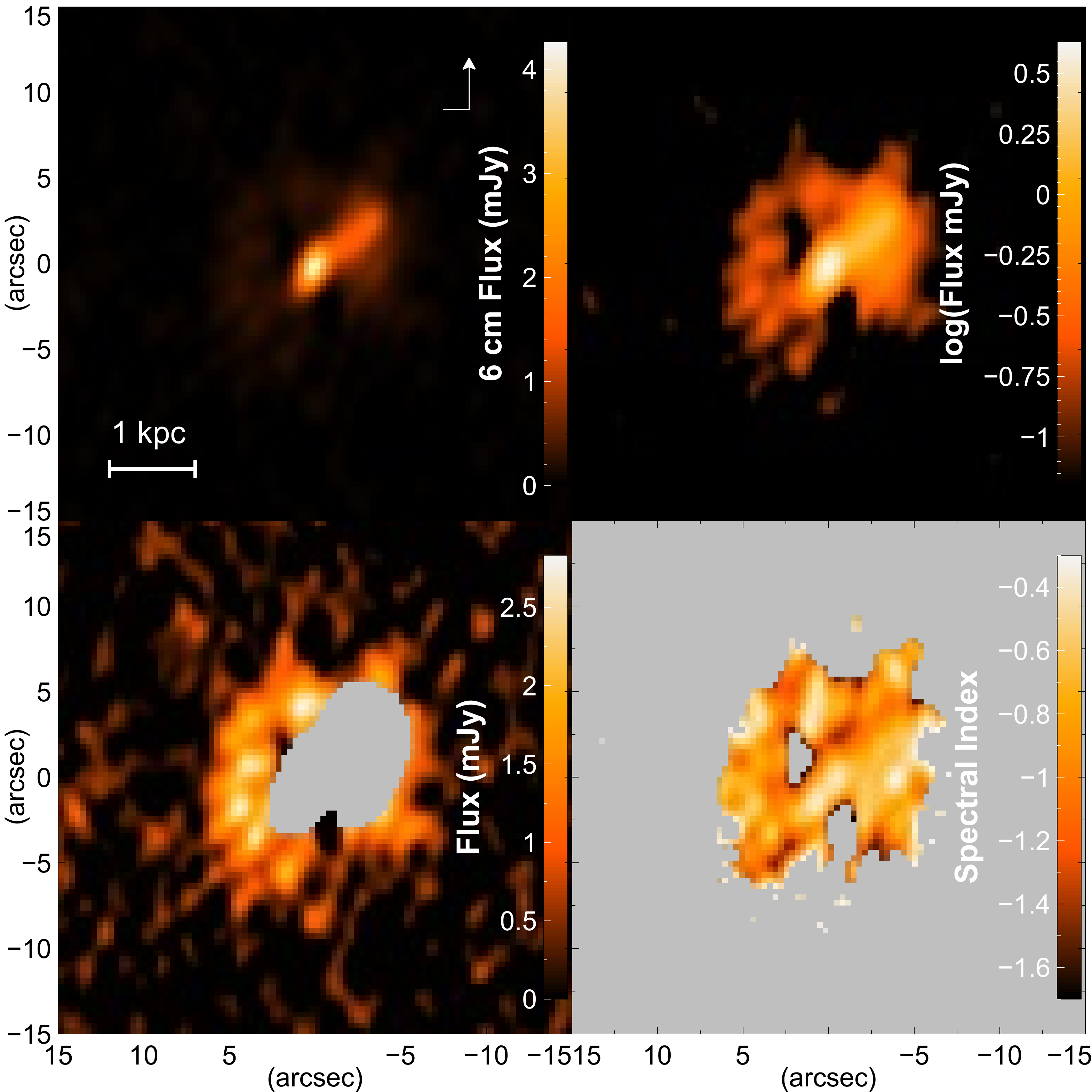}
	\caption{AGN and circumnuclear SF activity from VLA 6 cm images. Top left: linear scaling (to enhance the jet). Top right: logarithmic scaling (to enhance star-forming ring). Bottom left: Jet masked to show point-source supernova remnant emission. All flux values are in mJy. Bottom right: Spectral index derived from 6 cm and 20 cm fluxes; the negative index indicates non-thermal (i.e. synchrotron) emission.}
	\label{fig:ngc5728images21}
\end{figure*}

Fig. \ref{fig:ngc5728images21}  displays the VLA 6 cm fluxes at three different intensity scalings, to respectively enhance the jet and star-forming ring. Masking out all pixels that have a value $>$300 mJy (to remove the jet) (bottom left plot), the SF ring is seen as a string of emission features, each with roughly the same luminosity ($\sim$2 mJy), which are probably individual supernova remnants. The VLA 20 cm high-resolution map presents a very similar structure (not mapped here). Aligning the 6 cm and 20 cm high-resolution maps allows measurement of the spectral index; this is also plotted in Fig. \ref{fig:ngc5728images21}. The spectral index is computed $\alpha =\Delta \log(S_\nu)/\Delta \log(\nu)$, and is in the range -1.7 -- -0.3, with a median value of -0.8, indicating non-thermal (i.e. synchrotron) emission.

\begin{table}[!htbp]
	\centering
	\footnotesize
	\caption{VLA observations for NGC~5728}
	\label{tbl:ngc5728VLAData}
	\begin{tabular}{@{}lcccc}
		\toprule
		Obs. Date   & Wavelength & Freq. & Pixel Scale & Beam Size \\
		            &       (cm) & (GHz) &     (\arcsec) &   (\arcsec) \\ \midrule
		18 Apr 1994 &         20 &   1.4 &      0.3144 &       1.3 \\
		23 Mar 1988 &       20.1 &  1.49 &           5 &        14 \\
		14 Jan 1984 &          6 &  4.86 &      0.3259 &       1.0 \\ \bottomrule
		            &            &
	\end{tabular}
\end{table}

\begin{figure*}[!htbp]
	\centering
	\includegraphics[width=.7\linewidth]{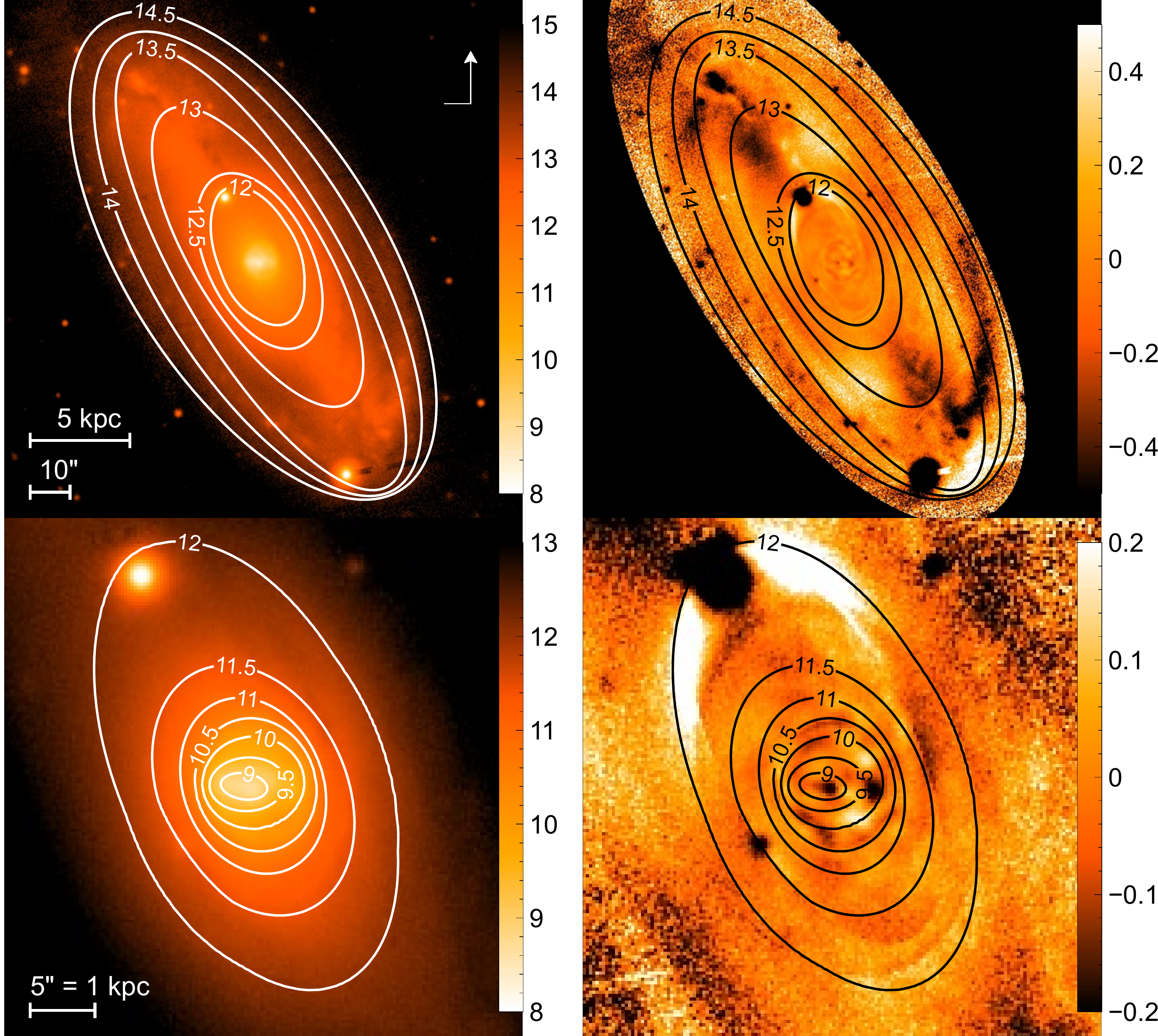}
	\caption{Elliptical model fit to the Pan-STARRS isophotes, showing the complex structure of outer disk, inner ring and nuclear bar. Left panel:  (top) outer, (bottom) inner parts of the galaxy, showing the centrally disturbed morphology. Right panel: Model residual (image--model); (top) outer, (bottom) inner. All color-bars and contours for isophote models have values in magnitudes arcsec\pwr{-2}.}
	\label{fig:ngc5728images12}
\end{figure*}

Fig. \ref{fig:ngc5728images12} shows the elliptical model fit to the Pan-STARRS image, using the IRAF \textit{stsdas.analysis.isophote ellipse} and \textit{bmodel} tasks. The parameters for the various features are given in Table \ref{tbl:ngc5728PSfeatures}. The `ellipticity' is $\epsilon = 1 - b/a$, where \textit{a} is the major and \textit{b} is the minor axis size (a smaller value means closer to circular); from which, assuming circular symmetry, the inclination \textit{i} of the ellipse to the LOS can be derived. This will become important in determining the inclination to the LOS of the jet. 

\begin{table}[!htbp]
	\centering
	\caption{Model fit to features. Col. 2 - the approximate limiting magnitude of the feature. The semi-major axis (SMA), position angle (PA) and ellipticity ($\epsilon$) are from the \textit{ellipse} model fit. PA is N=0\degr{}, E=90\degr. \textit{i}=0\degr{} is face-on, 90\degr{} is edge-on to LOS.}
	\label{tbl:ngc5728PSfeatures}
	\begin{tabular}{@{}lrrrrr@{}}
		\toprule
		Feature    &  Mag & SMA (pc) & PA(\degr) & $\epsilon~~~$ & \textit{i}(\degr) \\ \midrule
		Inner Bar  &  9.0 &      370 &      82 &      0.446 &              56 \\
		Inner Ring & 10.5 &     1045 &      20 &      0.128 &              29 \\
		Main Disk  & 12.5 &     4800 &      27 &      0.452 &              57 \\
		Outer Ring & 14.0 &    13200 &      32 &      0.580 &              65 \\ \bottomrule
	\end{tabular}
\end{table}

Fig. \ref{fig:ngc5728images24} shows the the \OIII{} MUSE flux image (see below) overlaid with the Chandra X-ray contours (acquired from the Chandra Source Catalog, \cite{Evans2010a}, using the \texttt{CSCView} tool\footnote{\url{http://cda.cfa.harvard.edu/cscview/}, Dataset ID ADS/Sa.CXO\#CSC/Reg/4077-1-P-2-0003}), and the VLA 6 cm contours. 

\begin{figure}[!htbp]
	\centering
	\includegraphics[width=.9\linewidth]{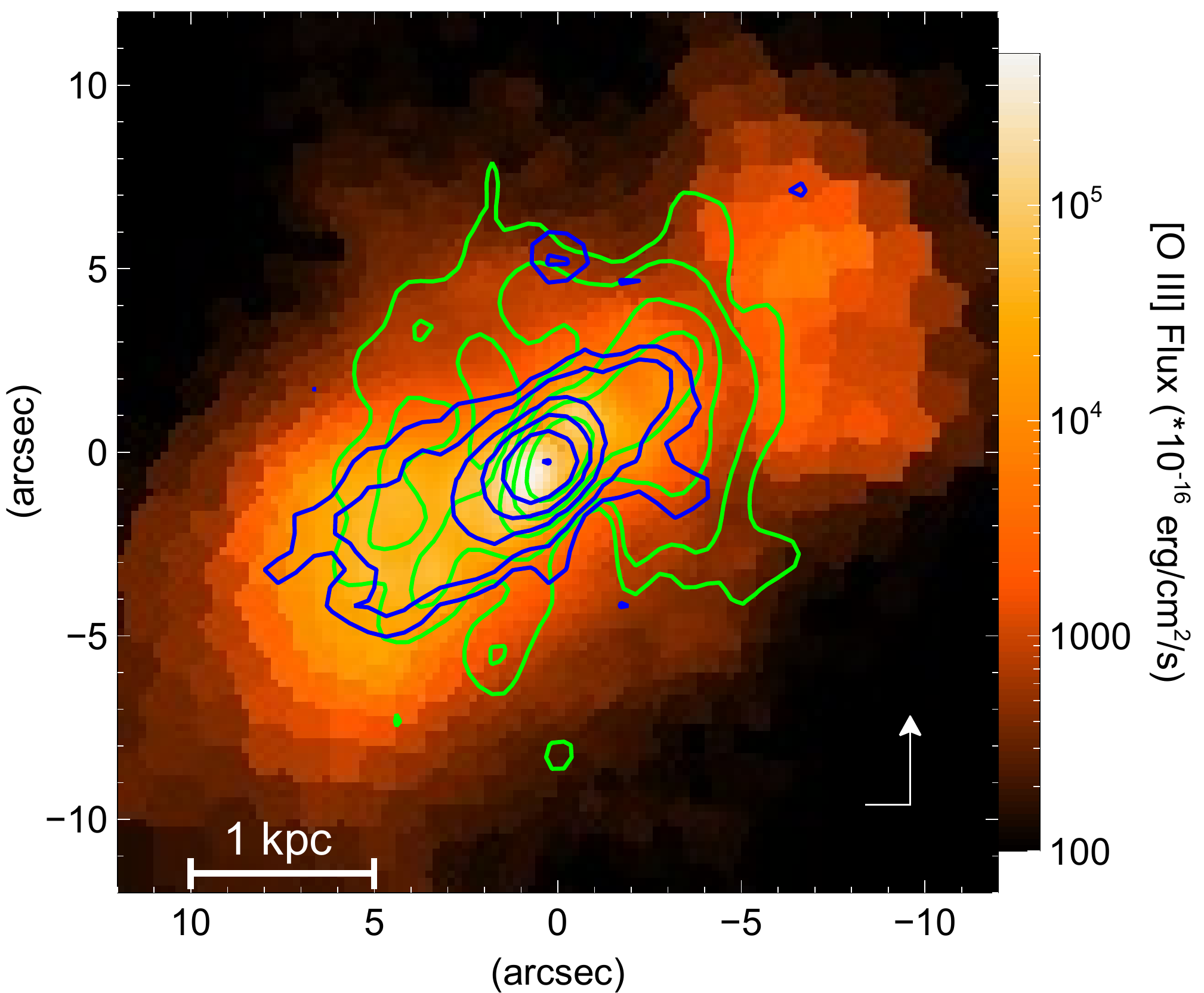}
	\caption{Chandra+VLA+\OIII. \OIII{} flux image (logarithmic scaling) from MUSE data (flux values as per color-bar), showing photo-ionized and shocked gas in the cones. Chandra (0.5--7 keV) smoothed flux contours (blue) with values of 0.1, 0.2, 0.5, 1, 2, and 5 $\times$ 10\pwr{-6} photons cm\pwr{-2} s\pwr{-1}; this shows the reflected AGN emission plus production from shocked gas. VLA 6 cm contours (green) flux levels at 0.2, 0.5, 1 and 2 mJy, delineating the synchrotron emission from the relativistic jet. The X-ray, radio, and emission-line gas structures are clearly aligned.}
	\label{fig:ngc5728images24}
\end{figure}

The X-ray and \OIII{} fluxes are co-spatial. The X-ray data have been smoothed by a Gaussian with a 2 pixel FWHM, to prevent pixellation. The X-ray orientation has the highest flux in the SE direction, counter to the radio jet; this can be explained by obscuration of the NW X-ray jet, plus relativistic beaming of the NW radio jet over the SE counter-jet. \cite{Rodriguez-Ardila2017}, studying NGC~1368, deduce that the extended X-ray emission in that object is caused by shocks greater than 200 \kms{} producing free-free emission; as will be seen in Paper II, the outflow velocities of NGC~5728 certainly exceed that value.
\subsection{The NIR and Optical Nuclear Spectrum}
\begin{figure*}[!htbp]
	\centering
	\includegraphics[width=1\linewidth]{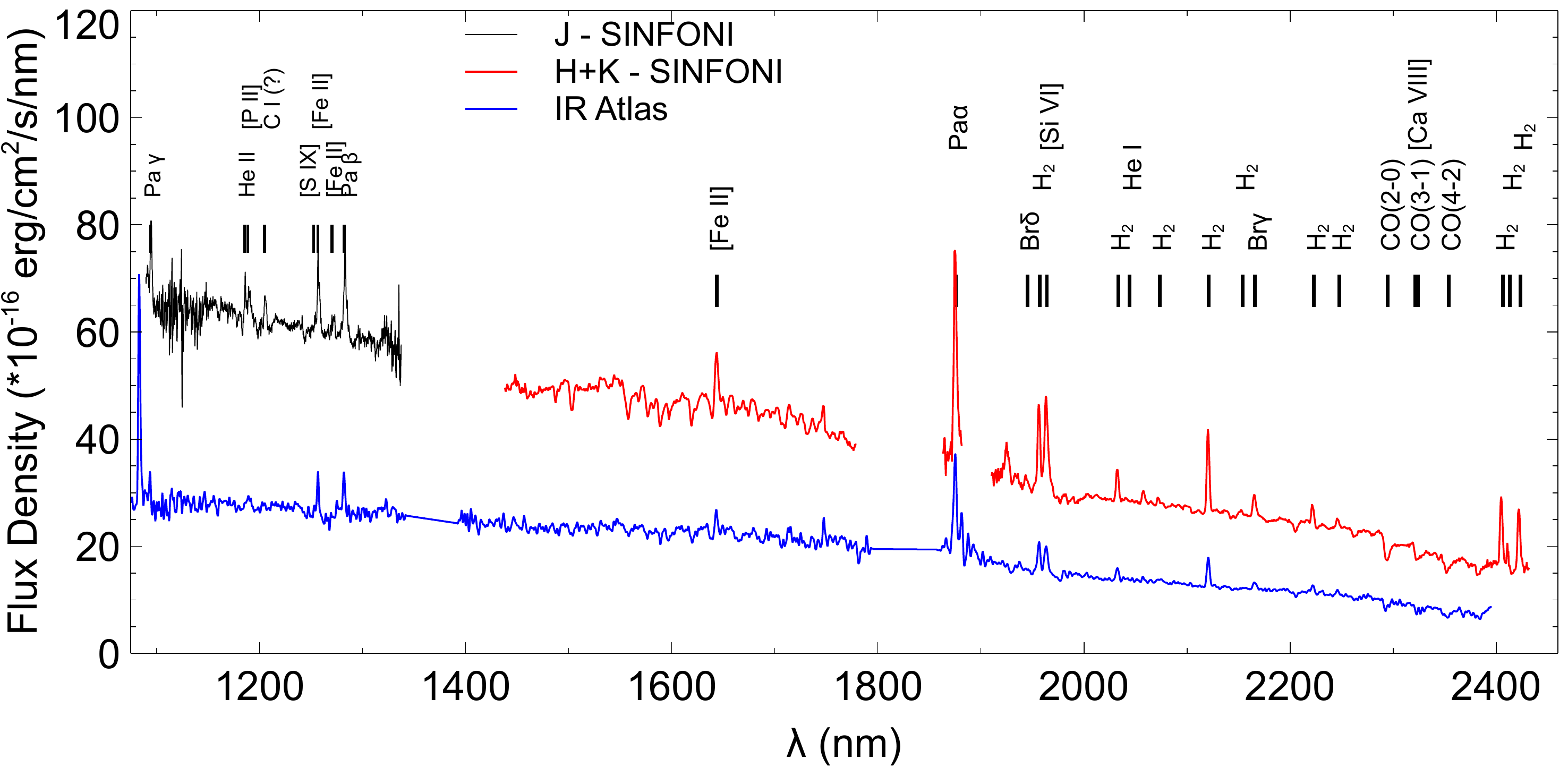}
	\caption{Nuclear spectrum from central 1\arcsec{} from SINFONI (\textit{J} and \textit{H+K}), plus the NIR atlas of \citep{Riffel2006}. Emission lines and CO absorption band-heads are marked. The NIR atlas spectrum is rescaled and offset for clarity. All spectra are reduced to rest-frame.}
	\label{fig:ngc5728nuclearspectrum1}
	\centering
	\includegraphics[width=1\linewidth]{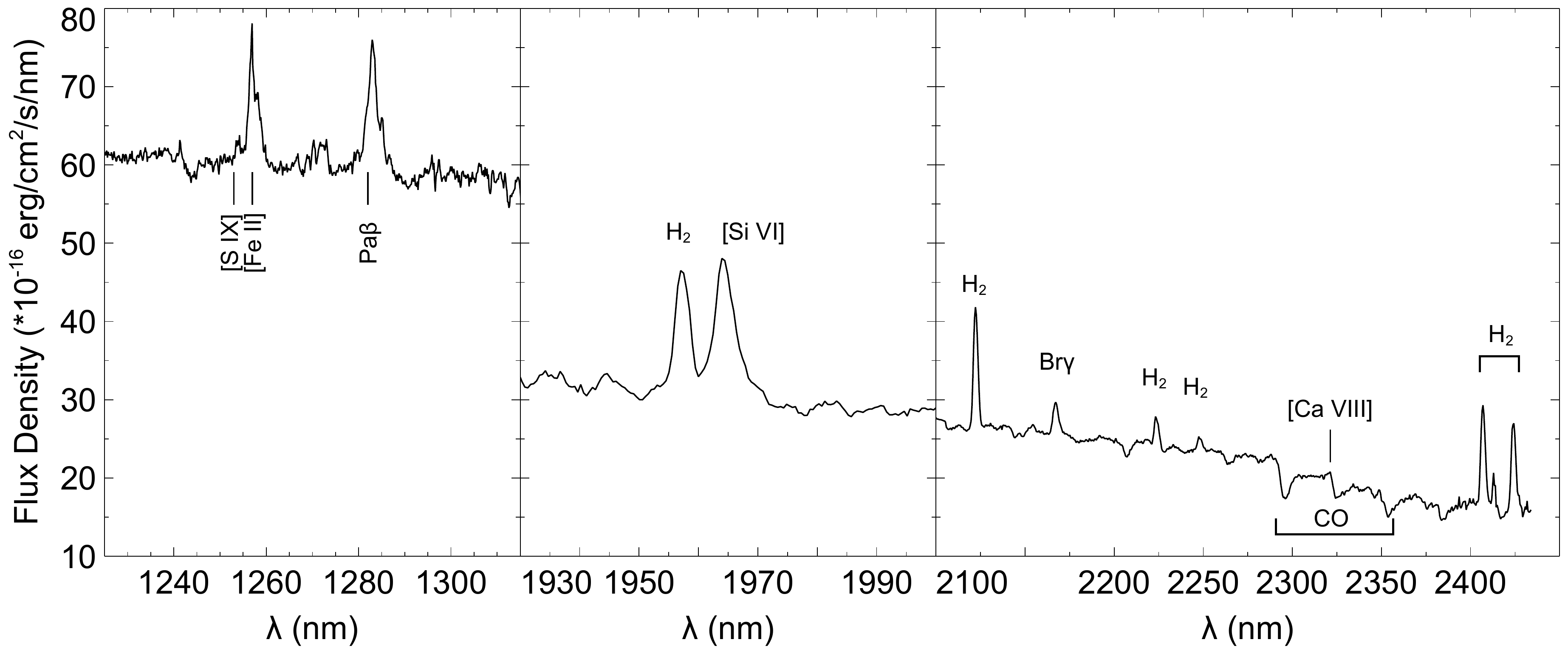}
	\caption{Enlarged SINFONI spectra from Fig. \ref{fig:ngc5728nuclearspectrum1}, showing the main lines.  Spectra are reduced to rest-frame.}
	\label{fig:ngc5728nuclearspectrum2}
\end{figure*}

\begin{figure*}[!htbp]
	\centering
	\includegraphics[width=1\linewidth]{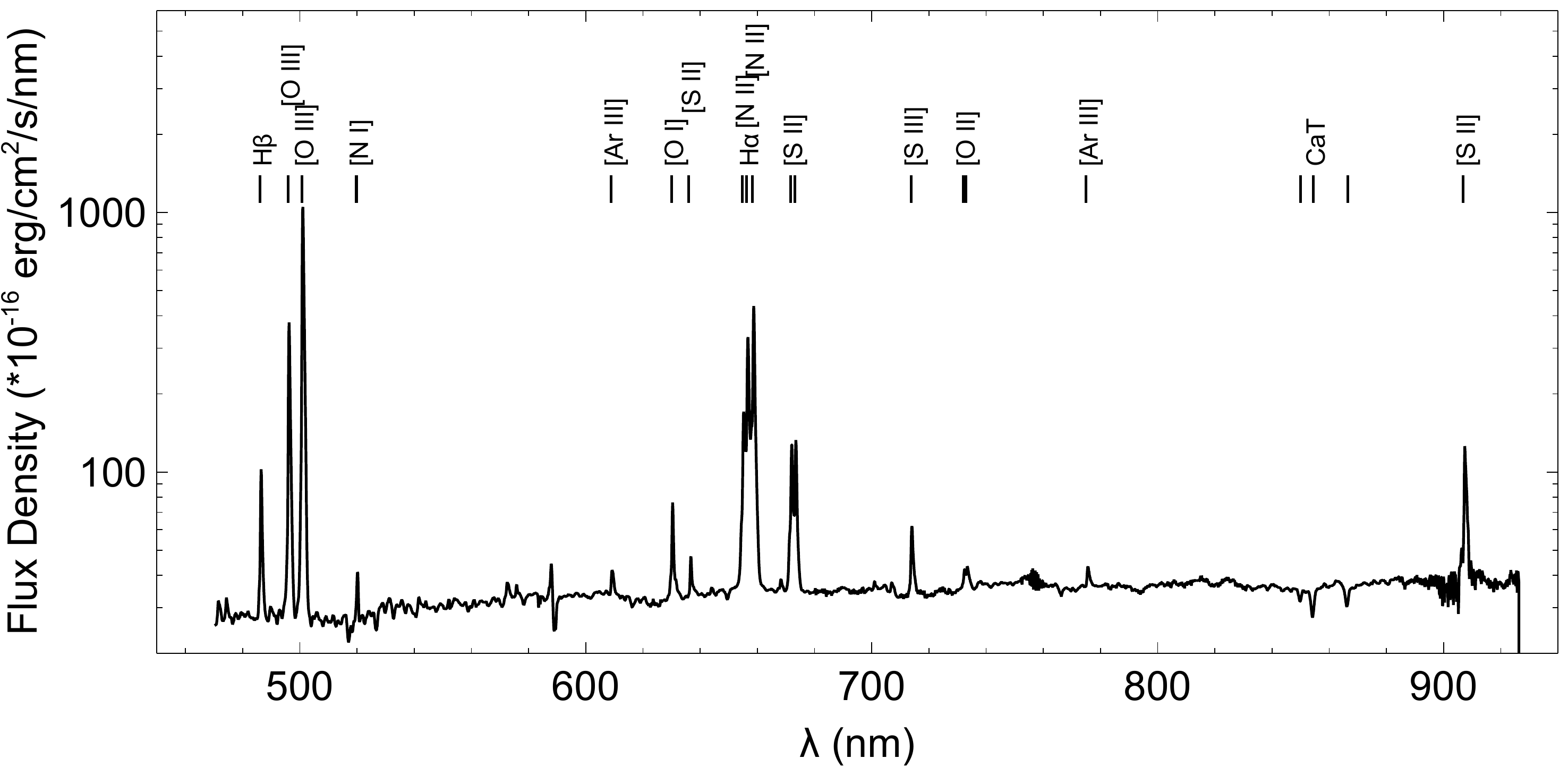}
	\caption{MUSE spectra of central 1\arcsec, showing the main lines. The spectrum is reduced to rest-frame. The flux is log-scale plotted because of the large dynamic range and to enhance weak lines. The \SII{} (679 nm) and the \OII{} (740 nm) are doublets. The Ca triplet (CaT) is also indicated.}
	\label{fig:ngc5728nuclearspectrum3}
\end{figure*}

The nuclear spectrum was obtained by integrating the flux in a circular aperture of radius 0\arcsec.5 around the brightest pixel in the continuum image for the SINFONI and MUSE data cubes. Fig. \ref{fig:ngc5728nuclearspectrum1} gives the spectra for the whole NIR wavelength range, showing the good flux calibration between the data cubes, taken at different dates and instruments. For comparison, the spectrum from the 0.8--2.4 \um{} atlas of AGN \citep{Riffel2006} is also plotted (rescaled for clarity). Fig. \ref{fig:ngc5728nuclearspectrum2} shows the detail for the main emission lines of interest. The optical MUSE nuclear spectrum is presented in Fig. \ref{fig:ngc5728nuclearspectrum3}, showing the Seyfert 2-like narrow emission lines of hydrogen, oxygen, nitrogen, and sulfur. It also shows smooth flux continuity with the NIR spectrum.


\subsection{Continuum Emission}
\label{sec:ngc5728continuum}
\begin{figure*}[!htbp]
	\centering
	\includegraphics[width=1\linewidth]{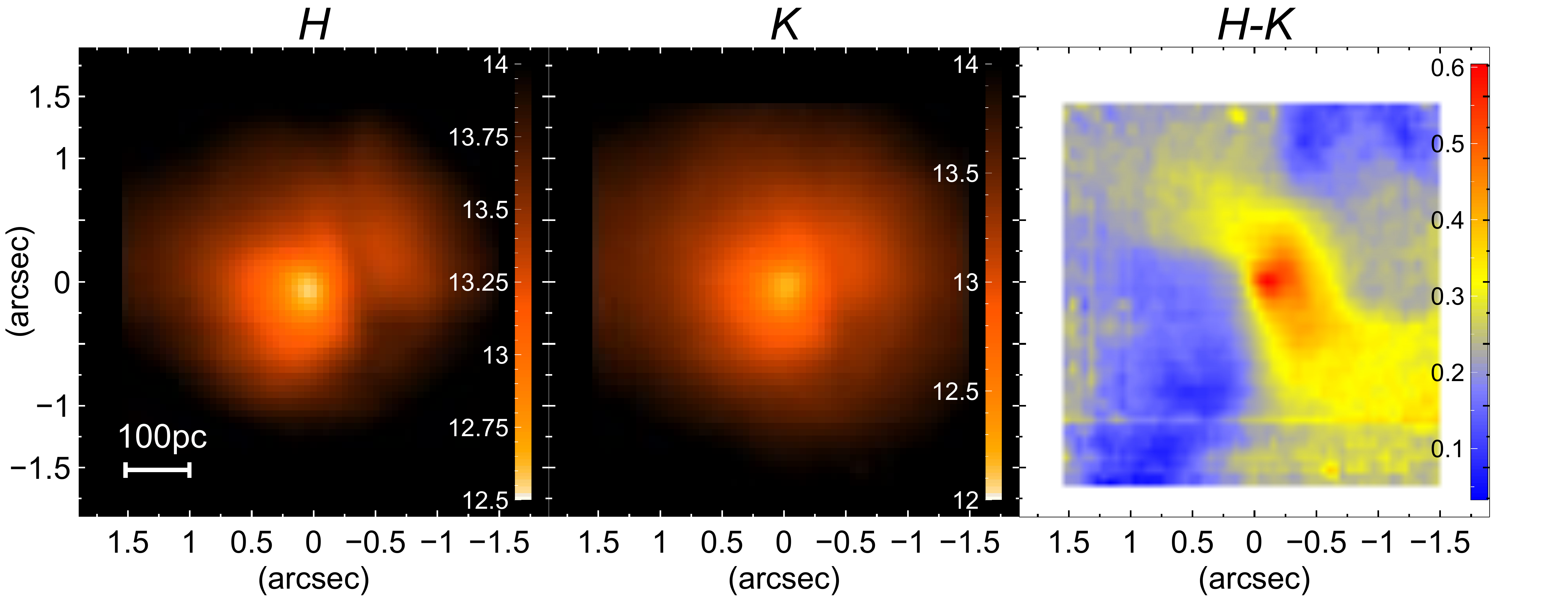}
	\caption{\textit{H} and \textit{K} surface magnitude and \hk{} color map, as labeled. The magnitude colors are plotted with square-root scaling, to enhance contrast. The color-bar values are in mag arcsec\pwr{-2}. The \hk{} plot shows the AGN equatorial obscuration.}
	\label{fig:ngc5728continuum}
\end{figure*}

We can examine the nuclear structure, stellar populations, and obscuring dust using continuum imagery. As the \textit{J} band cube has poorer observational resolution (0\arcsec.25 with no AO), we will only use the \textit{H+K} cube (0\arcsec.05 with AO). The \textit{H} and \textit{K} continuum magnitude images are derived from the cube by measuring the average flux over 10 nm around the 2MASS filter effective wavelengths (\textit{H} = 1662 nm, \textit{K} = 2159 nm), converting to mag arcsec\pwr{-2} for each pixel, and deriving the color \hk{}. Fig. \ref{fig:ngc5728continuum} presents the magnitude and color maps for the central 3\arcsec.8 ($760\times760$ pc).

The dust-lane (starting at $\Delta$RA=+1\arcsec, $\Delta$Dec=0\arcsec, ending at $\Delta$RA=-0\arcsec.5, $\Delta$Dec=-0\arcsec.5) shows up well in the \textit{H}-band magnitude map; as expected, it is less prominent in the \textit{K}-band. The central region is very red, \hk{} = 0.6 mag. The location of the peak of the central object appears to shift by $\sim$0\arcsec.085 between \textit{H} and \textit{K}. This is because of the increased dust penetration at longer wavelengths. The AGN is presumed to be located behind this dust lane; the precise location is described in Paper II and is plotted in this paper.

The \hk{} image shows a broad bar aligned at PA = 40\degr--220\degr, roughly 90\degr{} to the line of the cones. This is caused by the hot dust in the equatorial plane of the AGN, which also is the source of the very red colors at the nucleus; this is the equatorial toroidal obscuration component. The \hk{} image also shows reduced obscuration in the SE and NW quadrants; the edges of these (at $<0.2$ mag) align with the edges of the cones (as revealed by the channel maps presented in Paper II), especially in the SE.

\begin{figure*}[!htbp]
	\centering
	\includegraphics[width=.8\linewidth]{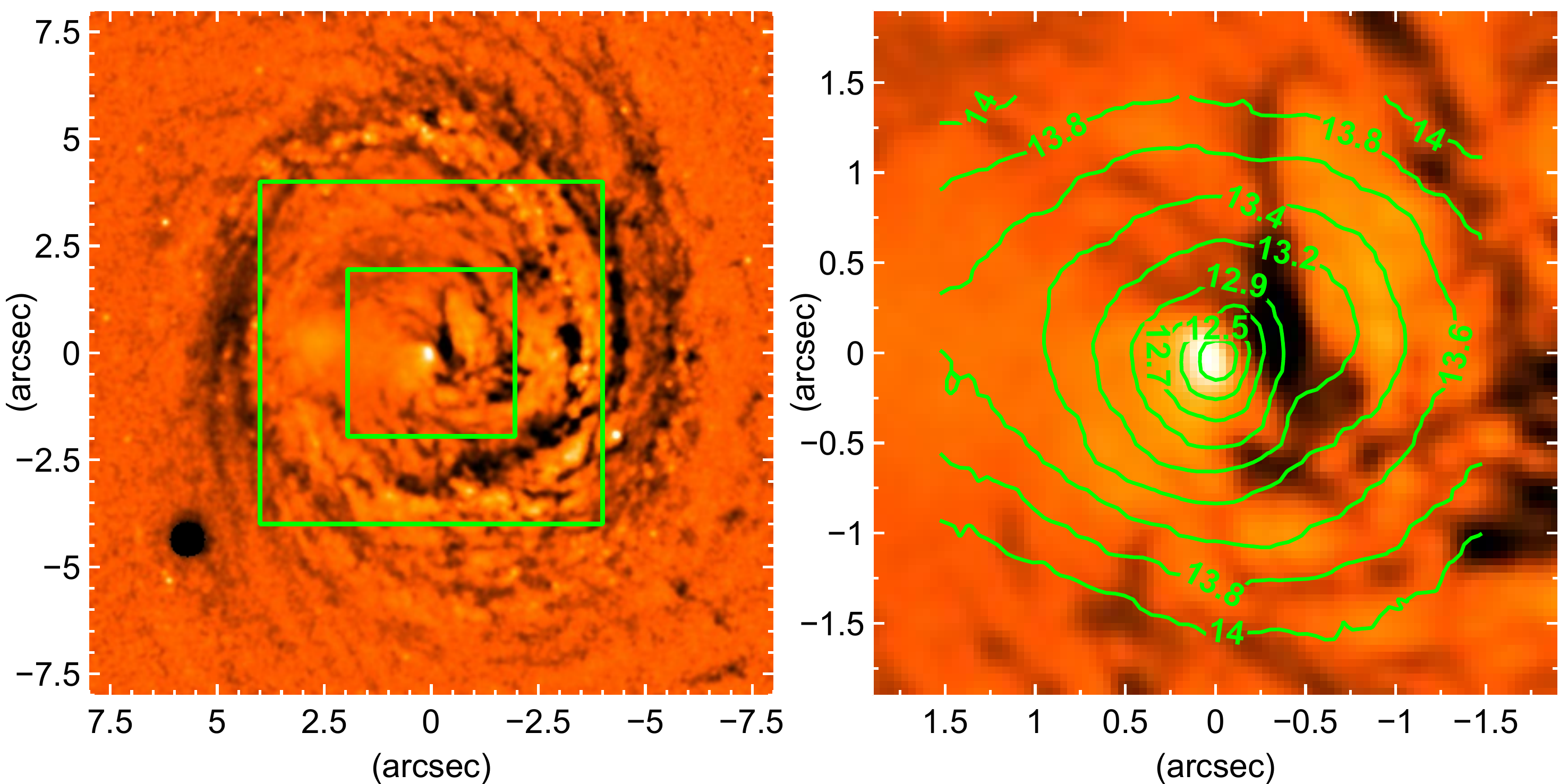}
	\caption{Left panel: \textit{HST} F814W structure map, showing the feeding spirals, the bright central nucleus, and the star-forming regions (bright points aligned with the outer edges of the dust lanes). The light green rectangles are the fields of view of the SINFONI \textit{H+K} (inner) and \textit{J} (outer) observations. Right panel: Central 3\arcsec.8 (760 pc) of the structure map overlaid with the \textit{K} magnitude contours; the \hk{} obscuration matches the feeding filament.}
	\label{fig:ngc5728images3}
\end{figure*}

Fig. \ref{fig:ngc5728images3} (left panel) shows the structure map for the central 16\arcsec{} of the \textit{HST} F814W image, clearly showing the feeding filaments in the inner spiral structure. The right panel then overlays the \textit{K}-band magnitudes contour over the central 3\arcsec.8. The dark lanes are seen as the extension into the center of the feeding filaments.

\subsection{Gas Line Emission}
Emission lines diagnose various physical parameters in the interstellar medium, including star formation and outflow dynamics, as well as the physical processes of excitation. Table \ref{tbl:ngc5728_Line_Flux} shows the flux (with uncertainty) for the observed emission lines in the central 1\arcsec{} of the nucleus, from the \textit{J}, \textit{H+K} and MUSE cubes. It will be noted that the observed central wavelengths do not exactly correspond to the red-shifted air wavelengths, especially in the NIR. This comes from the biased wavelengths of the multiple kinematic components, where the receding gas dominates over the approaching gas; with the ionized species lines in the cones (hydrogen recombination, \Fe, and \SiVI) show a recession velocity 150--250 \kms{} over systemic. The \Htwo{} lines, which are not kinematically or spatially associated with the ionization cones, have wavelengths closer to the systemic values.
\begin{table}[!htbp]
	\centering
	\caption{Integrated flux for emission lines for the central 1\arcsec~(0\arcsec.5~radius) aperture, with their respective uncertainties ($\Delta$F). Flux values are in 10\pwr{-16} \ecs. Measurements are grouped by data source.}
	\label{tbl:ngc5728_Line_Flux}
	\begin{tabular}{@{}lrrrr@{}}
		\toprule
		\textbf{Data Source}          &                 &                 &         &           \\
		Species                       & $\lambda$ (Obs) & $\lambda$ (Air) &       F & $\Delta$F \\ \midrule
		\textbf{MUSE}                 &                 &                 &         &           \\
		\Hb                           &           491.1 &           486.1 &   75.72 &      3.11 \\
		\OIII                         &           500.9 &           495.9 &  349.15 &     20.90 \\
		\OIII                         &           505.8 &           500.7 & 1039.59 &     53.67 \\
		\NI                           &           525.0 &           519.8 &   11.99 &      1.60 \\
		\ArIII                        &           615.1 &           608.9 &   12.91 &      1.27 \\
		\OI                           &           636.3 &           630.0 &   41.17 &      1.40 \\
		\SII                          &           642.8 &           636.0 &   13.07 &      0.94 \\
		\NII                          &           661.4 &           654.8 &  129.80 &      12.0 \\
		\Ha                           &           662.9 &           656.3 &  357.30 &     13.31 \\
		\NII                          &           664.9 &           658.3 &   462.6 &      12.3 \\
		\SII                          &           678.3 &           671.6 &  103.63 &      3.23 \\
		\SII                          &           679.3 &           673.1 &  101.66 &      2.97 \\
		\ArIII                        &           720.8 &           713.8 &   36.57 &      1.78 \\
		\OII                          &           739.3 &           732.0 &    5.81 &      0.88 \\
		\OII                          &           740.4 &           733.0 &   11.81 &      1.16 \\
		\ArIII                        &           783.0 &           775.0 &   10.49 &      0.78 \\
		\SIII                         &           916.1 &           906.9 &  128.00 &      6.20 \\ \midrule
		\textbf{SINFONI \textit{J}}   &                 &                 &         &           \\
		\pag                          &          1105.3 &          1094.1 &   13.94 &      1.01 \\
		\HeII                         &          1197.2 &          1185.5 &    7.78 &      1.28 \\
		\PII                          &          1201.3 &          1188.6 &   19.01 &      2.20 \\
		C I (?)                       &          1216.7 &          1204.9 &    9.77 &      0.98 \\
		\Fe                           &          1269.4 &          1256.7 &   39.47 &      2.57 \\
		\pab                          &          1295.3 &          1282.2 &   40.88 &      1.64 \\ \midrule
		\textbf{SINFONI \textit{H+K}} &                 &                 &         &           \\
		\Fe                           &          1660.4 &          1643.6 &   35.88 &      3.97 \\
		\Htwo{} 2-1 S(5)              &          1964.2 &          1944.9 &    8.26 &      2.06 \\
		\Htwo{} 1-0 S(3)              &          1976.3 &          1957.6 &   48.96 &      2.18 \\
		\SiVI                         &          1983.5 &          1964.1 &   77.10 &      2.72 \\
		\Htwo{} 1-0 S(2)              &          2053.2 &          2033.8 &   18.10 &      0.62 \\
		\HeI                          &          2078.8 &          2059.5 &    7.45 &      0.71 \\
		\Htwo{} 2-1 S(3)              &          2093.3 &          2073.5 &    3.27 &      0.60 \\
		\Htwo{} 1-0 S(1)              &          2142.1 &          2121.8 &   45.56 &      2.13 \\
		\Htwo{} 2-1 S(2)              &          2174.5 &          2154.2 &    4.37 &      1.12 \\
		\brg                          &          2187.3 &          2166.1 &   17.94 &      1.09 \\
		\Htwo{} 3-2 S(3)              &          2221.5 &          2201.4 &    0.56 &      0.25 \\
		\Htwo{} 1-0 S(0)              &          2244.5 &          2223.5 &   10.58 &      0.90 \\
		\Htwo{} 2-1 S(1)              &          2269.1 &          2247.7 &    6.44 &      0.51 \\
		\Htwo{} 1-0 Q(1)              &          2429.6 &          2406.6 &   41.15 &      1.79 \\
		\Htwo{} 1-0 Q(2)              &          2435.6 &          2413.4 &    7.80 &      1.36 \\
		\Htwo{} 1-0 Q(3)              &          2446.8 &          2423.7 &   36.49 &      1.82 \\ \bottomrule
	\end{tabular}
\end{table}

We derive maps of the gas emission fluxes from the \textit{velmap} (velocity map) procedure in \texttt{QFitsView} on the SINFONI data cubes (\textit{J} and \textit{H+K}). This procedure fits a Gaussian curve for every spaxel at the estimated central wavelength and full width half maximum (FWHM). This generates maps of the best fit continuum level (C), height (H), central wavelength ($\lambda$), and FWHM. These are readily converted to the dispersion ($\sigma$), the flux (\textit{F}) and the emission equivalent width (\textit{EW}). Maps were obtained for the species \pab~(1282 nm), \brg~(2166 nm), \Fe~(1257 and 1644 nm), \Htwo~(2121 nm) and \SiVI~(1964 nm). We note that the \brg{} and \SiVI{} maps have already been presented in \cite{Shimizu2018}, in the context of reevaluating the broad \Ha{} component for Seyfert 1.9 classifications, where misclassification is caused by the presence of strong, AGN-driven, outflows.

Map values are accepted or rejected based on ranges on each of the derived parameters produced by the procedure, based on visual inspection of the fluxes and kinematic structures i.e. the derived values of continuum, line flux, LOS velocity or dispersion have to be within certain values. Single pixels may be rejected because of large excursions from their neighbors; the source of anomalous values is usually noise spikes in the data or low flux values, causing poor model fits. Rejected pixels are then interpolated over from neighbors or masked out, as appropriate.

The \SiVI{} maps were difficult to produce due to the close and strong \Htwo~(1957 nm) line; since the range of velocities over the whole field is fairly large, the \textit{velmap} routine was `locking on' to the wrong line in places. This was solved by creating the maps for the \Htwo{} line, then subtracting a Gaussian  fit at each spaxel from the fitted parameters; this had the effect of removing the \Htwo{} line from the data cube. Additionally the fitting procedure was done independently for the north and south halves of the data cube and then combined, as the \Htwo{} velocity field is oriented in that direction. The \SiVI{} line was then fitted successfully.

We also obtained emission-line maps from the MUSE data cube in a similar manner. To increase the signal-to-noise ratio (S/N), especially in the outer regions were the gas emission is weak, we use the `Weighted Voronoi Tessellation' (WVT) method \citep{Cappellari2003}, using the \textit{voronoi} procedure in \texttt{QfitsView}. This aggregates spatial pixels in a region to achieve a target S/N. This needs both signal and noise maps; these are obtained at each spaxel by the data and the square root of the variance from the MUSE data cube, both averaged along the spatial axis. After some experimentation, the target S/N was set to 60. We obtained maps of the species \Ha, \Hb, \OIII{} (500.7 nm), \SII{} (671.6, 673.1 nm), \NII{}(654.8 nm) and \OI{} (630.0 nm). These maps were masked to a minimum flux level for each species and to remove the bright star some 19\arcsec NE of the nucleus.

\subsubsection{Emission-line Morphology}
\label{sec:ngc5728elid}
\begin{figure*}[!htbp]
	\centering
	\includegraphics[width=.9\linewidth]{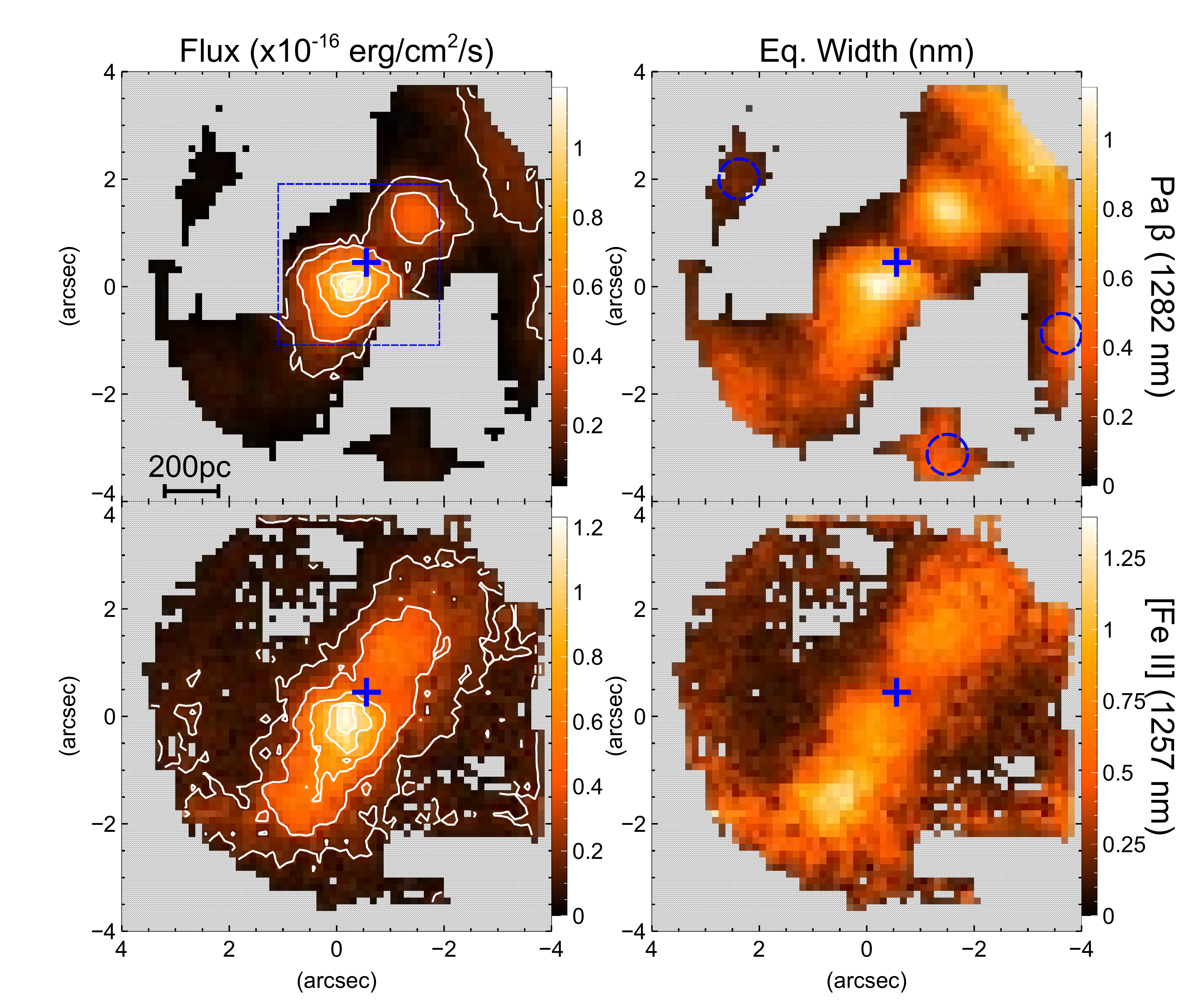}
	\caption{Ionization cone structure as shown by the flux and equivalent width values for \pab{} and \Fe{} (1257 nm). Flux values (shown in color-bar) in units of 10\pwr{-16} \ecs. Contours are 10, 25, 50, 75 and 90\%  of the maximum flux. Equivalent width values in nm. Blue circles on the \pab{} EW map are the locations for SF age measurement. The outer ring in the \pab{} maps traces the star-forming ring around the nucleus, \Fe{} traces the ionization cone boundaries. \added{The blue rectangle on the \pab{} flux map shows the FOV of the SINFONI \textit{H+K} filter observations.}}
	\label{fig:ngc5728gaskinematics1}
\end{figure*}

\begin{figure*}[!htbp]
	\centering
	\includegraphics[width=.75\linewidth]{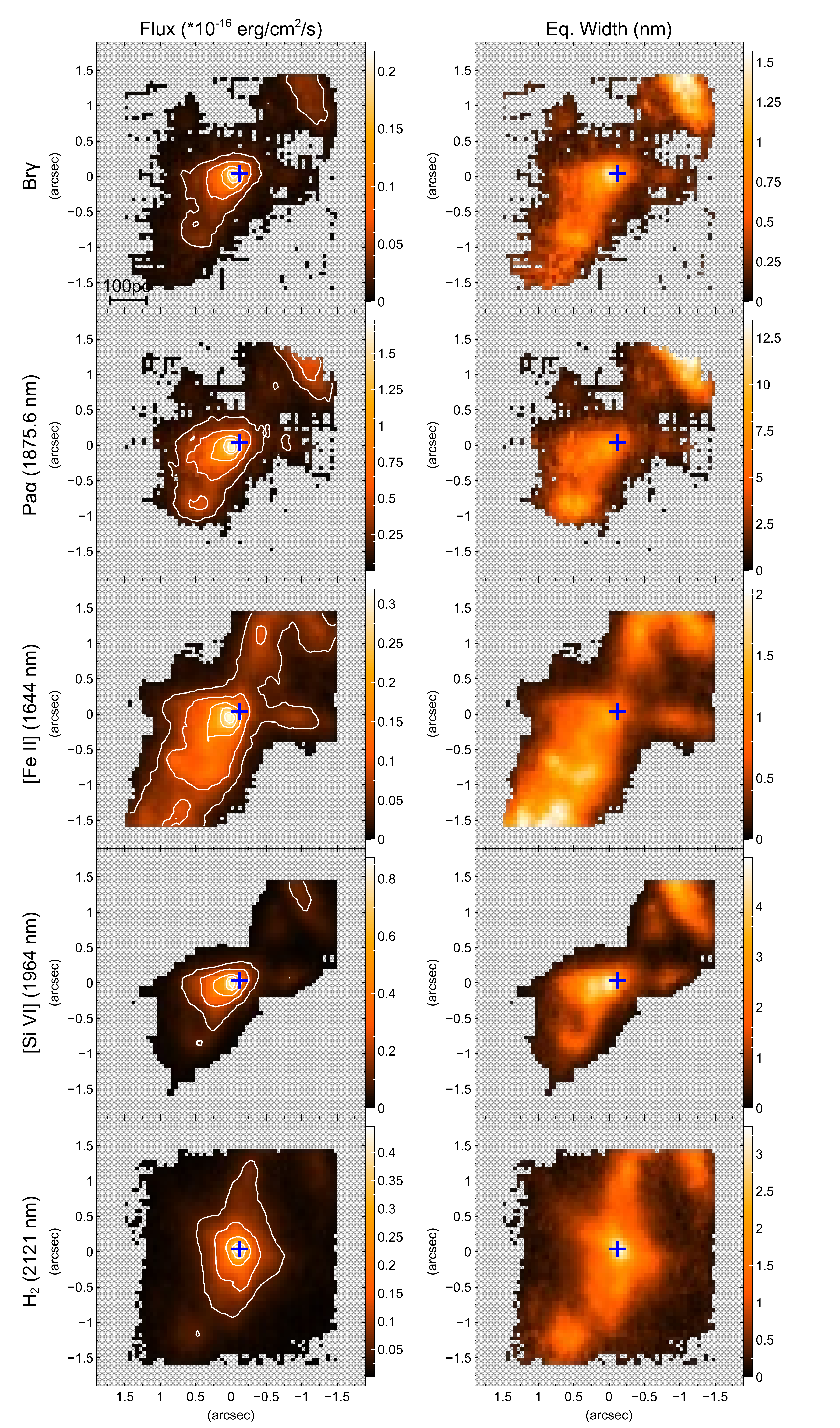}
	\caption{As for Fig. \ref{fig:ngc5728gaskinematics1}, but for \brg, \added{\paa,} \Htwo{} (2121 nm), \Fe{} (1644 nm), and \SiVI. The \SiVI{} has nearly the same extent as the hydrogen recombination emission. The star-forming ring is outside the scale of the maps.}
	\label{fig:ngc5728gaskinematics2}
\end{figure*}

\begin{figure*}[!htbp]
	\centering
	\includegraphics[width=0.9\linewidth]{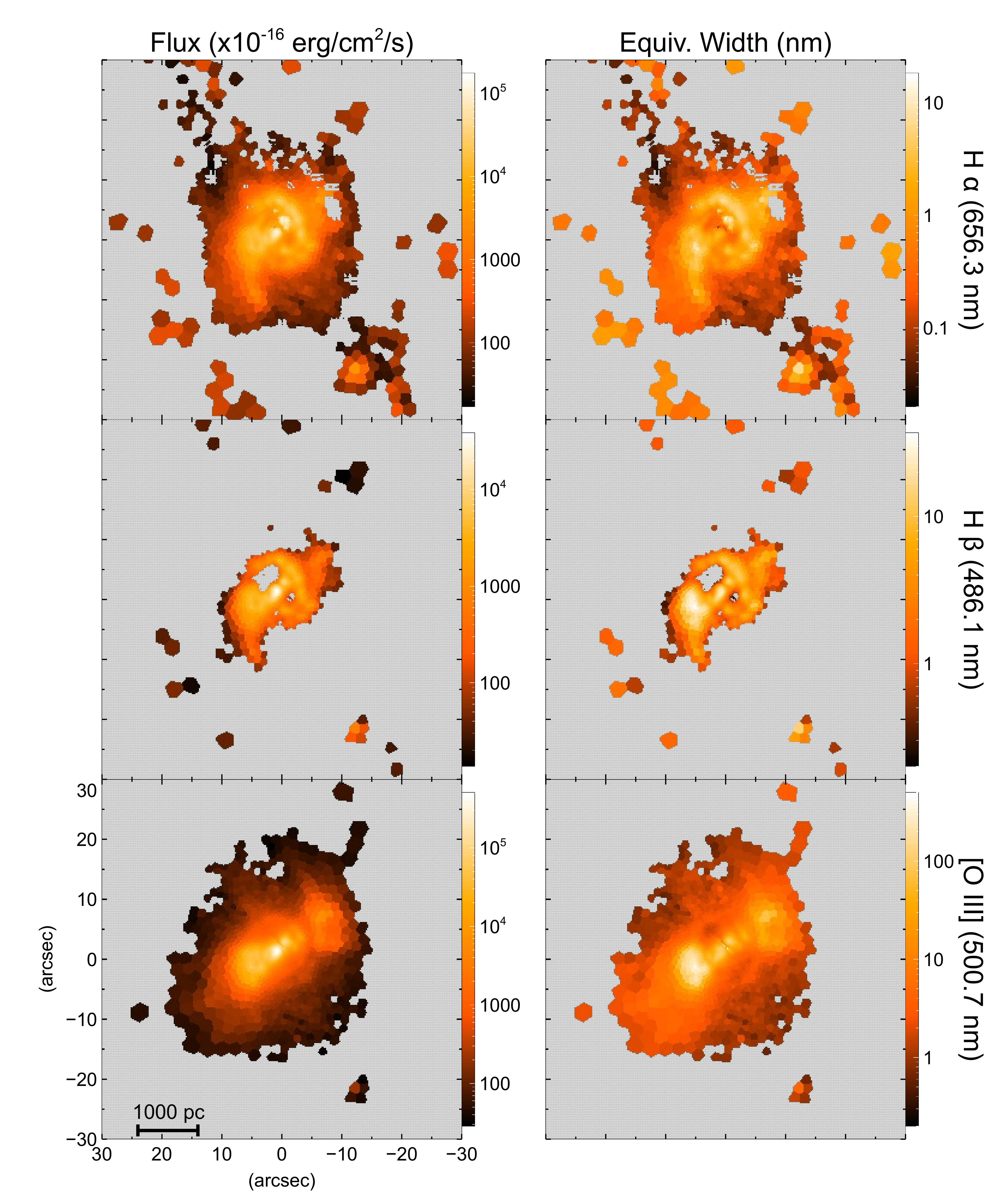}
	\caption{MUSE flux and EW maps for \Ha, \Hb{} and \OIII. The star-forming ring is prominent in \HII{} but is absent in \OIII. The flux values are plotted with logarithmic scale to enhance details. The ionization cones (most prominent in \OIII) can be traced up to 2.5 kpc from the nucleus.}
	\label{fig:ngc5728gaskinematics8}
\end{figure*}

\begin{figure*}[!htbp]
	\centering
	\includegraphics[width=0.9\linewidth]{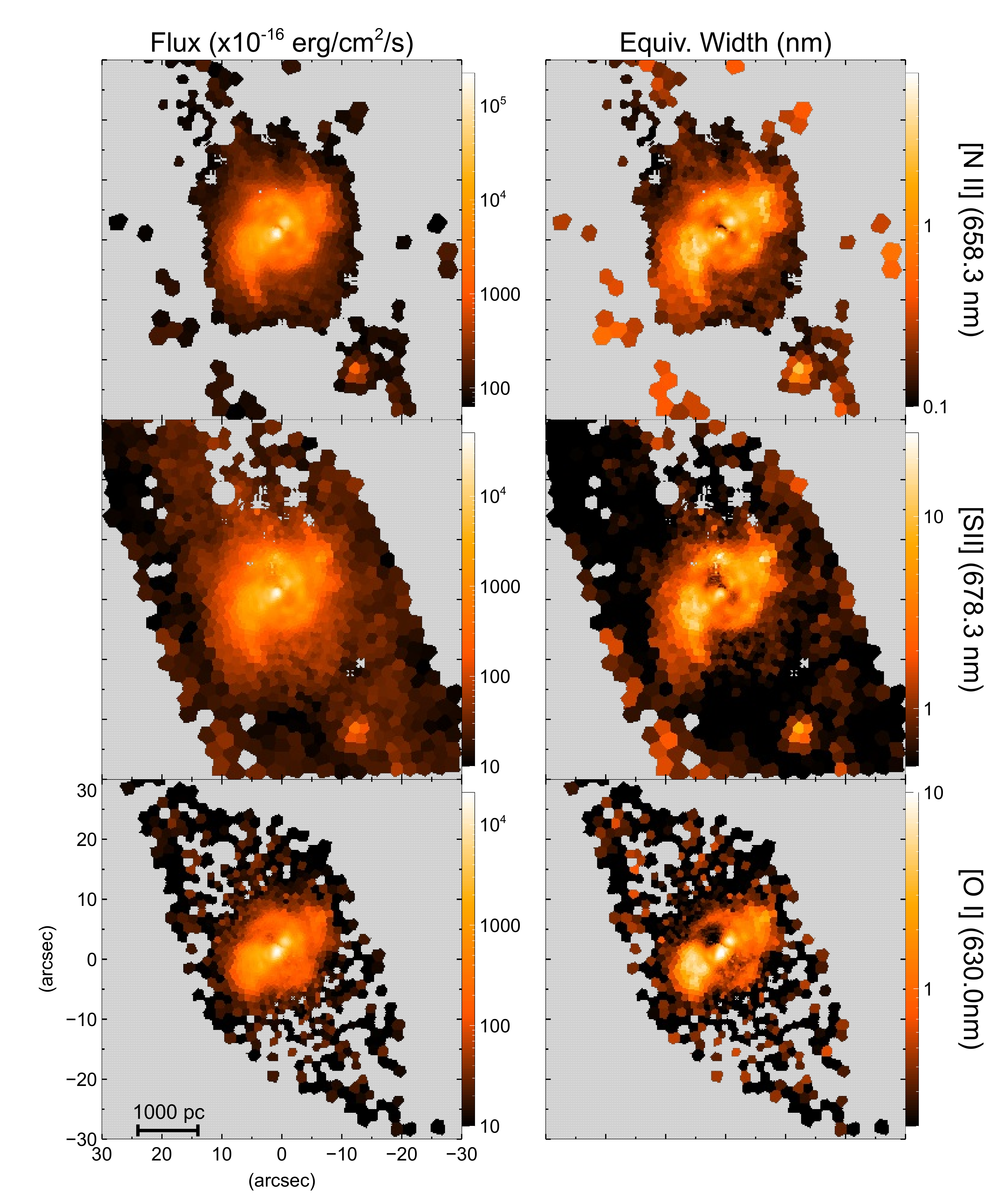}
	\caption{MUSE flux and EW maps for \NII, \SII{} and \OI. The star-forming ring is prominent in \HII{} but is absent in \OIII. The flux values are plotted with logarithmic scale to enhance details. The ionisation cones (most prominent in \OIII) can easily be traced up to 2.5 kpc from the nucleus.}
	\label{fig:ngc5728gaskinematics8a}
\end{figure*}

Figs. \ref{fig:ngc5728gaskinematics1}, \ref{fig:ngc5728gaskinematics2}, \ref{fig:ngc5728gaskinematics8} and \ref{fig:ngc5728gaskinematics8a} shows the flux and EW for each species. The \textit{J}-band maps have a spatial extent of 7\arcsec.5 square, as against the \textit{H+K}-band of 3\arcsec{} (1.5 vs. 0.6 kpc)\added{ (the FOV of \textit{H+K} is overplotted on the \pab{} flux map.)}. The hydrogen recombination emission, \Fe{} and coronal \SiVI{} lines all show similar structure of a biconal outflow with a PA = 140--320\degr{}; the equivalent width maps show this clearly (the EW is invariant to obscuration), displaying a distinct `waist' at the cone intersections \added{; this shows that there is significant stellar continuum in front of the emission and obscuration along our LOS.}. By contrast, the \Htwo{} is spatially distinct, presenting as a disk oriented roughly NS with `arms' trailing to the NW and SE.\added{ As we will show in Paper II, it is also kinematically distinct.}

The MUSE \Ha{} and \OIII{} maps show a similar picture, with a greater FOV of $30\arcsec\times30\arcsec$ ($6\times6$ kpc) (Fig. \ref{fig:ngc5728gaskinematics8}). The ionization cones can easily be traced with the \OIII{} emission up to 2.5 kpc from the nucleus. The \Ha{} emission also clearly shows the star-forming ring. Many \HII{} regions are visible in a ring of approximately 7 kpc radius around the nucleus; these trace the faint spiral arms from the outer ring to where they join the nuclear structure, especially SE of the SF ring, where there is a short extension pointing SW. 

\subsubsection{Coronal Line Emission}
\label{sec:ngc5728clr}
Emission lines of highly ionized species (e.g. \SiVI, \CaVIII, and \SIX) trace the direct photo-ionization either from EUV and soft X-rays from the AGN accretion disk, or by fast shocks. These species have ionization potentials (IP) up to several hundred eV, and are called coronal lines (CLs) and the emission locations are called coronal line regions (CLRs). This emission has the advantage of not being contaminated by photo-ionization from star formation i.e. it is diagnostic of AGN activity. For coronal species with an IP of $\sim100-150$ eV, the emission exhibits a similar morphology and kinematics to lower ionization lines such as \OIII; however for species with IP $>$ 250 eV, the emission is very compact.

\cite{Mazzalay2013} found for NGC~1068 that the morphology and kinematics were consistent with being driven by the radio jet, and that shock mechanisms were favored over photo-ionization. A similar conclusion was deduced for NGC~1368 by \cite{Rodriguez-Ardila2017}.

\begin{figure}[!htbp]
	\centering
	\includegraphics[width=.8\linewidth]{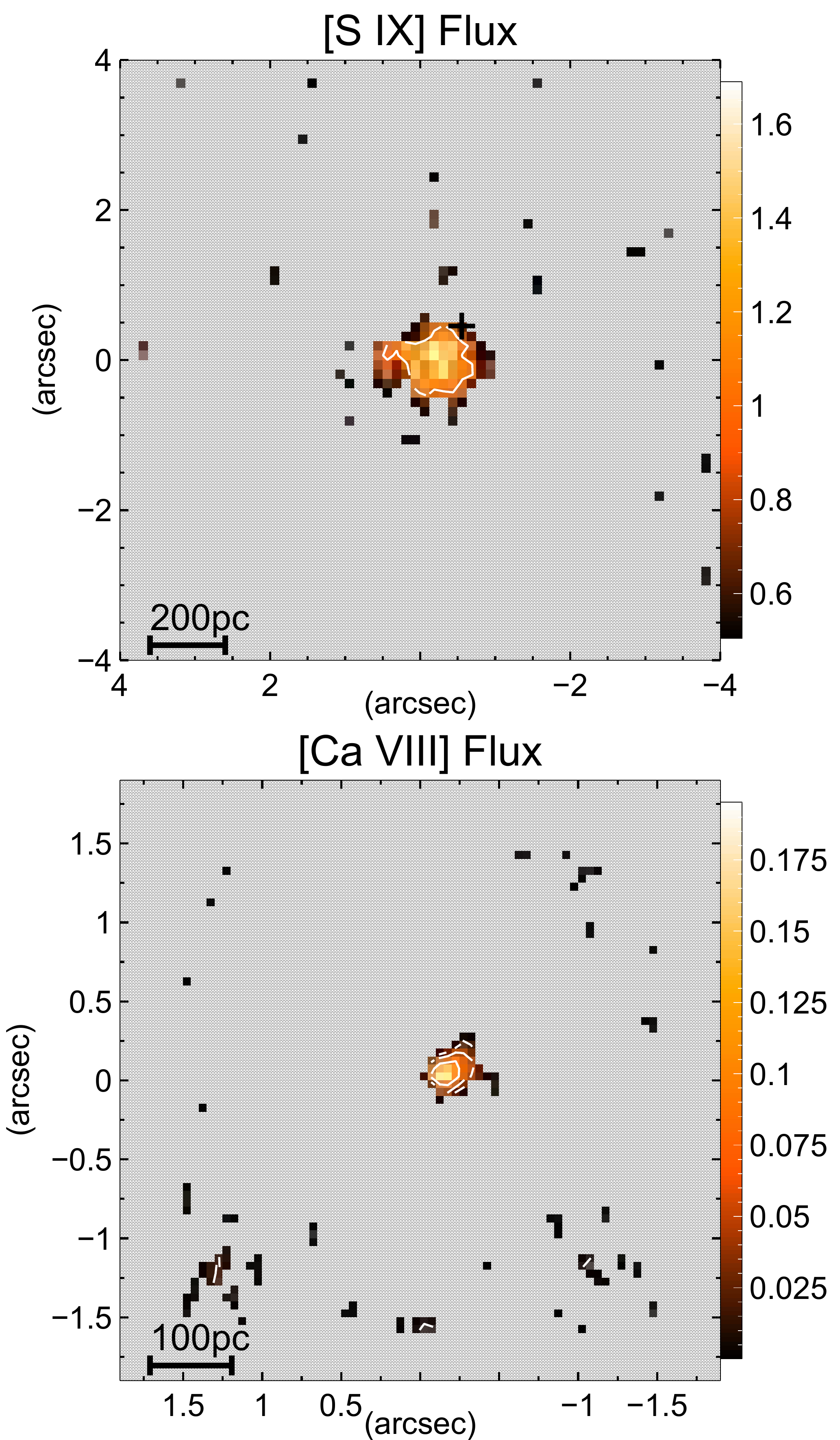}
	\caption{\SIX{} and \CaVIII{} emission line flux. \SIX{} contour levels at 0.2, 0.5, and 1, \CaVIII{} contour levels at 0.02, 0.05, and 0.1; all units are 10\pwr{-15} \ecs. The AGN is not marked on the \CaVIII{} map, as it is located at the flux maximum. Note that the plots are at different spatial scales. The very high excitation species are confined around the AGN.}
	\label{fig:ngc5728gaskinematics6}
\end{figure}

Apart from the \SiVI{} emission, we also detect weak \SIX{} and \CaVIII{} coronal lines. This is in contrast to  \cite{Rodriguez-Ardila2011}, who reported no detectable flux for those species for NGC~5728, in their coronal line study of 54 local AGN. We constructed flux maps these species by averaging the data cube along the spectral axis over the emission line profile, then subtracting the average continuum in a neighboring featureless spectral region; the result was multiplied by the line profile width to produce the flux. The weakness of the flux precluded using the standard \textit{velmap} procedure. Fig. \ref{fig:ngc5728gaskinematics6} plots the flux maps; the weakness of the lines will cause some uncertainties, so the flux values should not be taken as very accurate, however the structure is clear. 

The CLR, as delineated by the \SiVI{} emission, extends almost to full length of the NLR, out to 300 pc in each direction from the AGN. This indicates strongly collimated hard UV--soft X-ray flux from the AGN. While the \SiVI{} flux scales reasonably well with the \brg{} flux, there is an impression from the morphology that the production mechanism is different between the two cones; the SE emission is in a conal form, as would be expected from direct ionization from the AGN EUV field. By contrast, the NW emission outlines, rather than infills, the cone; this is clearer in the equivalent width plots rather than the flux plots. This indicates that shock mechanisms predominate in the NW.

The \SIX{} emission (IP = 379 eV, $\lambda~=~83.3$ nm) showing a small extension along the line of the SW cone, while the \CaVIII{} (IP = 147 eV, $\lambda~=~8.4$ nm) exhibits a more compact structure around the AGN location, with a possible extension along the NW cone; both of these extensions are close to the observational resolution limits. Neither of these species shows the same spatial extent as \SiVI{}; in the case of \CaVIII{} (which has a similar IP to \SiVI), the low line strength means that the flux level drops below the detectable threshold; in the case of \SIX{} an additional factor could be that the ISM fully absorbs the ionizing radiation closer to the AGN. We can also posit that the higher IP species are ionized directly by the AGN, while the \SiVI{} emission is also generated by shock excitation in the ionization cones, which do not have enough energy to excite the higher IP species.

\subsection{Locating the AGN}
\label{sec:ngc5728AGNLocn}
In Seyfert 2-type galaxies, the AGN is obscured by the dusty torus, so the small, unresolved BLR is not visible; this torus can extend from \replaced{1--100 pc}{0.1 to tens of parsecs \citep{Hopkins2006, Netzer2015,Elitzur2008}, depending on AGN luminosity}. The position of the AGN is usually taken as the brightest pixel in the continuum (on the assumption that there is a nuclear cluster), but in this case there is a dust lane that obscures this location, as seen on the \hk{} color map (Paper I). This lane is in addition to the supposedly compact dusty torus, as it extends across the field for $\sim400$ pc, and connects to the spiral feeding filaments. 

Presumably, the ionization cones have their origin (both positionally and kinematically) at the AGN central engine; however there are issues that will affect the symmetry of the observed flux and velocity fields:
\begin{itemize}
	\item Obscuration; the dust lane is on the approaching side; this will reduce the flux (and corresponding flux-weighted velocity).
	\item ISM impact; the SW velocity field shows an increase out from the AGN to a maximum of about 350 \kms{} at 300 pc from the AGN and then a deceleration to 150 \kms{} at 760 pc (projected distances); if the outflow encounters a more dense ISM (associated with increased obscuration), this will both reduce the distance to and the velocity of the maximum.
	\item Velocity bias; even without obscuration, an outflow that has a significant component of velocity along the LOS will have the approaching velocities biased towards zero, as described by \cite{Lena2015}, where approaching clouds preferentially show their non-ionized face to the observer; thus their line emission is attenuated by dust embedded in the cloud and the average velocity measured for the blue-shifted clouds is skewed toward smaller values.
\end{itemize}

\begin{figure}[!htpb]
	\centering
	\includegraphics[width=1\linewidth]{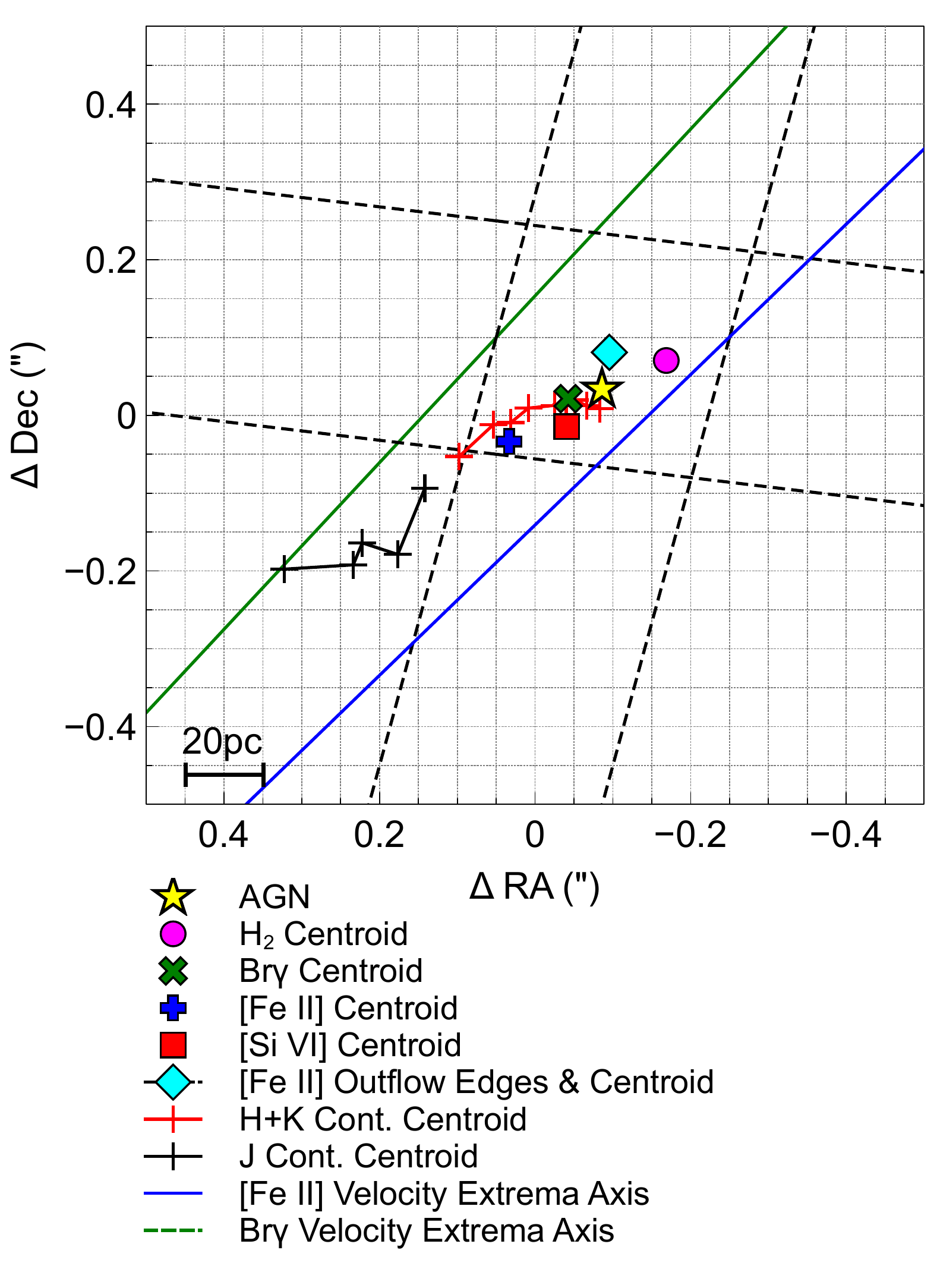}
	\caption{AGN location derived by different methods: see text for explanation. The outflow edges as delineated by the \Fe{} zero velocity channel maps are shown as black dotted lines, with the centroid shown as the cyan diamond. The centroids for the flux maxima for \textit{H+K} and \textit{J} cubes with wavelength are shown as read and black trajectories, respectively. The axes are the data cube spatial pixels; 1 pixel = 10 pc = 0\arcsec.05 projected distance; the grid lines are 1 pixel. The (0,0) position in RA, DEC is the brightest pixel of the \textit{H+K} image from the wavelength collapsed cube. The AGN location is taken as the average of the centroids of the \Htwo, \brg, \SiVI{} and continuum flux peaks, plus the outflow boundary intersection.}
	\label{fig:ngc5728velocitycenter}
\end{figure}

We deduce the position of the AGN by 3 different methods, each of which is plotted on Fig. \ref{fig:ngc5728velocitycenter}, which shows the magnified central region (1\arcsec~=~20 pixels a side). These are derived from:
\begin{itemize}
	\item The centroid/maximum of the \Htwo{} flux, assuming this is kinematically cold and settled around the center of mass of the nucleus. The \Fe{} (1644 nm), \Htwo{} and \brg{} flux maxima locations are also plotted for comparison.
	\item The \SiVI{} flux is presumably generated by direct photo-ionization from the AGN, and thus the peak value will be on the AGN location.
	\item The \textit{H+K}-band flux ratio of \Fe{} and \brg{} shows that the \Fe{} jackets the \brg{}, thus defining the edges of the cones. The cone edges are plotted as lines; these are derived from the \Fe{} zero velocity channel map (see Paper II).
	\item The continuum centroid positions are plotted for \textit{H+K} wavelengths at 1460~nm, then in 100~nm increments from 1600--2400~nm. This assumes that the flux peak moves closer to the AGN as the obscuration becomes less important at longer wavelengths.
	\item The lines that joins the peaks of the velocity extrema (maximum and minimum) for \Fe{} (1644 nm) and \brg{} are also plotted - there is some uncertainty in defining these lines, as the extrema locations are not well defined.
\end{itemize} 

All points are assumed to have a positional error of $\pm$1 pixel. The \Htwo{} flux, the continuum centroid trajectory and the outflow edges all seem to align to within $\pm2$ pixels. 
The final AGN location is taken to be the centroid of the locations of the \Htwo, \brg{} and \SiVI{} flux maxima centroids and the \Fe{} bicone center, plus the location of the maximum of the image at 2400 nm; this is marked as the yellow star in Fig. \ref{fig:ngc5728velocitycenter}. We did not include the \Fe{} flux maximum, as this is only excited in low ionization regions, i.e. away from the central AGN source. If we assume the AGN location is the same as the \textit{Chandra} compact, 2--7 keV X-ray source, this is RA~=~14h~42\arcmin~23.883\arcsec,  DEC~=~-17\degr~15\arcmin~11.25\arcsec{} \citep[with typical 0\arcsec.6 precision]{Evans2010a}.

We can match the AGN location on the \textit{J}-band data cube by following the trajectory of the flux peak with wavelength. This is plotted at 1140~nm, then in 50~nm increments from 1200--1350~nm. We equate the continuum centroid at 1350~nm in \textit{J} as the same as 1460~nm in \textit{H+K}, displaced to follow the trend of the trajectory with wavelength. These positions are shown on previous and subsequent maps with a `+' symbol. 

\added{Further support for the determination of the AGN location is given by the close alignment with the stellar velocity kinematics and the maximum velocity gradient of the gas kinematics.}
\section{Discussion}
\subsection{Gaseous Excitation}
\label{sec:ngc5728excitation}
\subsubsection{Infrared Diagnostics}
\label{sec:ngc5728diagir}
Excitation mechanisms broadly fall into two categories (1) photo-ionization by a central, spectrally hard radiation field (from an AGN accretion disk) or by young, hot stars (UV ionization and recombination or collisional heating) or (2) thermal heating, which can be either by shocks (from AGN outflows, supernova remnants, star formation, asymptotic giant branch stellar winds or radio jet interaction) or UV/X-ray heating of gas masses. \Htwo{} is also excited by UV pumping and fluorescence \citep{Black1987}. The emission line spectra flux ratios enable determination of the particular or mixture of mechanisms. We derive excitation mode diagnostic diagrams from the emission line ratios \Fe{} 1257 nm/\pab{} and \Htwo/\brg. Nuclear activity for NIR emission line objects can be categorized by a diagnostic diagram \citep{Larkin1998,Rodriguez-Ardila2005}, where the log of the flux ratio of \Htwo/\brg{} is plotted against that of \Fe (1257 nm)/\pab{}. This is analogous to the BPT diagrams \citep{Baldwin1981} commonly used in the optical regime \citep[e.g.][]{Kewley2006}. Following the updated limits from \cite{Riffel2013a}, the diagram is divided into three regimes for star-forming (SF) or starburst (SB) (dominated by \HII{} regions), AGN i.e. subjected to the radiation field from the accretion disk and low-ionization nuclear emission line region (LINER) excitations, where shocks from SN and evolved stellar outflows dominate. Transition Objects (TOs) have a mixture of excitations, and can be sub-divided by the diagnostic ratios into those where \Fe{} dominates and those where \Htwo{} dominates.

As we do not have the \textit{J}- and \textit{K}-band data at the same spatial scale, the \textit{J}-band diagnostic ratio (\Fe{} 1257 nm/\pab) is re-calibrated to the  ratio (\Fe{} 1644 nm/\brg) in the \textit{H+K}-band, using the ratios (\Fe{} 1257/1644) and (\pab/\brg) as given in Table \ref{tbl:NGC5728ExcitationDiag}. 
\begin{table}[!htbp]
	\centering
	\caption{Infrared excitation mode diagnostic diagram regimes flux ratios.}
	\label{tbl:NGC5728ExcitationDiag}
	\begin{tabular}{lccc}
		\toprule
		Excitation Mode & \Htwo{}/\brg & \Fe/\pab & \Fe/\brg \\
		&              &  (1257 nm)  & (1644 nm)  \\ \midrule
		SF/SB           &   $<$ 0.4    &   $<$ 0.6   &  $<$ 2.6   \\
		AGN             &   0.4 -- 6   &  0.6 -- 2   & 2.6 -- 8.6 \\
		LINER           &    $>$ 2     &    $>$ 2    &  $>$ 8.6   \\ \bottomrule
	\end{tabular}
\end{table}
The diagnostic emission lines are convenient; the pairs are close together in wavelength, removing the dependency on calibration accuracy and differential extinction. The table has been supplemented with the expected ratio regimes for \Fe{} 1644 nm/\brg{} for \textit{H} and \textit{K} combined spectra, taking the ratios \pab/\brg{} ~=~5.88 and \Fe{}~1257/1644~nm~=~1.36. If this ratio is used, the fluxes must be corrected for extinction.

\begin{figure*}[!htbp]
	\centering
	\includegraphics[width=.9\linewidth]{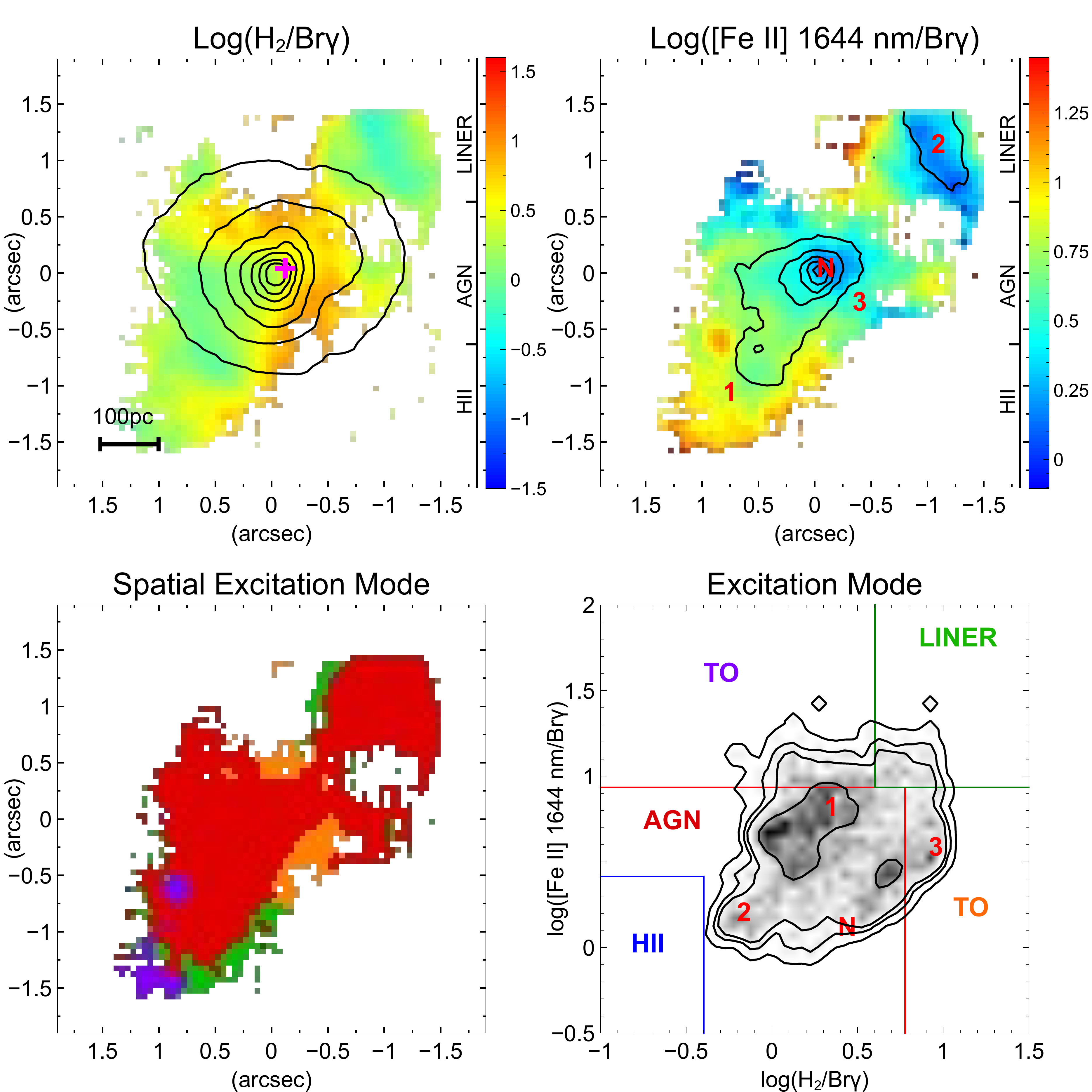}
	\caption{Excitation diagnostic maps from SINFONI \textit{H+K} data. Top left: \Htwo/\brg{} flux ratio map (log ratio). Black contours are \textit{K}-band continuum flux, normalized to 10, 30, 50, 70, and 90\% of the maximum. Color-bar labels show the excitation mode (\HII--AGN--LINER). Top right: \Fe{}1644 nm/\brg{} flux ratio, corrected for reddening, black contours are the \brg{} flux; normalized to 10, 30, 50, 70, and 90\% of the maximum. Bottom left: Spatial excitation mode; colors are red-AGN, green-LINER, orange-TO (\Htwo{} predominating), purple-TO (\Fe{} predominating). Bottom right: Excitation pixel density map for log(\Htwo/\brg) vs. log(\Fe~1644 nm/\brg). Contour levels at 1, 5, 10, and 50\% of maximum. The labels (`N',`1',`2',`3') correspond to locations on the \Fe/\brg{} ratio plot, as described in the text.}
	\label{fig:ngc5728excitationir}
\end{figure*}

The \textit{H+K}-band diagnostic ratios are shown in Fig. \ref{fig:ngc5728excitationir} (top left and right panels). Normally, extinction correction is not required as the line pairs are close together, however the \Fe{} 1644 nm flux values must be multiplied by 1.17 to correct for the extinction relative to \brg{} in this case, taking a value of \Av = 2.6 from Section \ref{sec:ngc5728Extinction}, below. Fig. \ref{fig:ngc5728excitationir} also shows (bottom right) the density of all pixels that have a measurement of both ratios on the excitation diagnostic diagram; the labels correspond to the locations on the \Fe{} 1644 nm/\brg{} map, i.e. the nucleus(`N'), the two cones (`1' and `2')  and an off-axis location (`3'). 

Overall, the excitation mode diagram show no `pure' photo-ionization/SF mode, with the vast majority of pixels in the AGN mode region, with some LINER and TO excitation. Fig. \ref{fig:ngc5728excitationir} (bottom left) shows the predominating mode at each spatial location, as defined by the regions outlined in the bottom right panel. The colors denote the mode, as defined in the caption; the `TO' regions have been divided into that which is dominated by excess \Htwo{} and that dominated by excess \Fe{} emission. The plot shows the predominance of the AGN mode, with a TO (\Fe{} excess)/LINER at the cone end caps and TO (\Htwo{} excess) at the `waist'.

The correlation between the two sets of ratios on a spaxel-by-spaxel basis is weak or non-existent ($R^2 = 0.044$), unlike that found for e.g. IC~630 \citep{Durre2017} and for IC~4687, but similar to NGC~7130 \citep[both][]{Colina2015}. This is due to the multi-component nature of this object; IC~630 is a starburst galaxy and IC~4687 a prototype star-forming luminous infrared galaxy (LIRG), whereas NGC~7130 has a mixture of star-forming regions and compact AGN excitation. \cite{Davies2016} have presented computations of AGN and star-forming mixing ratio for NGC~5728 from WiFeS optical observations on a 1\arcsec{} scale (the S7 survey), where there is a clear trajectory of excitation. From our observations, this diagnostic is not possible for the nuclear region from infrared observations at high resolution because of the mixture of ionizing and thermal processes at any one LOS.

The standard division of the NIR excitation diagram into \HII{} regions, AGNs, LINERs, and TOs is a little misleading, as what is being plotted is excitation mechanism, i.e. photo-ionization and recombination versus thermal processes (heating plus shocks). Especially in the bottom-left corner of the diagram, the source can be either photo-ionization by young hot stars (UV) or accretion disk radiation (EUV/soft X-rays). In the \Fe{}1644 nm/\brg{} ratio plots, the cones shows low ratios, however these are not generated by hot, young stars, but by the AGN radiation field, as shown by the \SiVI{} flux. \SiVI{} has an IP = 167 eV (extreme UV/X-ray -- 7.4 nm), which fully ionizes \Fe{} to \FeIII{} (which has an IP 16.2 eV), therefore the ratio of \Fe/\pab{} will be reduced in the ionization cones, which is exposed directly to the AGN radiation. This is also illustrated by the X-ray image (Fig. \ref{fig:ngc5728images24}). 

We distinguish between excitation sources at a particular location by examination of the continuum and emission line maps, as well as the diagnostic ratios. From Fig. \ref{fig:ngc5728excitationir} (panel 1), for the \Fe/\pab{} ratios, we can see that there is almost no `pure' star formation, except in the ring around the edge of the field (the 1.6 kpc ring). The ionization cones, unsurprisingly, show mainly AGN excitation, with a mixture of AGN and LINER excitation surrounding it. The actual mode in the periphery will be somewhat uncertain, as the flux values are low, with correspondingly larger uncertainties. By contrast, the \Htwo/\brg{} ratios have a different structure; this is due to the molecular hydrogen being decoupled both spatially and kinematically from the cones. This is also shown by the velocity and dispersion histograms; the \Fe{} and hydrogen recombination absolute velocity and dispersions peaks are higher than \Htwo (these results will be presented in Paper II).

We can use the forbidden lines \Fe{} and \SiVI{} as diagnostics for density, as above a critical density ($n_{cr}$) for each species the emission is suppressed because of rapid collisional de-excitation \citep{Rodriguez-Ardila2011}. For \Fe, this is of the order $n_{cr} = 10^5$ cm\pwr{-3}; for \SiVI{} this is $n_{cr} = 6.3 \times 10^8$ cm\pwr{-3}. This indicates that the densities inside and outside the cones could be different by several orders of magnitude.
 
\subsubsection{Optical Diagnostics}
\label{sec:ngc5728diagopt}
\begin{figure*}[!htbp]
	\centering
	\includegraphics[width=1\linewidth]{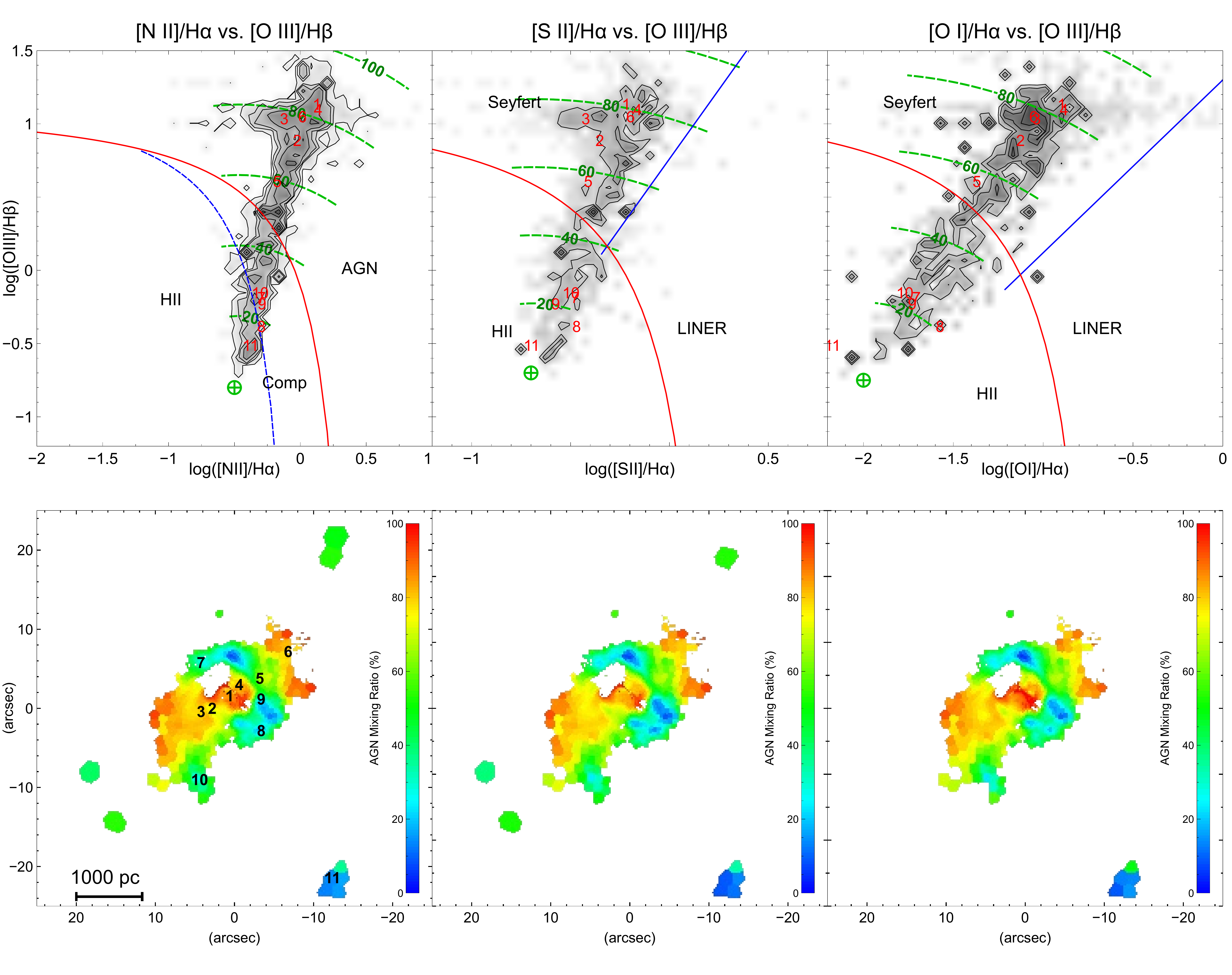}
	\caption{Optical excitation (BPT) diagnostic diagrams from MUSE data. Top row: Excitation pixel density map for log(\NII/\Ha), log(\SII/\Ha) and log(\OI/\Ha) vs. log(\OIII/\Hb). Contour levels at 1, 5, 10, and 50\% of maximum density. The labels (`1'..`11') correspond to the representative locations, as described in the text. The basis point for the star-formation/AGN mixing ratio is plotted as a green symbol, with the ratio shown by green dotted contours. Bottom row: Spatial excitation mode for each set of diagnostics, color-coded by star-formation/AGN mixing distance. }
	\label{fig:ngc5728excitationopt}
\end{figure*}

We also constructed BPT diagrams diagrams from the MUSE data cube optical emission-line maps (see Section \ref{sec:ngc5728elid}). The pixel excitation density maps for \NII/\Ha, \SII/\Ha{} and \OI/\Ha{} vs. \OIII/\Hb{} are shown in Fig. \ref{fig:ngc5728excitationopt} (top row). We use the classification lines from \cite{Kewley2006}, dividing up the diagrams into \HII-region-like, Seyferts, LINERs, and composite \HII–AGN types. The tight distribution of the points shows a smooth transition of star formation to AGN excitation. Following \cite{Davies2014b,Davies2014a}, this is parameterized on each diagram using a star-formation/AGN mixing ratio measure, where the 0 and 100\% AGN fraction is assigned to the extrema of the measured mixing sequence. The 0\% mixing ratio point is shown on each plot as a green symbol, with the green dotted contour lines delineating the mixing percentage along the sequence.
	
The bottom row of Fig.  \ref{fig:ngc5728excitationopt} shows the excitation mode maps for each diagnostic set, color-coded by the mixing ratio of each spaxel; there is complete consistency between the plots. Star-formation/composite excitation is coded as blue, cyan and green (with the mixing ratio $\lesssim$50\%), with Seyfert/AGN excitation coded yellow and red (with the mixing ratio $\gtrsim$50\%).
	
\begin{table*}[!htbp]
	\centering
	\caption{Basis locations for SF/AGN mixing ratios.}
	\label{tbl:ngc5728basispts}
	\begin{tabular}{lccc}
	\toprule
	                & log(\NII/\Ha) & log(\SII/\Ha) & log(\OI/\Ha) \\ \midrule
	Basis Point SF  &  (-0.5,-0.8)  &  (-0.7,-0.7)  & (-2.0,-0.75) \\
	Basis Point AGN &  (0.2,1.45)   &  (0.05,1.45)  &  (0.7,1.45)  \\ \bottomrule
	\end{tabular}
\end{table*}

The excitation values for several representative locations were computed for a 1\arcsec diameter aperture; the flux values (with errors) were measured by Gaussian fits to the emission line. These locations are at the nucleus, two in the SE and three in the NW ionization cones, three on the star-forming ring, and two in the spiral arm that intersects the SF ring from the south (one near the intersection point and one at a star forming site \squiggle26\arcsec SW of the nucleus).  The values are given in Table \ref{tbl:ngc5728excitdiag}. For the nucleus, where the line of sight intersect both approaching and receding gas, a multi-component Gauss fit was required to de-blend the 6 overlapping components (two each for \Ha{} and \NII{} $\lambda\lambda$ 654.9, 658.3 nm). These locations are shown as numbers on the \NII/\Ha{} mixing ratio map; each location is plotted on the respective excitation density map (red numbers). 

\begin{table*}[!htbp]
	\centering
    \footnotesize
	\caption{Excitation diagnostic values and estimated electron temperature and density at locations plotted in Fig. \ref{fig:ngc5728excitationopt}. Locations are labeled as ``IC'' - ionization cone, ``SFR'' - star-forming ring and ``S Arm'' - southern spiral arm.}
	\label{tbl:ngc5728excitdiag}
	\begin{tabular}{@{}lcccccc@{}}
		\toprule
		Location     &  log(\OIII/\Hb)  &  log(\NII/\Ha)   &  log(\SII/\Ha)   &   log(\OI/\Ha)   &         T$_e$         &       N$_e$       \\ \midrule
		1 - Nucleus  & 1.13 $\pm$ 0.04  & 0.13 $\pm$ 0.02  & -0.22 $\pm$ 0.03 & -0.90 $\pm$ 0.03 &    8045 $\pm$ 361     & 50450 $\pm$ 12958 \\
		2 - IC SE 1  & 0.89 $\pm$ 0.02  & -0.02 $\pm$ 0.02 & -0.35 $\pm$ 0.02 & -1.13 $\pm$ 0.02 &    9080 $\pm$ 802     & 21590 $\pm$ 10430 \\
		3 - IC SE 2  & 1.04 $\pm$ 0.01  & -0.12 $\pm$ 0.02 & -0.42 $\pm$ 0.02 & -1.04 $\pm$ 0.02 &    9559 $\pm$ 659     & 15280 $\pm$ 7516  \\
		4 - IC NW 1  & 1.10 $\pm$ 0.03  & 0.14 $\pm$ 0.03  & -0.16 $\pm$ 0.04 & -0.88 $\pm$ 0.04 &    8059 $\pm$ 307     & 37473 $\pm$ 14909 \\
		5 - IC NW 2  & 0.61 $\pm$ 0.02  & -0.17 $\pm$ 0.01 & -0.41 $\pm$ 0.01 & -1.37 $\pm$ 0.02 &    8042 $\pm$ 853     & 15422 $\pm$ 5327  \\
		6 - IC NW 3  & 1.05 $\pm$ 0.01  & 0.02 $\pm$ 0.02  & -0.20 $\pm$ 0.02 & -1.05 $\pm$ 0.04 &    8311 $\pm$ 790     &  9262 $\pm$ 6604  \\
		7 - SFR 1    & -0.18 $\pm$ 0.04 & -0.29 $\pm$ 0.01 & -0.48 $\pm$ 0.01 & -1.70 $\pm$ 0.05 & 8000\tablenotemark{*} & 11605 $\pm$ 3374  \\
		8 - SFR 2    & -0.38 $\pm$ 0.04 & -0.30 $\pm$ 0.01 & -0.47 $\pm$ 0.01 & -1.57 $\pm$ 0.04 & 8000\tablenotemark{*} & 10562 $\pm$ 3670  \\
		9 - SFR 3    & -0.23 $\pm$ 0.03 & -0.29 $\pm$ 0.01 & -0.57 $\pm$ 0.01 & -1.73 $\pm$ 0.03 & 8000\tablenotemark{*} & 19503 $\pm$ 3589  \\
		10 - S Arm 1 & -0.15 $\pm$ 0.02 & -0.30 $\pm$ 0.01 & -0.50 $\pm$ 0.01 & -1.77 $\pm$ 0.07 & 8000\tablenotemark{*} &  8874 $\pm$ 3931  \\
		11 - S Arm 2 & -0.51 $\pm$ 0.02 & -0.38 $\pm$ 0.01 & -0.70 $\pm$ 0.01 & -2.17 $\pm$ 0.08 & 8000\tablenotemark{*} &  6963 $\pm$ 2788  \\ \bottomrule
	\end{tabular}
\tablenotetext{*}{Estimated temperature.}
\end{table*}

The electron density $N_e$ \citep{Acker1986} and temperature $T_e$ \citep{Osterbrock2006} can also be estimated from optical diagnostics, using the \NII{} and \SII{} emission lines for the MUSE spectrum. The relationships are:

\begin{align}
R_N&= I(\lambda654.8+\lambda658.4)/I(\lambda575.4)\\
R_S&= I(\lambda671.7)/I(\lambda673.1)\\
T_e&= \dfrac{25000}{\ln\left(R_{N}/8.3\right)}\\
N_e&=10^{2}~T_{e} \left(\dfrac{R_S - 1.49}{5.62 - 12.8 R_S}\right)
\end{align}

where $R_N$ is the ratio for the \NII{} lines and $R_S$ is the ratio for the \SII{} lines. The alternate diagnostic for $T_e$ using the \OIII{} lines \citep{Kwok2007} is not available in the MUSE spectral range. For the SF-dominated regions (locations `7' to `11'), the \NII{}$\lambda$575.4 was not measurable; we used an estimated temperature of T$_e$=8000 K, somewhat lower than the AGN-dominated locations. The calculated values are given in Table \ref{tbl:ngc5728excitdiag}; while the electron temperatures are reasonably consistent (with error bars of the order of 5-10\%) in the AGN-dominated regions, the electron density has larger errors (from 25 to 70\%, mainly due to the uncertainties in the flux of the weak \NII{}$\lambda$575.4 nm line), but there is a clear trend with lower density being at a greater distance from the AGN. $N_e$ linearly scales with a lower assumed $T_e$.

\subsubsection{\Htwo{} Excitation}
Molecular hydrogen (\Htwo) is very important in the star-formation context of AGN activity, since it is the basic building block for stars. In the \textit{K} band, there are a whole series of rotational-vibrational emission lines, which can be used to examine the excitation mechanism for \Htwo{}, which can be either:
\begin{itemize}
	\item UV photons (fluorescence) from star formation and/or AGN continuum emission \citep{Black1987}.
	\item Shocks from supernovae, AGN outflows or star-formation winds \citep{Hollenbach1989}.
	\item X-rays from the AGN irradiating and heating dense gas \citep{Maloney1996}.
\end{itemize}
In reality, all these different mechanisms occur together; however, the dominating mechanism can be estimated and the contributing fractions of different mechanisms can be constrained \citep{Busch2016}. 

Following the method outlined in \cite{Wilman2005}, for gas with density $n_T > 10^{5}$~cm\pwr{-3}, the thermal (collisional) temperatures can be estimated. The occupation numbers of the excited ro-vibrational levels of the \Htwo{} molecule will be in thermal equilibrium at a temperature $T_{exc}${} equal to the kinetic temperature of the gas. This leads to the relationship:
\begin{equation}\label{eqn:ngc5728H2Temp}
\ln\left( \dfrac{F_{i}~\lambda_{i}}{A_{i}~g_{i}}\right)  = constant - \frac{T_{i}}{T_{exc}}
\end{equation}
where $F_i$ is the flux of the \textit{i}th \Htwo{} line, $\lambda_i$ is its wavelength, $A_i$ is the spontaneous emission coefficient, $g_i$ is the statistical weight of the upper level of the transition and $T_i$ is the energy of the level expressed as a temperature. The left-hand side of this equation is equivalent to $\ln(N_{upper})$, the occupation number of the upper level of the \Htwo{} transition. This relation is valid for thermal excitation, under the assumption of an \textit{ortho:para} abundance ratio of 3:1. The $A_i$, $g_i$, and $T_i$ for each line was obtained from on-line data `Molecular Hydrogen Transition Data'\footnote{\url{www.astronomy.ohio-state.edu/~depoy/research/observing/molhyd.htm}}. 

At a particular location, plotting $\ln(N_{upper})$ vs. $T_i$~gives a linear relationship, where the negative inverse slope is the excitation temperature $T_{exc}$.

The \Htwo{} 2–-1 S(3) (2073.5 nm) is diagnostic for X-ray excitation, which is expected to suppress the line, as the upper level of this transition is depopulated by a resonance with photons around the wavelength of Ly$\alpha$ at 1216 \AA, which are readily generated by X-ray heating \citep{Black1987, Krabbe2000, Davies2005}.

The flux for the \Htwo{} \textit{K}-band lines was measured at several locations, P1--P6, as shown in Fig. \ref{fig:ngc5728excitationh22} (top left panel). The spectrum at each location was taken from a circular region 4 pixels wide and the flux measured from a Gaussian fit to the line, using \texttt{QFitsView} fitting facility. The locations were chosen to be the nucleus and 0\arcsec.5 N and S, in line with the rotation, plus other features visible in the flux map.

\begin{figure}[!htbp]
	\centering
	\includegraphics[width=1\linewidth]{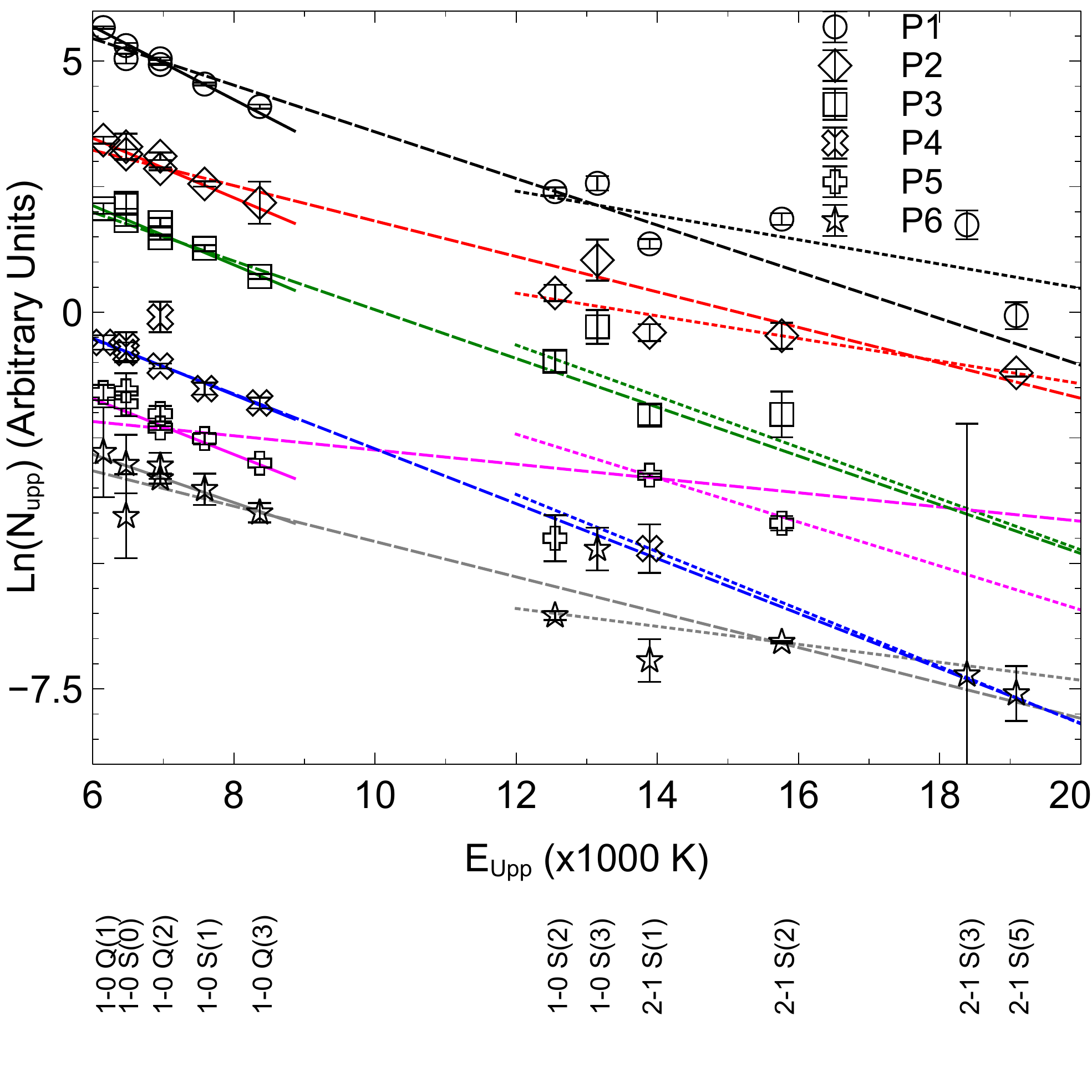}
	\caption{\Htwo{} temperature plots. The value of the inverse slope of the relationship between ln(N$_{upper}$) and E$_{upper}$ is the excitation temperature of \Htwo{} for LTE. The values are plotted for each location, each with an offset for clarity. The transitions are labeled below the plot. The $\nu=1-0$ transition fits are plotted with solid lines, the $\nu=2-1$ and $\nu=3-2$ fits are plotted with dotted lines, those for all transitions with dashed lines. Including the higher-excitation increases the derived temperatures, but these may not be in LTE.}
	\label{fig:ngc5728excitationh21}
\end{figure}

\begin{figure}[!htbp]
	\centering
	\includegraphics[width=1\linewidth]{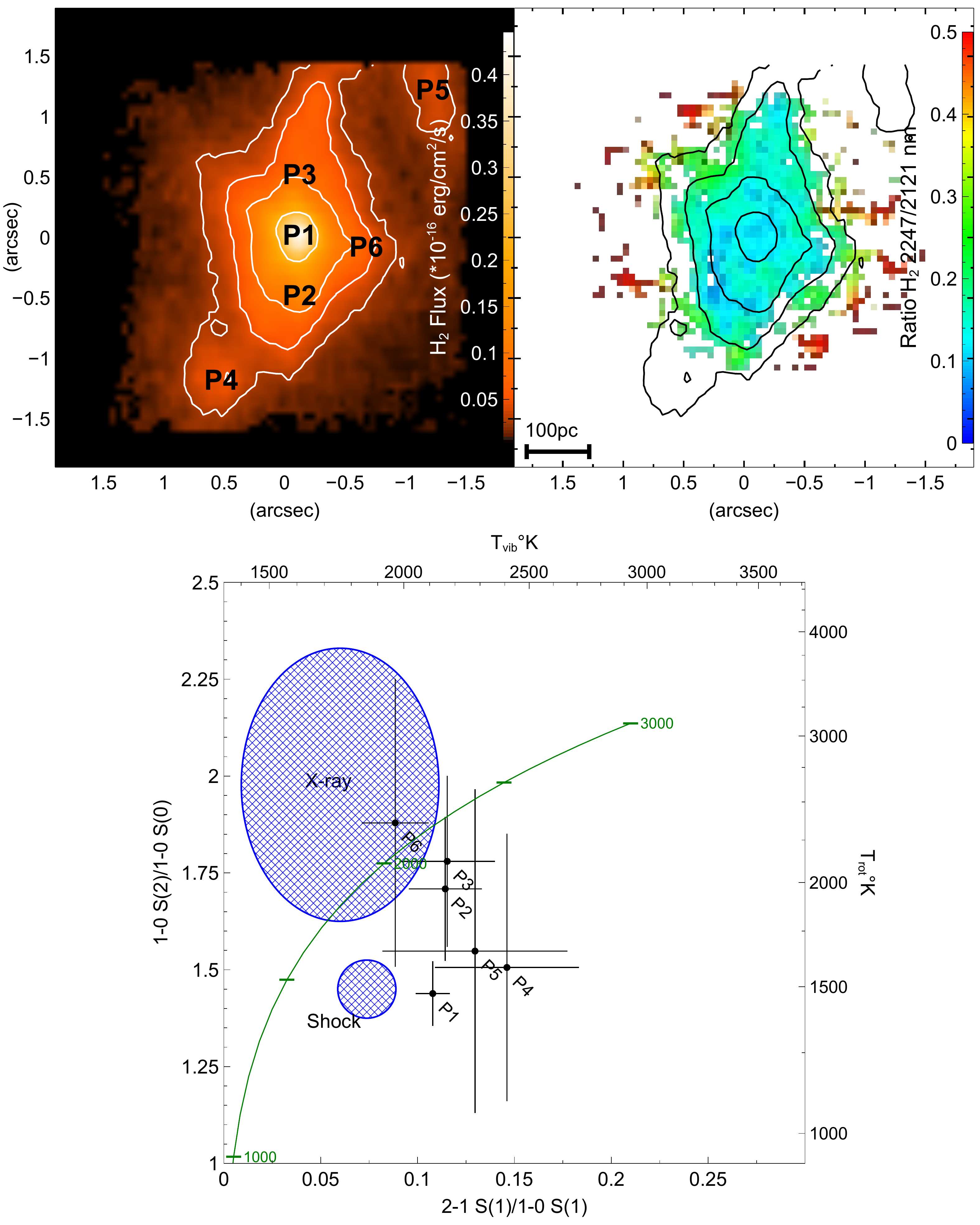}
	\caption{Left panel: \Htwo{} measurement locations on flux map. Contours are at 5, 10, 30, and 50 \% of maximum flux. Right panel: \Htwo{} 2248/2121 nm flux ratio. Over-plotted contours are \Htwo{} 2121 nm flux, at levels of 5, 10, 20, and 50\% of maximum. Thermal processes dominate at high \Htwo{} flux, with an increasing contribution from UV fluorescence at lower fluxes. Bottom panel: Diagnostic diagram \citep{Mouri1994} for \Htwo{} locations. The green trajectory is for gas at LTE with the temperature indicated. The values for points P1--P6 are plotted. Thermal processes (X-ray heating and shocks) predominate.}
	\label{fig:ngc5728excitationh22}
\end{figure}

We could measure all of the low excitation lines ($\nu=1-0$); however the high excitation line fluxes ($\nu=2-1$ and $\nu=3-2$) were more uncertain. At location P4, only the 2--1 S(3) line could be measured (see Fig. \ref{fig:ngc5728excitationh21}). The results for the derived parameters are given in Table \ref{tbl:ngc5728H2Temps} for the excitation temperatures. 
\begin{table}[!htbp]
	\begin{center}
		\caption{\Htwo{} excitation temperatures.}
		\footnotesize
		\label{tbl:ngc5728H2Temps}
		\begin{tabular}{ccccc}
			\toprule
			Location &     Name     & $T_{exc}$~(K) &     $T_{exc}$~(K)      & $T_{exc}$~(K) \\
			         &              &   $\nu=1-0$   &     $\nu=2-1/3-2$      &      All      \\ \midrule
			   P1    &   Nucleus    &  $1375\pm35$  &      $4165\pm510$      &  $2155\pm30$  \\
			   P2    & 0\arcsec.5 S & $1685\pm160$  &      $4445\pm445$      &  $2840\pm50$  \\
			   P3    & 0\arcsec.5 N & $1700\pm100$  &      $1960\pm560$      &  $2060\pm80$  \\
			   P4    &      SE      & $1775\pm210$  & \dots\tablenotemark{*} & $1830\pm165$  \\
			   P5    &      NW      & $1800\pm185$  &      $2275\pm400$      & $7040\pm310$  \\
			   P6    &      W       &  $2085\pm60$  &      $5620\pm875$      &  $2840\pm75$  \\ \bottomrule
		\end{tabular}
	\end{center}
	\tablenotetext{*}{No slope could be determined; only 1 data point}
\end{table}

Independently, the rotational and vibrational temperatures can be determined from two ortho/para lines that belong to the same vibrational level, e.g. 1--0 S(0) (2223.5 nm) and 1--0 S(2) (2033.8 nm), whereas the vibrational excitation temperature can be determined by connecting two transitions with same J but from consecutive $\nu$ levels, e.g. 2--1 S(1) (2247.7 nm), and 1--0 S(1) (2121.8 nm) \citep{Riffel2014a,Busch2016}. The formulas are:

\begin{align}
T_{rot(\nu=1)}&=\dfrac{1113 K}{1.130+\ln\left(\dfrac{F_{1-0 S(0)}}{F_{1-0 S(2)}}\right)}\label{eqn:ngc5728H2Trot}\\
T_{vib}&=\dfrac{5594 K}{0.304+\ln\left(\dfrac{F_{1-0 S(1)}}{F_{2-1 S(1)}} \right)}\label{eqn:ngc5728H2Tvib}
\end{align}

The computed values are given in Table \ref{tbl:ngc5728H2Temps2}. Fig. \ref{fig:ngc5728excitationh21} shows the excitation temperature plot derived from the method of \cite{Wilman2005} and Fig. \ref{fig:ngc5728excitationh22} (bottom panel) shows the rotation and vibrational temperatures plotted on the \cite{Mouri1994} excitation mechanism diagram. These results show that the gas is close to the local thermodynamic equilibrium (LTE) line, indicating predominantly thermal processes. The 2-1 S(3) line is present at all locations, however in most locations (P1, P2, P3, and P6) it is below the fitted line, indicating that X-rays make up a minor component of the excitation  \cite[e.g.][for NGC~1275]{Krabbe2000}, with shocks probably the main contributor to the excitation.
\begin{table}[!htbp]
	\centering
	\caption{\Htwo{} rotational and vibrational temperatures.}
	\label{tbl:ngc5728H2Temps2}
	\begin{tabular}{ccc}
		\toprule
		Location & $T_{Rot}$~(K) & $T_{Vib}$~(K) \\ \midrule
		   P1    &  $1450\pm10$  &  $2210\pm72$  \\
		   P2    & $1870\pm394$  & $2262\pm152$  \\
		   P3    & $2007\pm463$  & $2270\pm198$  \\
		   P4    & $1542\pm515$  & $2512\pm289$  \\
		   P5    & $1604\pm674$  & $2384\pm379$  \\
		   P6    & $2225\pm973$  & $2050\pm147$  \\ \bottomrule
	\end{tabular}
\end{table}

We calculate the excitation temperatures from the inverse slope of the low-excitation ($\nu=1-0$), high-excitation ($\nu=2-1/3-2$), and all lines, respectively. These are given in  Table \ref{tbl:ngc5728H2Temps}. The low-excitation temperatures are very similar (1740$\pm$50K), except for the nucleus and location P6. It can be hypothesized that the \Htwo{} in the line of sight of the nucleus is shielded from the some of the heating effect of the AGN by the circumnuclear torus or dusty bar that is visible in the \hk{} magnitude maps (Fig. \ref{fig:ngc5728continuum}). The radio jet impact on the ISM at location P6 could increase the excitation in that region. 

If we include all the lines, the derived temperatures increase somewhat; however the high-excitation flux values are more uncertain. These temperatures now compatible with results for other Seyfert galaxies in the literature, in the range 2100--2700~K \citep[e.g.][]{Riffel2015, Storchi-Bergmann2009, Riffel2014a, Riffel2011, Riffel2010a}. The temperature derived for P5 for all lines is would seem to be unphysical, being above the disassociation temperature for \Htwo{}. However, \cite{Davies2003,Davies2005} shows that the lower $\nu=1-0$ levels may be thermalised, but the $\nu=2-1$ and $\nu=3-2$ levels can be overpopulated due to fluorescent excitation by far-ultraviolet photons. If we fit just the $\nu=2-1$ and $\nu=3-2$ levels, we can see that the P5 location has a very high temperature ($\sim7000$~K); this is in the NW ionization cone at the hypothesized point of impact of the radio jet with the ISM and is also presumably illuminated by the accretion disk radiation field where the fluorescent excitation is highest. Alternately, there may be a component of much hotter gas close to the \Htwo{} dissociation temperature of $\sim4000$~K, or there is non-thermal emission due to excitation of the molecule by secondary electrons deep in the cloud.

The flux ratio for the 2--1 S(1) (2247.7 nm) and 1--0 S(1) (2121.8 nm) lines is diagnostic for excitation by soft-UV photons (from star formation) vs. thermal processes (from shocks or X-ray heating), with a value of $\sim0.1-0.2$ for thermal and $\sim0.55$ for fluorescent processes \citep[following][in their excitation study of NGC~1068]{Riffel2014a}. Fig. \ref{fig:ngc5728excitationh22} (top right panel) shows this plot; where there is strong \Htwo{} flux, the ratios indicate purely thermal processes. Higher values at the periphery reflect an increasing contribution of excitation by hot stars away from the X-ray heating and shocks in the nuclear region, plus uncertainties in the flux measurements.

It is noted that there is a difference between the optical and IR excitation diagnostic diagrams, with the optical diagram showing a tight relationship along the mixing sequence, but the IR diagram shows a broad spread. This difference can be ascribed to the use of \Htwo{} in the IR diagnostic, which incorporates different excitation mechanisms (X-ray heating and UV fluorescence) to the that of the metal ions (AGN power-law radiation field).

\subsubsection{Other Diagnostics}
The \Fe{} and \PII{} 1188.6 nm emission lines can be used to diagnose the relative contribution of photo-ionization and shocks \citep{Oliva2001,Storchi-Bergmann2009}, where ratios $\sim$ 2 indicate photo-ionization (as the \Fe{} is locked into dust grains), with higher values indicating shocked release of the \Fe{} from the grains (up to 20 for SNRs). This ratio was determined at three locations with a 1\arcsec{} aperture; the AGN ($5.9\pm1.2$), the SE cone maximum flux ($5.8\pm1.6$), and the NW cone maximum flux ($3.9\pm0.7$). This shows the excitation is a mixture of shock and photo-ionization, with the latter predominating.

This result is confirmed by the diagnostics from \cite{Mouri2000}, who calculated photo-ionization and shock heating models, and generated diagnostic diagrams for the ratios of \Fe{} 1257 nm/\pab{} vs. \OI/\Ha{} for both power-law and blackbody photo-ionization, as well as shock heating models. From Table \ref{tbl:ngc5728_Line_Flux} values, the ratios within the inner 1\arcsec{} are \Fe{} 1644 nm/\brg{} = 2.0 (which translates to \Fe{} 1257 nm/\pab{} = 0.46) and \OI/\Ha{} = 0.115 (from the MUSE data). These values are compatible with both power-law photo-ionization and shock models (favoring photo-ionization), where the metal abundances are sub-solar and the ionization parameter is high, $U\ge10^{-1.5}$ \citep[see Fig. 3 of ][]{Mouri2000}.

We attempted to measure the electron density from the \Fe{} line ratio, from the methods described in \cite{Storchi-Bergmann2009}, from the ratios of \textit{H}-band emission lines for \Fe{} 1533/1644, 1600/1644, 1664/1644, and/or 1677/1644 nm. However, all fluxes, other than the 1644 nm line, were all very weak even at the location of the peak \Fe{} 1644 nm flux. The only one that could be measured with reasonable certainty (the 1533 nm line) gave a low flux ratio ($0.062\pm0.01$) which indicated a low electron density ($<1000$ cm\pwr{-3}). This is a firm upper limit, as the 1644 nm flux is strong. This low density is measured in the partially ionized \Fe{} line emitting region, and not close to the central engine, where it would be much higher.
\subsection{Extinction}
\label{sec:ngc5728Extinction}
We can derive the extinction by dust from emission line flux ratios, where the known ratio of a pair of emission lines is compared against observations and extrapolated to the $\bv$ extinction and $A_V$ (the absolute extinction in the $V$ band). In general, the formulae are as follows:
\begin{align}
E{(B-V)} &= \alpha_{\lambda_1,\lambda_2} log\left( \dfrac{R_{\lambda_1,\lambda_2}}{F_{\lambda_1}/F_{\lambda_2}}\right) \label{eqn:ngc5728Xtn}\\
A_V&=E(B-V) \times R_{V} 
\end{align}
where $R_{\lambda_1,\lambda_2}$ is the intrinsic emissivity ratio of the two lines and $\alpha_{\lambda_1,\lambda_2}$ extrapolates from the emission line wavelengths to $\bv$. $R_V$ is the total-to-selective extinction ratio; we use the fiducial value of 3.1. We use the \cite{Cardelli1989} empirical relationships to derive the $\alpha_{\lambda_1,\lambda_2}$ values. The hydrogen line ration are from \cite{Hummer1987}, assuming Case B recombination, an electron temperature $T_e = 10^4$ K and a density $n_e=10^3$~cm\pwr{-3}. The \Fe{} ratios in the literature are discrepant, we will use the commonly accepted value of 1.36 \citep[see the discussion in ][]{Durre2017}. 

\begin{table}[!htbp]
	\centering
	\caption{Extinction constants for Equation \ref{eqn:ngc5728Xtn}.}
	\label{tbl:ngc5728ExtinctionCalc}
	\begin{tabular}{crrrr}
		\toprule
		Line Pair  & $ \lambda_1 $ & $ \lambda_2 $ & $\alpha_{\lambda_1,\lambda_2}$ & $R_{\lambda_1,\lambda_2}$ \\ \midrule
		\pab--\brg &        1282.2 &        2166.1 &                           6.07 &                      5.88 \\
		   \Fe     &        1256.7 &        1643.6 &                           8.22 &                      1.36 \\
		\pag--\pab &        1094.1 &        1282.2 &                          10.24 &                      0.55 \\
		\paa--\brg &        1875.6 &        2166.1 &                          26.55 &                     12.15 \\
		 \Hb--\Ha  &         486.1 &         656.3 &                           2.33 &                      2.86 \\ \bottomrule
	\end{tabular}
\end{table}

For this object, the \Fe{} 1257/1644 nm ratio will be somewhat uncertain, even with the corrected telluric feature in the 1257 nm emission line. The \pag/\pab{} ratio has the advantage that it is measured in the same band, but the \pag{} gas kinematics are of poor quality due to the line being near the short end of the spectrum and close to the edge of the atmospheric window. An attempt at calculating the extinction using this ratio produced unphysical values over the whole field. We also suspect that there is some telluric contamination, as the line width is broader than \pab{}. The \pab/\brg{} ratio can only be mapped over part of the whole central ring feature, as the \textit{H+K}-band observations are at a smaller scale than the \textit{J}-band ones; uncertainties are also introduced by the rescaling and any flux calibration uncertainties. The \paa/\brg{} ratio has the advantage that it is in the one data cube; however the \paa{} line is in the atmospheric absorption region between the \textit{H} and \textit{K} bands. On examination of the cube, it was seen that the \paa{} line is very strong, and the continuum is piece-wise smooth between the bands, so a definitive measurement can be made.

\begin{figure}[!htbp]
	\centering
	\includegraphics[width=1\linewidth]{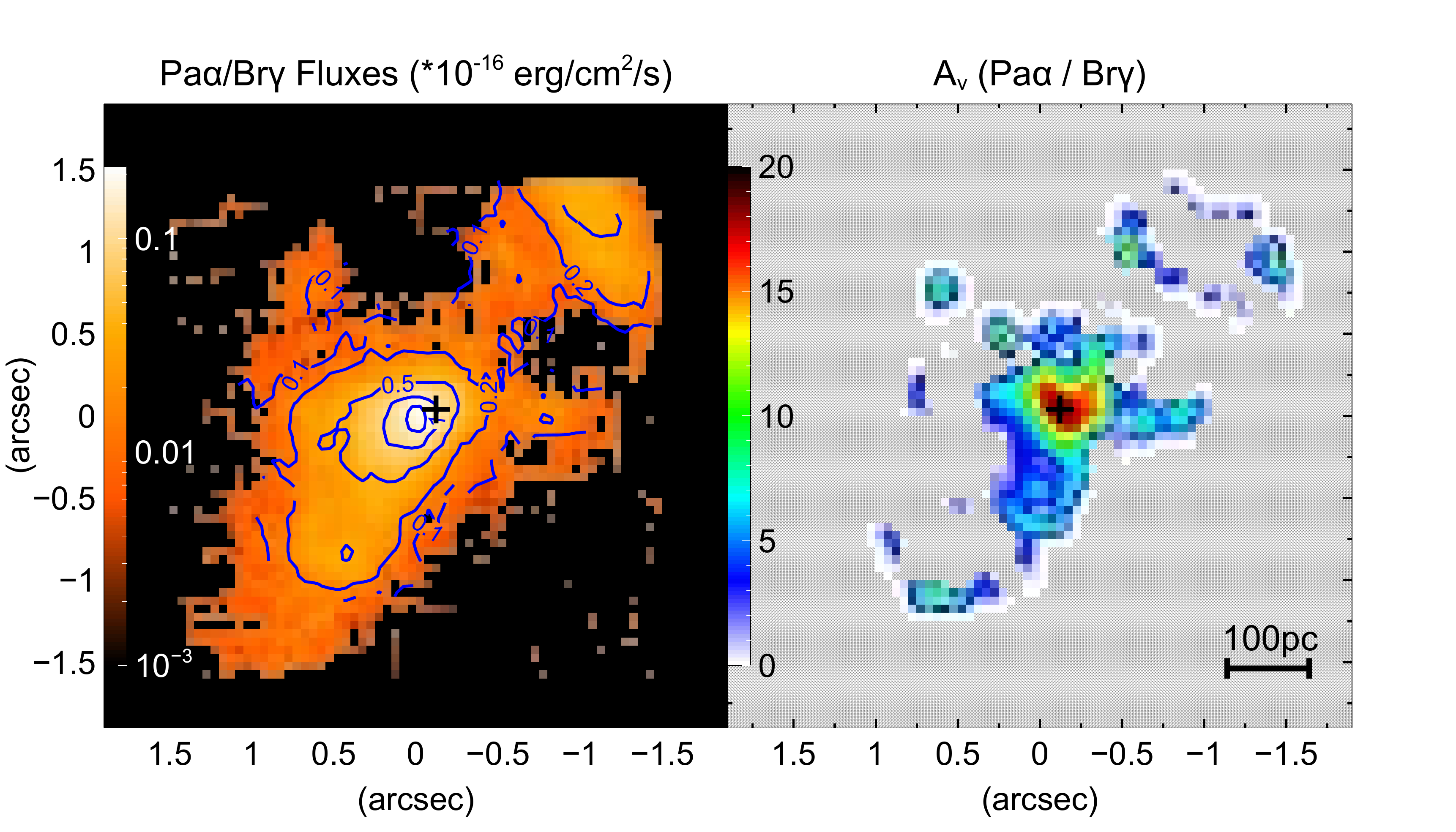}
	\caption{Extinction map from hydrogen recombination (\paa--\brg). Left panel: \paa{} and \brg{} fluxes; color map and color-bar is \brg{} flux, contours and labels are \paa{} flux, both in units of 10\pwr{-16} \ecs. Right panel: Visual extinction \Av{} in magnitudes.}
	\label{fig:ngc5728extinction2}
\end{figure}

We derived the extinction value \Av{} from the flux maps of \paa{} and \brg{}, with the constants as given in Table \ref{tbl:ngc5728ExtinctionCalc}; unphysical values are masked out. This is shown in Fig. \ref{fig:ngc5728extinction2}, where the left panel shows the color map of the \brg{} and the contour giving the \paa{} flux; the extinction map is plotted in the right panel. The \Av{} values range up to 19 magnitudes, and the extinction is strongly concentrated around the AGN, presenting as a roughly elliptical region or bar of size $64 \times 28$ pc (where the extinction is $>16$ mag), aligned at right angles to the ionization cones. This is comparable in size to the obscuring dust bar found for NGC~2110 at $55 \times 27$ pc \citep{Durre2014}. The average extinction over all valid pixels is 3.7 magnitudes. 

Determining the extinction from the \Fe{} line ratio was more difficult, given the uncertainties mentioned above. The \textit{J} flux map was sub-sampled to the same scale and aligned to the \textit{H+K} map by matching the flux maxima; the fluxes of the lines are shown in Fig. \ref{fig:ngc5728extinction2} (left panel - color map is \Fe{} 1257 nm flux, contours 1644 nm flux). The values range up to 15 magnitudes of extinction, and the morphology is comparable with the \paa/\brg{} maps, with some differences (Fig. \ref{fig:ngc5728extinction3}, right panel). The peak location is for \Fe{} extinction is not on located on the AGN, but somewhat to the SE ($\sim30$ pc). The morphology is also more even, with the interiors of the ionization cones having greater extinction than the \paa/\brg{} extinction. These values and conclusions should be treated with some caution, as it depends on the exact alignment of the two \Fe{} flux maps (e.g. a 2-pixel shift in the RA co-ordinate places the peak extinctions at the same location).  

The extinction was also measured by taking a spectrum from the data cubes in an aperture of 1\arcsec{} diameter around the AGN location, showing \Av{} = 14.9 mag for \paa/\brg, and 6.45 mag for \Fe. The lower values, in general, supports the conclusion that the \Fe{} flux originates from a skin around the ionization cones, and thus the extinction does not probe the full depth of the cones.

Apart from the high extinction around the AGN, there is an indication that the hydrogen recombination extinction also traces the boundaries of the ionization cones; one can hypothesize that the dust (which causes the extinction) is sublimated in the AGN radiation field. To check this, the \SiVI{} flux (which traces the high-ionization region and outflow shock boundaries) is plotted with the extinction from hydrogen recombination (Fig. \ref{fig:ngc5728extinction3}, left panel); the extinction traces the boundaries of the \SiVI{}, providing support for the hypothesis. Around the AGN, the obscuring dusty toroidal structure overlays the \SiVI{} emission along our LOS.
\begin{figure}[!htbp]
	\centering
	\includegraphics[width=1\linewidth]{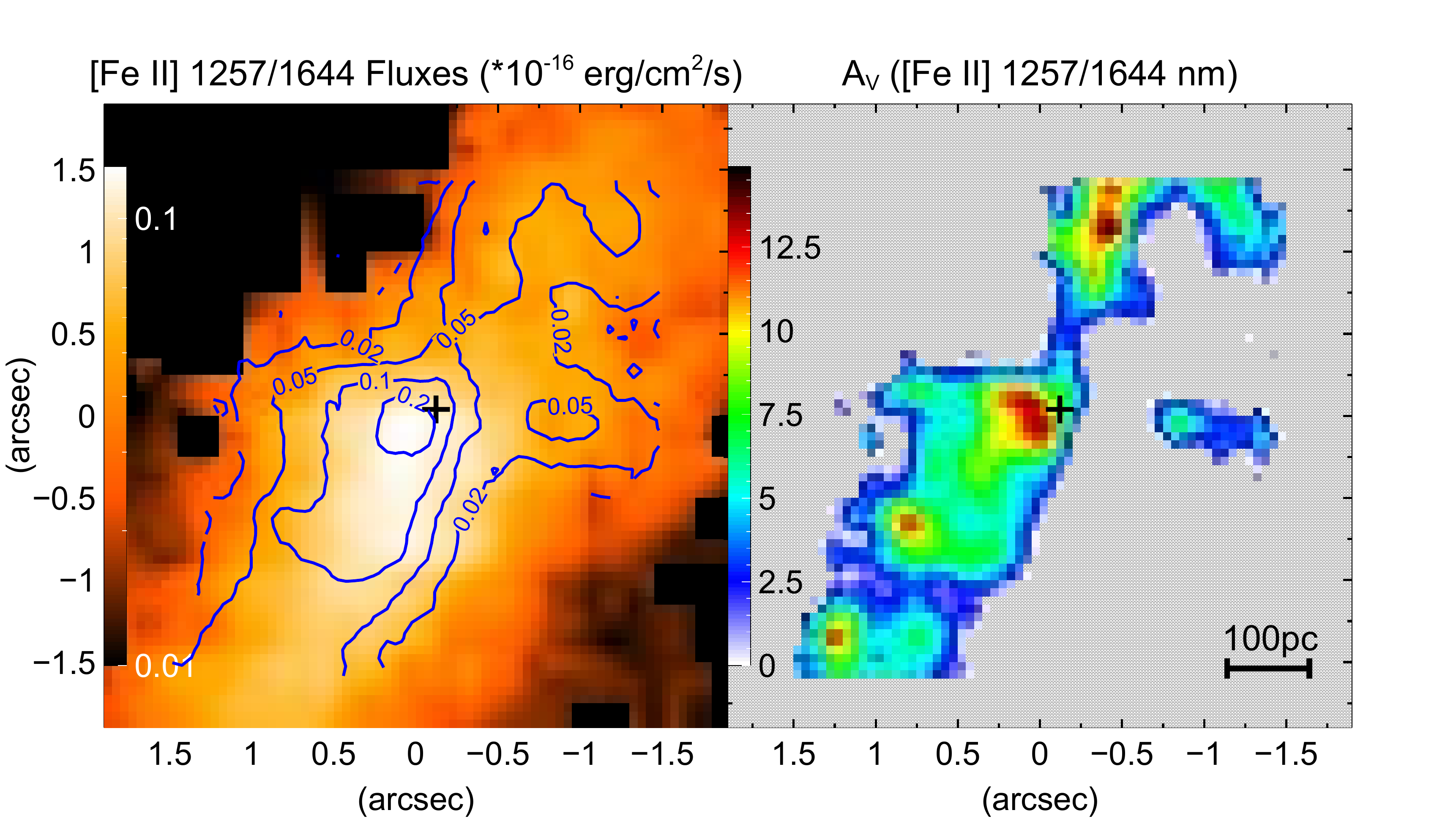}
	\caption{Extinction map from \Fe. Left panel: \Fe{} fluxes; color map and color-bar is 1257 nm flux, contours and labels are 1644 nm flux, both in units of 10\pwr{-16} \ecs. Right panel: Visual extinction \Av{} in magnitudes.}
	\label{fig:ngc5728extinction1}
\end{figure}
\begin{figure}[!htbp]
	\centering
	\includegraphics[width=1\linewidth]{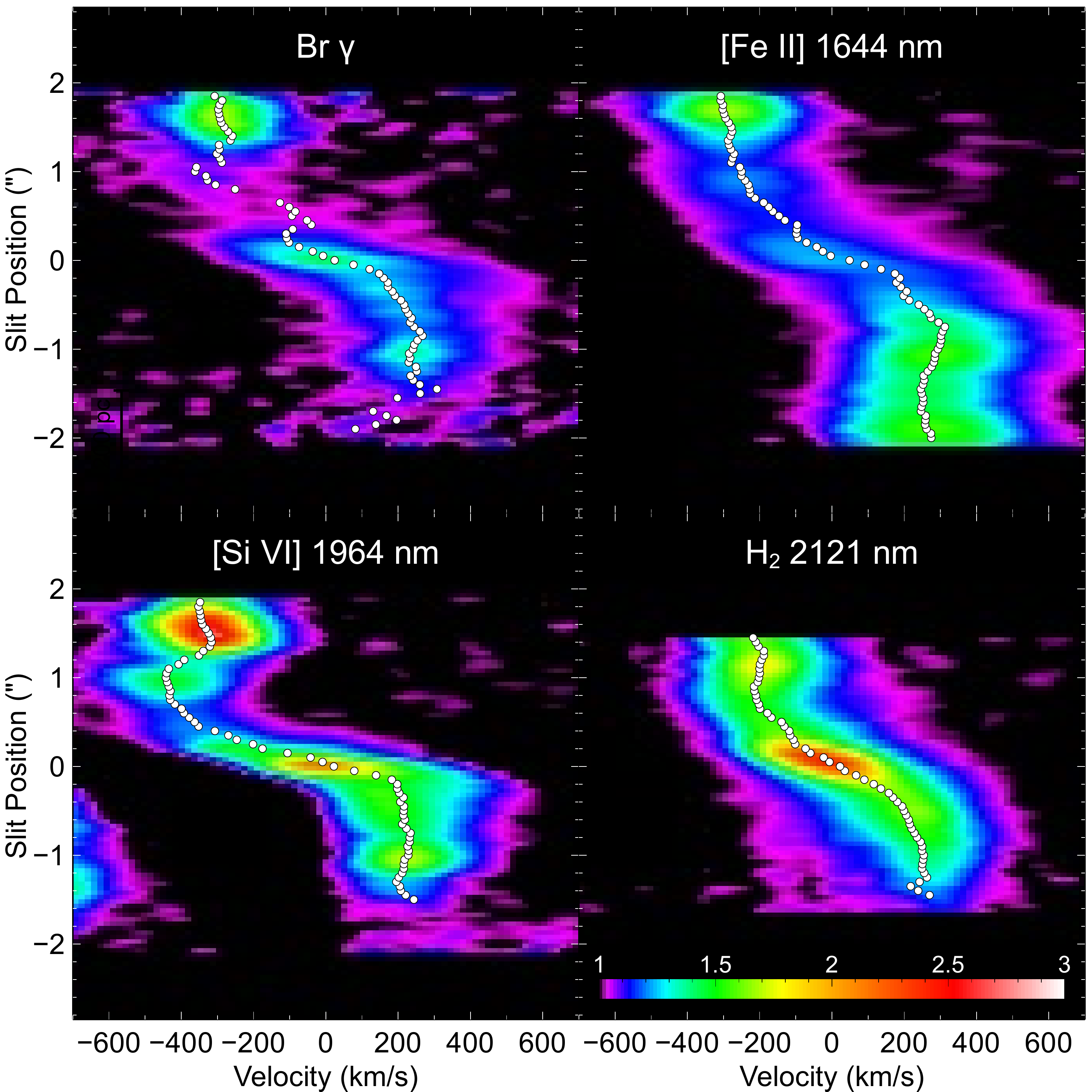}
	\caption{Left panel: Ionization cone extinction, as shown by \Av{} from \pab/\brg (color map - values given in color-bar) compared with \SiVI{} flux (contours at 1, 2, 5, 10, 20, and 50\% of maximum). Right panel: \Av{} from \Fe{} 1257/1644 (color map - values given in color-bar) compared to \Av{} from \paa/\brg{} (contours at 2, 7, 12, and 17 mag).}
	\label{fig:ngc5728extinction3}
\end{figure}

\begin{figure}[!htbp]
	\centering
	\includegraphics[width=1\linewidth]{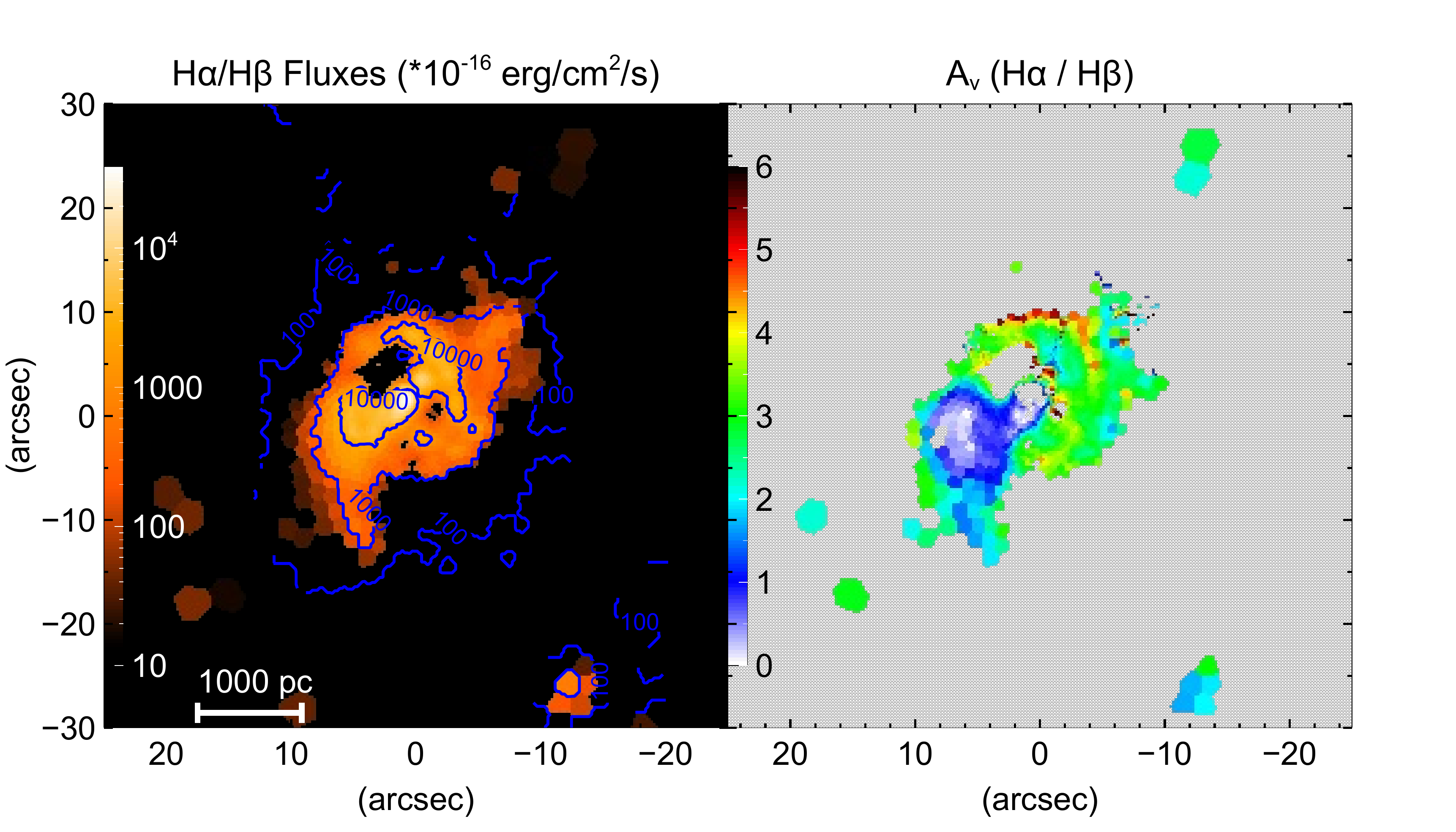}
	\caption{Extinction map from hydrogen recombination (\Ha--\Hb). Left panel: \Ha{} and \Hb{} fluxes; color map and color-bar is \Hb{} flux, contours and labels are \Ha{} flux, both in units of 10\pwr{-16} \ecs. Right panel: Visual extinction \Av{} in magnitudes.}
	\label{fig:ngc5728extinctionopt}
\end{figure}

The extinction can also be calculated at a broader scale from the \Ha/\Hb{} ratio, using the MUSE data. The intrinsic ratio for AGN NLRs is quoted as 3.1 \citep[][following]{Davies2017a} instead of the fiducial value of 2.86. The extinction map (Fig. \ref{fig:ngc5728extinctionopt}) uses this value where the AGN mixing ratio is greater than 50\%. (Section \ref{sec:ngc5728diagopt}). The two values produces a difference of $\sim0.27$ mag in the calculated \Av. We subtract the foreground galactic extinction of 0.279 \citep{Schlafly2011}.

The striking feature of this map is the low extinction in the SE ionization cone. This supports the previously mentioned hypothesis of dust sublimation in the ionization cone; the star-forming ring (which has extinction values in the range 2--5) obscures the NW cone. The high values near the center have some uncertainties, caused by errors in the flux measurements associated with the multiple kinematic components.

We also note the increased extinction that borders the star-forming ring; this will produce the drop in EW for the ionized gas species in the NW ionization cone, as ring is in front of the cone along our line of sight.

\added{The obscuration caused by the dust lane is seen in Fig. \ref{fig:ngc5728continuum} in the \textit{H-K} map and Fig. \ref{fig:ngc5728gaskinematics1} in the equivalent width maps. Since the EW measure is supposedly independent of obscuration, we posit that there is significant stellar continuum in front of the gas emission and obscuration along our LOS. In the same manner, the \Htwo{} emission is also in front of this obscuration.}
\subsection{Gas Masses}
\label{sec:ngc5728gasmass}
The cold gas column density can be derived from the visual extinction value. The gas-to-extinction ratio, $N_{H}/A_{V}$, varies from 1.8 \citep{Predehl1995} to 2.2 $\times$ 10\pwr{21} cm\pwr{-2} \citep{Ryter1996}. \cite{Zhu2017} finds the Milky Way ratio for solar abundances to be $\sim 2.1 \times$ 10\pwr{21} cm\pwr{-2}; this increases to $\sim 2.5 \times$ 10\pwr{21} cm\pwr{-2} for sub-solar abundances. We will use a value of 2.0 $\times$ 10\pwr{21} cm\pwr{-2}. Using an average atomic/molecular weight of 1.4, we thus derive the relationship:
\begin{equation}
\sigma_{Gas} = 22.1~A_V~\mmsun{}~pc^{-2} \label{eqn:ngc5728ISM}
\end{equation}
The extinction value towards the AGN implies a cold gas column density of 420 \msun~pc\pwr{-2}, equivalent to $N_H = 4 \times 10^{22}~$cm\pwr{-2}. \cite{Davies2015} give the column density based on X-ray modeling as $N_H = 1.6 \times 10^{24}~$cm\pwr{-2}, a factor of 40 times higher. This absorption is likely to be at the parsec scale in the broad-line region (BLR) \citep{Burtscher2016}, i.e. it is probing a very narrow LOS, while the recombination lines are measured over a much more extended area; also, these lines will not penetrate very deep into the obscuration, as the high extinction will attenuate the emission lines so much that they will not contribute to the measured total.

We also derive the ionized hydrogen plus warm/hot and cold \Htwo{} gas masses, using the formulae from \cite{Riffel2013b,Riffel2015}:
\begin{align}
M_{HII} &\approx 3 \times 10^{17} F_{Br\gamma}D^2~\mmsun  \label{eqn:ngc5728HII}\\
M_{H_2} (Warm) &\approx 5.0776 \times 10^{13} F_{H_2}~D^2~\mmsun  \label{eqn:ngc5728H2H}\\
M_{H_2} (Cold) &\approx R~L_{H_{2}}~\dfrac{\mmsun}{\mlsun} \nonumber\\
&\approx 3.12 \times 10^{16}~R~F_{H_2}~D^2~\mmsun  \label{eqn:ngc5728H2C}
\end{align}
where \textit{D} is the distance in Mpc and \textit{F} is measured in erg cm\pwr{-2} s\pwr{-1}. which gives $M_{Cold}/M_{Warm}$ as $7.2\times10^5$ and $2.7\times10^6$, respectively.   For the \brg{} emission, a standard electron temperature $T = 10^4$~K and density $n_e = 100$~cm\pwr{-3} is assumed. The \Htwo{} flux is that of the 2121 nm line.

The constant $R$ is given variously as 1174 \citep{Mazzalay2013a} for 6 local galaxies, or 4000 \citep{MullerSanchez2006} from 17 LIRG/ULIRG galaxies; these give $M_{Cold}/M_{Warm}$ as $7.2\times10^5$ and $2.7\times10^6$, respectively. The \cite{Mazzalay2013a} value is originally derived from the observed CO radio emission with estimates of CO/\Htwo{} ratios (which can vary over a range 10\pwr{5} to 10\pwr{7}). At the centers of galaxies hosting AGN, this ratio could be substantially overestimated, as a greater proportion of the gas will be excited. Given this uncertainty, we will use the \cite{Mazzalay2013a} value of $R=1174$.

Table \ref{tbl:ngc5728gasmass} presents the results. The surface density values for the inner 100 pc radius are within the ranges of the results from \cite{Schonell2017} who summarize results for 5 Seyfert 1, 4 Seyfert 2 and 1 LINER galaxies observed by the AGNIFS group; the range of \HII{} surface densities is 1.5--125 \msun/pc\pwr{2} and of \Htwo{} (cold+warm) surface densities is 526--9600 \msun/pc\pwr{2}. Star formation rates and supernova rates have not been calculated, as it is clear that both the hydrogen and \Fe{} excitation is caused by the AGN photo-ionizing source plus associated outflow shocks, rather than star formation.

\begin{table*}[!htbp]
	\footnotesize
	\centering
	\caption{Gas masses for the cold ISM, \HII{}, and \Htwo, in the pixel with greatest flux/extinction and within a radius of 100 and 200 pc from the center and over the whole observed field (600$\times$600 pc), plus the 100 and 200 pc radius surface densities. The extinction/ISM is not measured outside the 100 pc radius.}
	\label{tbl:ngc5728gasmass}
	\begin{tabular}{@{}lcccccc@{}}
		\toprule
		&      Max Pixel      &         \msun          &       \msun        &       \msun        &  \msun{}pc\pwr{-2}  &  \msun{}pc\pwr{-2}  \\
		&  \msun{}pc\pwr{-2}  &         100 pc         &       200 pc       &      (Total)       &       100 pc        &       200 pc        \\ \midrule
		ISM ($\sigma_{Gas}$) &         420         & 4.9 $\times$ 10\pwr{6} &     $ \cdots $     &    $  \cdots $     &         155         &     $  \cdots $     \\
		\HII                 &        110.0        &   $7.93\times10^{5}$   & $1.31\times10^{6}$ & $1.92\times10^{6}$ &        25.2         &        10.4         \\
		\Htwo{} (warm)       & $3.83\times10^{-2}$ &         388.5          &       695.6        &       957.2        & $1.24\times10^{-2}$ & $5.54\times10^{-3}$ \\ \bottomrule
	\end{tabular}
\end{table*}

The mass of warm \Htwo{} over the whole observed field ($\sim960~\mmsun$) compares well with that found derived by \cite{Rodriguez-Ardila2005} (900~\msun);  the estimated total (cold) \Htwo{} mass is $7\times10^8~-~2.6\times10^9~\mmsun$. \cite{Combes2002}, from radio observations of CO(1-0), deduced a total \Htwo{} mass of $3.1\times10^9~\mmsun$, which covers the whole galactic disk, including the extended nuclear region and the SF loci at the tips of the main bar. 
\subsection{AGN Torus Properties}
\begin{figure}[!htbp]
	\centering
	\includegraphics[width=.7\linewidth]{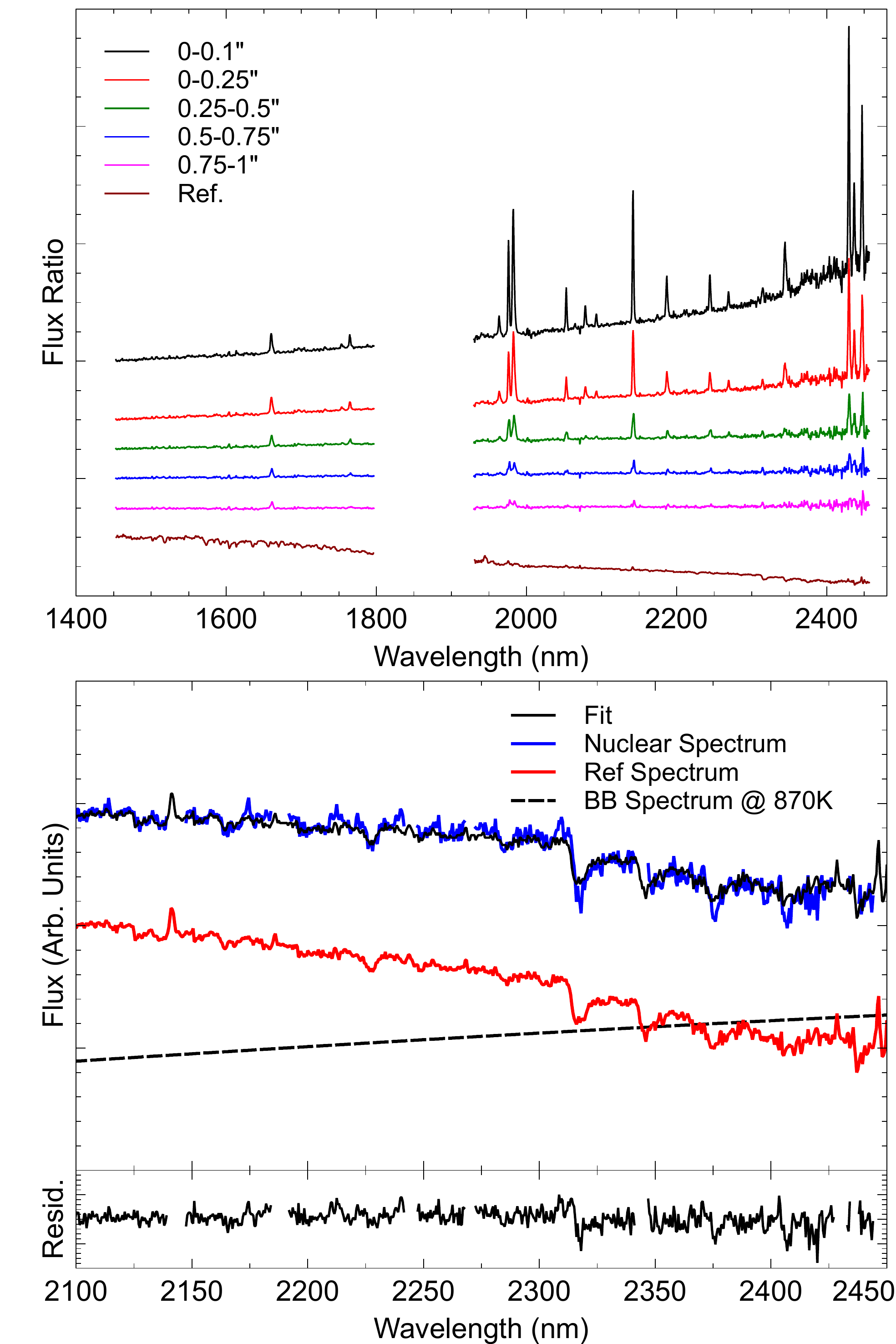}
	\caption{Hot dust contribution to the nuclear spectrum. Top panel: Ratio of continuum in annuli around the nucleus to a non-nuclear reference spectrum, showing the long-wavelength rise indicative of hot nuclear dust. The first and second plots are apertures of radius 0\arcsec.1 and 0\arcsec.25, then in annuli of 0\arcsec.25 up to 1\arcsec, around the nucleus. The reference spectrum is also plotted for comparison. The ratios have been offset for clarity. Bottom panel: Fit to nuclear flux from reference spectrum and black-body spectrum at 870~K. See text for an explanation of the fitting process. The vertical axis is in arbitrary flux units. The emission lines in the nuclear spectrum have been masked out. The residual flux is also plotted at the same scale.}
	\label{fig:ngc5728continuumratio}
\end{figure}

AGN tori typically have hot ($\sim600-1000$~K) dust emission; this can be seen as an increase in the continuum slope towards the end of the \textit{K} band. Fig. \ref{fig:ngc5728continuumratio} shows the ratio of the spectrum in annuli around the AGN location to that of a reference spectrum located 1\arcsec.3 NW of the nucleus, which is presumably representative of the underlying stellar population. This ratio is normalized at the shortest wavelength in the spectrum (1453 nm). The changing continuum slope is indicative of an increasing hot dust contribution at smaller radii. It should be emphasized that the actual continuum slopes are always negative; the inner slopes are relatively less negative. 

The inner torus directly around the AGN is encompassed within the central pixel; we can find the temperature by fitting the \textit{K}-band spectrum for this pixel with a linear combination of the reference spectrum described above and a black-body spectrum:
\begin{equation}
S_C = \alpha~S_R + \beta~S_{BB}(T) + \gamma
\end{equation}
where $S_C$ is the central pixel spectrum, $S_R$ is the reference spectrum, and $S_{BB}(T)$ is a black-body spectrum at temperature $T$; $\alpha$, $\beta$, and $\gamma$ are fitting constants. The equation was solved for $T$ using the generalized reduced gradient algorithm (`GRG Nonlinear') implemented in the \texttt{MS-Excel} add-on \textit{Solver}. The emission lines have been masked out from the central pixel spectrum. The best fit black-body spectrum is at a temperature of 870~K; this is shown in Fig. \ref{fig:ngc5728continuumratio}, bottom panel. \cite{Burtscher2015} report median hot dust temperatures of 1292~K for Seyfert 1 and 887~K for Seyfert 2 types for 51 local AGN; the value for NGC~5728 is in good agreement with this range. 

\subsection{High-Ionization Emission}
\cite{Murayama1998} proposed a model of a clumpy CLR associated with the NLR, where the individual clumps are `matter-bounded', i.e. the whole cloud is ionized and its extent is simply that of the gas cloud itself. Since, under this model, receding outflows (presumably the far side of the AGN) would preferentially show the highly-ionized face towards the observer, the emission should be more prominent on that side. However, our observations of the flux ratio \SiVI/\brg, as displayed in Fig. \ref{fig:ngc5728gasratios} do not show significant difference between the two ionization cones.

\cite{Rodriguez-Ardila2004a} deduced that photo-ionization alone is not enough to generate coronal lines at the distances and velocities from the source that our observations show. In 5 of the 6 objects studied, the coronal line emission was significantly broader and asymmetric towards the blue than low-ionization lines. In contrast to their findings, the \SiVI{} FWHM for our observations is virtually identical to that of \brg, both over the whole field and specifically along the the outflow axis. 

\begin{figure}[!htbp]
	\centering
	\includegraphics[width=.7\linewidth]{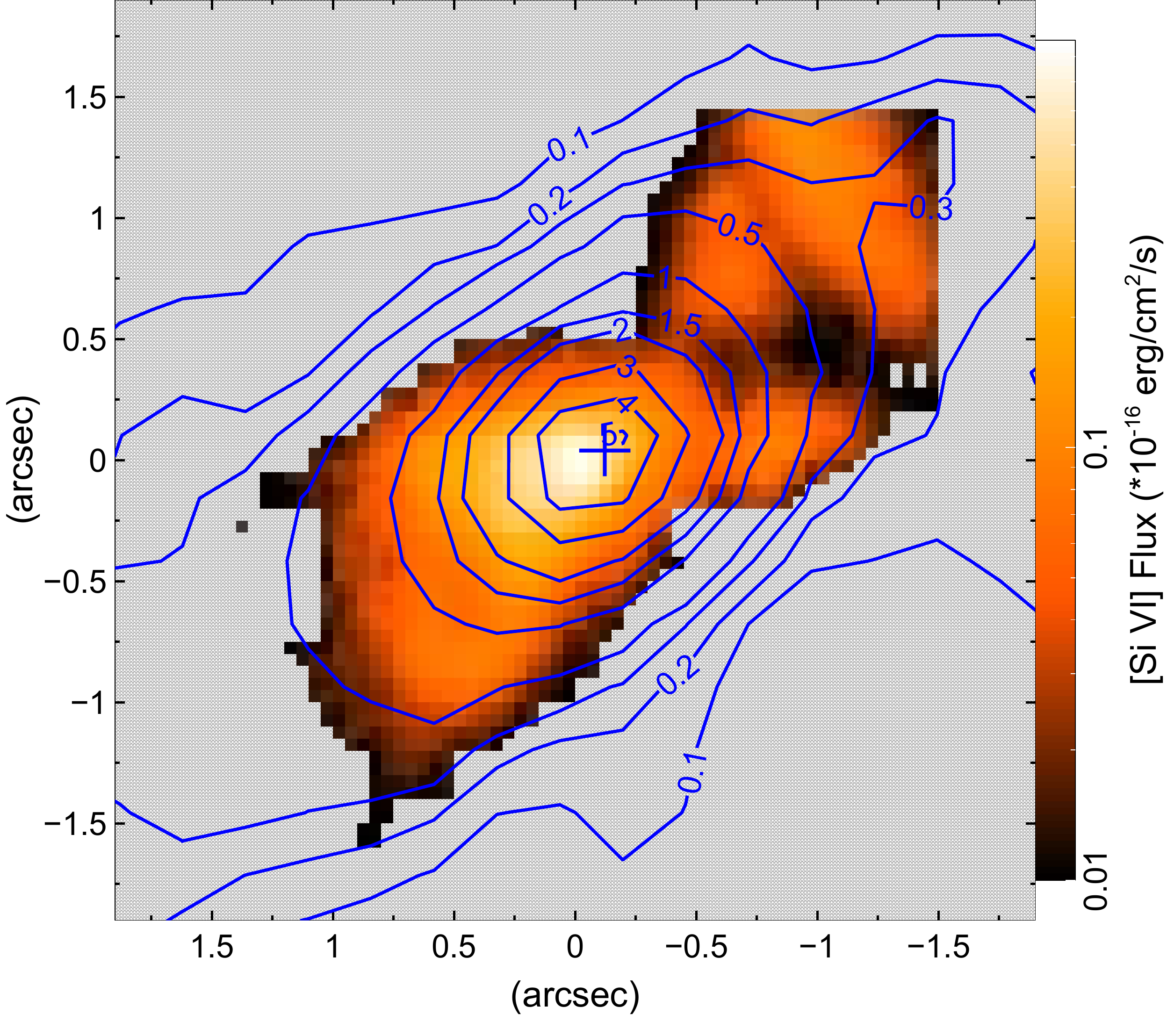}
	\caption{\SiVI{} emission-line flux (color map) overlaid by the \textit{Chandra} X-ray flux (contours). The color map is log scaled to bring out faint details; the X-ray flux contours are labeled, in units of 10\pwr{-6} photons cm\pwr{-2} s\pwr{-1}. Note the alignment of of the highly-ionized gas and the X-ray jet.}
	\label{fig:ngc5728sivix-ray}
\end{figure}

\begin{figure*}[!htbp]
	\centering
	\includegraphics[width=.8\linewidth]{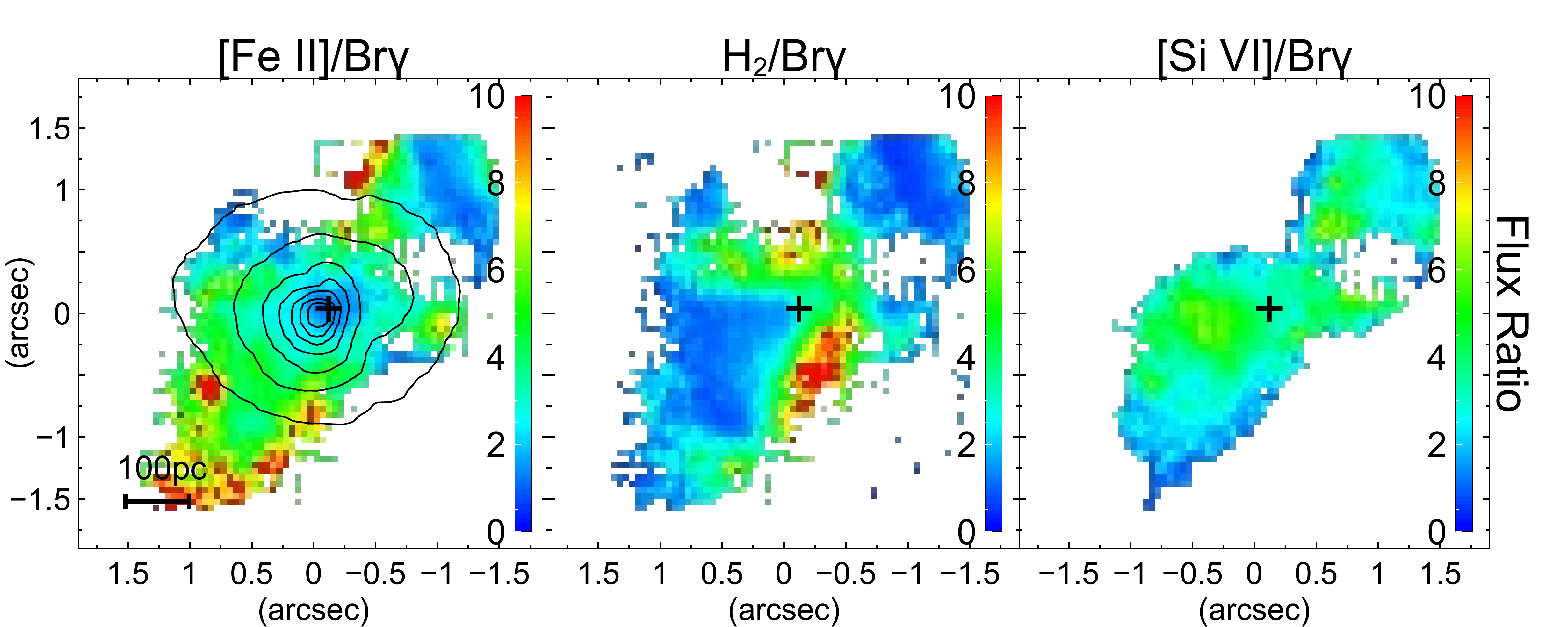}
	\caption{Flux ratios of \Fe{}, \Htwo{}, and \SiVI{} to \brg, showing relative excitation. Contours on the \Fe{}/\brg{} plot are the \textit{K}-band normalized flux, at 10, 25, 50, 75, and 90\%  of the maximum flux. \Fe{} is on the outflow boundaries, \SiVI{} is more prominent closer to the AGN, and \Htwo{} and \brg{} emissions are disjoint.}
	\label{fig:ngc5728gasratios}
\end{figure*}

The X-ray jet observed by Chandra (especially in the soft X-rays) can contribute to the photo-ionization of CL species. The X-rays can be generated by several mechanisms; direct photo-ionization from the AGN, shocks with velocities $>$200 \kms{} \citep{Rodriguez-Ardila2017}, and inverse Compton scattering of relativistic electrons associated with the radio jet. From the gas kinematics, velocities of the required magnitude are certainly present, and are aligned with the \SiVI{} emission generated by shocks. Fig. \ref{fig:ngc5728sivix-ray} shows this alignment, where the color map shows the \SiVI{} flux and the overlaid contours show the X-ray emission. The plots were aligned by equating the derived AGN location on the \SiVI{} flux map with the a Gaussian fit to the location of the central peak of the smoothed X-ray flux map. The absence of an X-ray jet in the NW ionization cone is most probably caused by higher gas column densities; this is certainly the region of the dust lane.

\cite{Rodriguez-Ardila2017} examined the powerful outflows in NGC~1386, and found that the \SiVI{} and \CaVIII{} emission extended $\sim150$ pc in one direction from the nucleus. A similar result was found by \cite{Muller-Sanchez2011} for 7 Seyfert 1.5 -- 2 galaxies; the \SiVI{} emission had radial extents from the nucleus ranging from 80--150 pc. The largest CLR so far is that of NGC~5135 \citep{Bedregal2009} of $\sim600$ pc. The CLR of NGC~5728 certainly approaches that size.

Overall, the CLR is generated by a complex mixture of photo-ionization (both directly from the AGN and from shock-generated X-rays further along the outflow cone) and shocks. The morphology and kinematics of CLR species for each galaxy will depend on the AGN properties; photo-ionizing flux from the accretion disk, outflow dynamics and collimation to generate shocks and secondary X-rays produced by those shocks. 

Assuming a uniform ISM, the \Fe/\brg{} and \SiVI/\brg{} ratios plot the relative ionizing conditions for those species. Fig. \ref{fig:ngc5728gasratios} plots the flux ratio of \Fe{} (1644 nm), \Htwo{}, and \SiVI{} to \brg. This shows that \Fe{} becomes relatively stronger towards the end and around the edges of the outflow. The \SiVI{} is more concentrated near the AGN, however it is present the full length of the outflow. The \Htwo{} emission is centrally concentrated, but as we will see, is kinematically disassociated from the outflow and ionized species. This will be explored in Sections \ref{sec:ngc5728clr} and \ref{sec:ngc5728excitation}.

\subsection{The Star-Forming Ring}
The stellar age in the SF ring was determined from the \pab{} EW, using the STARBURST99 models of \cite{Leitherer1999}. These produce EW data for a range of metallicities and initial mass functions; all showing a static EW until about 3--5 Myr, then a rapid decrease to about 10 Myr. Three locations in the ring, but out of the line of the outflows, were used, as shown in Fig. \ref{fig:ngc5728gaskinematics1} top right panel (circles); the EW was averaged within a radius of 3 pixels. These values were in the range 0.16 -- 0.42 nm, which is equivalent to 7.4 -- 8.4 Myr for the model a Solar metallicity, a mass cutoff of 100 \msun{}, and an IMF slope of 2.35.

We can obtain a measure of the SF rate in the ring, using the \Ha{} flux and the \cite{Kennicutt2009} relationship. We estimate the true, non-AGN contaminated flux by selecting pixels with the AGN mixing ratio $<$ 50\%. We also select pixels where the Balmer decrement-derived extinction is less than 4 mag, as the derived extinction is probably uncertain over that value. We then correct the flux at each pixel by the formula:
\begin{equation}
F(H\alpha)_{C} = F(H\alpha)_{O} \times 10^{0.4~A_{v}~k(\lambda)}
\end{equation} 
where $F(H\alpha)_{O}$ and $F(H\alpha)_{C}$ are the observed and corrected \Ha{} fluxes, respectively, and $k(\lambda) = A_{\lambda}/A_{v}$. For \Ha, $k(\lambda) = 0.818$, from the \cite{Cardelli1989} formulation. 

\begin{figure*}
	\centering
	\includegraphics[width=.9\linewidth]{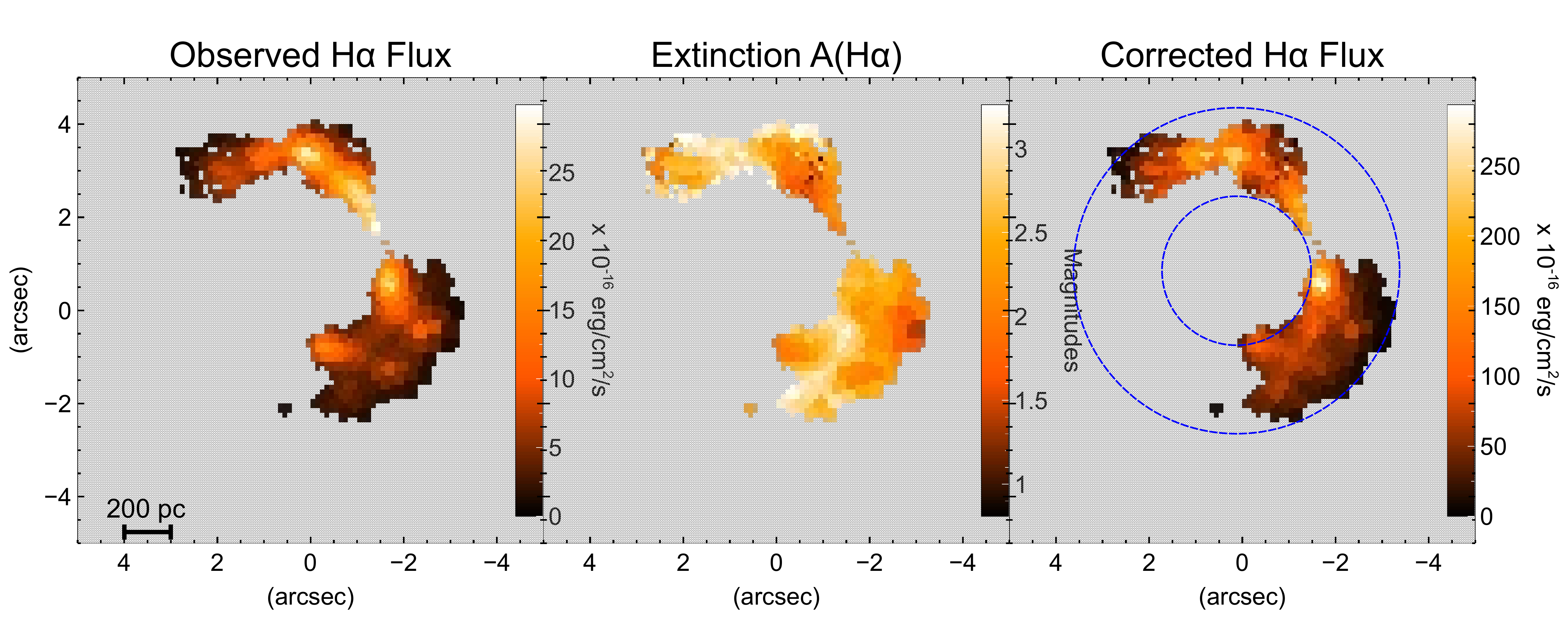}
	\caption{Star-forming ring \Ha{} flux for central 10\arcsec (note the scale in left panel). Left panel: observed flux. Middle panel: A(\Ha) extinction (magnitudes). Right panel: corrected flux; the blue dashed circles show the annulus used to estimate the total SF rate. Flux values in units of 10\pwr{-16} \ecs.}
	\label{fig:ngc5728_SFR_SFR}
\end{figure*}

Fig. \ref{fig:ngc5728_SFR_SFR} show the \Ha{} observed and corrected fluxes and the extinction map for the nuclear region (10\arcsec - note the scale). The average \Ha{} flux for the selected pixels is $7.41~\times~10^{-16}$ \ecs{} ($L(H\alpha)~=~1.5 \times 10^{38}$ \es{} at 41.1 Mpc), which gives a star-formation rate of $\sim 0.029$ \msy{} arcsec\pwr{-2} (0\arcsec.2/pixel). We can estimate the SF ring size by assuming an annular geometry around the nucleus; this is measured at $\sim$ 3\arcsec.2 for the inner and 7\arcsec for the outer radii. The total SFR over the ring is thus $\sim3.5$ \msy.

We can also estimate the stellar ages from the \Hb{} EW, using the same method as for \pab{}, above. The \Hb{} EWs for the flux peaks around the ring (interpreted as individual star clusters), is in the range 0.78 to 1.43 nm; this equates to stellar ages 6.15--6.45 Myr (broadly compatible with the values derived from \Hb).

If the total VLA 6 cm flux, as shown in Fig. \ref{fig:ngc5728images21} (bottom left panel), is summed on the NE sector of the circle (since the jet interferes with the SW sector), a value of $\sim70$ mJy is obtained. At the distance of 41.1 Mpc, this translates to a luminosity of $\sim1.4\times10^{22}$ W Hz\pwr{-1}. Using the indicator of \cite{Condon1992} (their equation 18) where $\nu~=~5$ GHz and $\alpha~=~-0.8$, this is equivalent to a supernova rate of $\sim$0.4 yr\pwr{-1}; this value may well be high, as the jet emission may be contributing to the measured flux.  The individual supernova remnant (SNR) luminosities are $\sim4.4\times10^{20}$ W m\pwr{-2}; again the unmasked jet emission will contribute some of this luminosity. This can be compared to the SNRs in M82, which have luminosities in the range $0.02 - 1.7\times10^{20}$ W m\pwr{-2} \citep{Muxlow1994}. The spacing of the SNRs is reminiscent of the nuclear rings of star clusters; examples are cited in \cite{Brandl2012} and \cite{Pan2013} for NGC~7552 and \cite{Boker2008} for 5 nearby spiral galaxies.

\section{Conclusion}
We have examined the nuclear region of NGC~5728 using data at multiple wavelengths and methods (X-ray, optical and NIR IFU spectra, and radio), revealing a highly complex object showing gas with multiple morphologies and excitation modes, driven by a powerful AGN. We have presented and analyzed emission line morphologies  and excitation diagnostics, estimating stellar populations and ages and obscuration from continuum emission and line ratios and also determined gas column densities.

In summary, the nuclear region has the following features:
\begin{itemize}
	\item Dust lanes and spiraling filaments feed the nucleus, as revealed by the \textit{HST} structure maps.
	\item A star-forming ring with a stellar age of 7.4 -- 8.4 Myr, with a SFR $\simeq 3.5$ \msy.
	\item A one-sided radio jet, impacting on the ISM at about 200 pc from the nucleus, combined with radio emission from the SN remnants in the SF ring.
	\item Biconal ionization, traced by gas emission of \HII, \Fe, and \SiVI, extending off the edges of the observed field. MUSE \OIII{} data shows that the full extent of the cones is over 2.5 kpc from the nucleus. \item The AGN and BLR is hidden by a dust bar of size $64 \times 28$ pc, with up to 19 magnitudes of visual extinction, with an estimated dust temperature at the nuclear position of $\sim870$~K. Extinction maps derived from both hydrogen recombination and \Fe{} emission lines show similar structures, with some indication that \Fe-derived extinction does not probe the full depth of the ionization cones. There is some evidence for dust being sublimated in the outflows, reducing the extinction.  This is supported by extinction measures derived from MUSE \Ha/\Hb{} emission, which show a  decreased absorption in the ionization cones.
	\item An extended coronal-line region traced by \SiVI, out to $\sim300$ pc from the nucleus. This is excited by direct photo-ionization from the AGN, plus shocks from the high-velocity outflows. Higher-ionization species (\CaVIII{} and \SIX) are closely confined around the AGN.
	\item Line-ratio active galaxy diagnostics show mostly AGN-type activity, with small regions of LINER and TO modes at the peripheries of the outflows and where the dusty torus obscures the direct photo-ionization from the BH.  MUSE optical diagnostics (BPT diagrams) show a clean star-formation/AGN mixing sequence.
	\item The \Htwo{} gas is dynamically and spatially independent of the outflows, concentrated in an equatorial disk in the star-forming ring, but also showing entrainment along the sides of the bicone. The warm \Htwo{} has a mass of 960 \msun in the central $600\times600$ pc, with an estimated $6\times10^8~\mmsun$ of cold \Htwo{} in the field of view. Line-ratio diagnostics indicate that the gas is excited by thermal processes (shocks and radiative heating of gas masses) to temperatures in the range 1400-- 2100~K, with an increased ratio of fluorescent excitation towards the SF ring.
	\item With 100 pc of the nucleus, the ionized gas mass (derived from the \brg{} emission) is $8 \times 10^5 \mmsun$ and the cold \HI{} gas mass (from extinction calculations) is $4.9 \times 10^6 \mmsun$; however this could be underestimated due to dust sublimation in the outflows.
\end{itemize}

To understand the complex nature of AGNs, their associated nuclear gas and stars must be studied at the highest resolution possible using multi-messenger observations. ALMA observations, revealing the cool dust component, would add materially to the physical map we have constructed here.
\acknowledgments
This study is based on observations made with ESO Telescopes at the Paranal Observatory. The respective program IDs are given in section \ref{sec:ngc5728Observations} above.

This research has made use of the NASA/IPAC Infrared Science Archive, which is operated by the Jet Propulsion Laboratory, California Institute of Technology, under contract with the National Aeronautics and Space Administration. The SIMBAD Astronomical Database is operated by CDS, Strasbourg, France. IRAF is distributed by the National Optical Astronomy Observatories, which is operated by the Association of Universities for Research in Astronomy, Inc. (AURA) under cooperative agreement with the National Science Foundation. The National Radio Astronomy Observatory is a facility of the National Science Foundation operated under cooperative agreement by Associated Universities, Inc.

The Pan-STARRS1 Surveys (PS1) and the PS1 public science archive have been made possible through contributions by the Institute for Astronomy, the University of Hawaii, the Pan-STARRS Project Office, the Max-Planck Society and its participating institutes, the Max Planck Institute for Astronomy, Heidelberg and the Max Planck Institute for Extraterrestrial Physics, Garching, The Johns Hopkins University, Durham University, the University of Edinburgh, the Queen's University Belfast, the Harvard-Smithsonian Center for Astrophysics, the Las Cumbres Observatory Global Telescope Network Incorporated, the National Central University of Taiwan, the Space Telescope Science Institute, the National Aeronautics and Space Administration under Grant No. NNX08AR22G issued through the Planetary Science Division of the NASA Science Mission Directorate, the National Science Foundation Grant No. AST-1238877, the University of Maryland, Eotvos Lorand University (ELTE), the Los Alamos National Laboratory, and the Gordon and Betty Moore Foundation.


We would like to thank our Triplespec colleagues at Caltech for their ongoing support of our survey. 
We acknowledge the continued support of the Australian Research Council (ARC) through Discovery project DP140100435. M.D. dedicates this paper (and Paper II) to the memory of Dr. Ross McGregor Mitchell (1954-2018), atmospheric scientist and astrophysicist.

\facilities{VLT:Yepun (SINFONI, MUSE)}
\software{TinyTim \citep{Krist2011}, QFitsView \citep{Ott2012}, gasgano \citep{ESO2012}}
\bibliographystyle{apj}
\bibliography{library.bib}

\begin{thebibliography}{99}
\expandafter\ifx\csname natexlab\endcsname\relax\def\natexlab#1{#1}\fi

\bibitem[{Acker \& Jaschek(1986)}]{Acker1986}
Acker, A., \& Jaschek, C. 1986, {Astronomical methods and calculations}
  (Chichester, New York: Wiley)

\bibitem[{Arribas \& Mediavilla(1993)}]{Arribas1993}
\href{http://adsabs.harvard.edu/doi/10.1086/172774}{{Arribas, S., \&
  Mediavilla, E.}} 1993, ApJ, 410, 552

\bibitem[{Bacon {et~al.}(2010)Bacon, Accardo, Adjali, Anwand, Bauer, Biswas,
  Blaizot, Boudon, Brau-Nogue, Brinchmann, Caillier, Capoani, Carollo, Contini,
  Couderc, Daguis{\'{e}}, Deiries, Delabre, Dreizler, Dubois, Dupieux, Dupuy,
  Emsellem, Fechner, Fleischmann, Fran{\c{c}}ois, Gallou, Gharsa, Glindemann,
  Gojak, Guiderdoni, Hansali, Hahn, Jarno, Kelz, Koehler, Kosmalski, Laurent,
  {Le Floch}, Lilly, Lizon, Loupias, Manescau, Monstein, Nicklas, Olaya, Pares,
  Pasquini, P{\'{e}}contal-Rousset, Pell{\'{o}}, Petit, Popow, Reiss,
  Remillieux, Renault, Roth, Rupprecht, Serre, Schaye, Soucail, Steinmetz,
  Streicher, Stuik, Valentin, Vernet, Weilbacher, Wisotzki, \&
  Yerle}]{Bacon2010}
\href{http://proceedings.spiedigitallibrary.org/proceeding.aspx?doi=10.1117/12.856027}{{Bacon,
  R., Accardo, M., Adjali, L., {et~al.}}} 2010, Proc. SPIE, 7735, 773508

\bibitem[{Baldwin {et~al.}(1981)Baldwin, Phillips, \& Terlevich}]{Baldwin1981}
\href{http://iopscience.iop.org/article/10.1086/130766}{{Baldwin, J.~A.,
  Phillips, M.~M., \& Terlevich, R.}} 1981, PASP, 93, 5

\bibitem[{Bedregal {et~al.}(2009)Bedregal, Colina, Alonso-Herrero, \&
  Arribas}]{Bedregal2009}
Bedregal, A.~G., Colina, L., Alonso-Herrero, A., \& Arribas, S. 2009, ApJ, 698,
  1852

\bibitem[{Binney \& Merrifield(1998)}]{Binney1998}
Binney, J., \& Merrifield, M. 1998, {Galactic Astronomy} (Princeton, NJ:
  Princeton University Press)

\bibitem[{Black \& van Dishoeck(1987)}]{Black1987}
\href{http://adsabs.harvard.edu/doi/10.1086/165740}{{Black, J.~H., \& van
  Dishoeck, E.~F.}} 1987, ApJ, 322, 412

\bibitem[{B{\"{o}}ker {et~al.}(2008)B{\"{o}}ker, Falc{\'{o}}n-Barroso,
  Schinnerer, Knapen, \& Ryder}]{Boker2008}
B{\"{o}}ker, T., Falc{\'{o}}n-Barroso, J., Schinnerer, E., Knapen, J.~H., \&
  Ryder, S. 2008, AJ, 135, 479

\bibitem[{Brandl {et~al.}(2012)Brandl, Mart{\'{i}}n-Hern{\'{a}}ndez, Schaerer,
  Rosenberg, \& van~der Werf}]{Brandl2012}
Brandl, B.~R., Mart{\'{i}}n-Hern{\'{a}}ndez, N.~L., Schaerer, D., Rosenberg,
  M., \& van~der Werf, P.~P. 2012, A{\&}A, 543, 61

\bibitem[{Burtscher {et~al.}(2015)Burtscher, {Orban de Xivry}, Davies, Janssen,
  Lutz, Rosario, Contursi, Genzel, Graci{\textperiodcentered}-Carpio, Lin,
  Schnorr-M¸ller, Sternberg, Sturm, \& Tacconi}]{Burtscher2015}
Burtscher, L., {Orban de Xivry}, G., Davies, R.~I., {et~al.} 2015, A{\&}A, 578,
  47

\bibitem[{Burtscher {et~al.}(2016)Burtscher, Davies, Graci{\'{a}}-Carpio, Koss,
  Lin, Lutz, Nandra, Netzer, {Orban de Xivry}, Ricci, Rosario, Veilleux,
  Contursi, Genzel, Schnorr-M{\"{u}}ller, Sternberg, Sturm, \&
  Tacconi}]{Burtscher2016}
Burtscher, L., Davies, R.~I., Graci{\'{a}}-Carpio, J., {et~al.} 2016, A{\&}A,
  586, A28

\bibitem[{Busch {et~al.}(2017)Busch, Eckart, Valencia-S., Fazeli,
  Scharw{\"{a}}chter, Combes, \& Garc{\'{i}}a-Burillo}]{Busch2016}
Busch, G., Eckart, A., Valencia-S., M., {et~al.} 2017, A{\&}A, 598, A55

\bibitem[{Buta {et~al.}(2015)Buta, Sheth, Athanassoula, Bosma, Knapen,
  Laurikainen, Salo, Elmegreen, Ho, Zaritsky, Courtois, Hinz,
  Mu{\~{n}}oz-Mateos, Kim, Regan, Gadotti, de~Paz, Laine, Menendez-Delmestre,
  Comeron, Ferrer, Seibert, Mizusawa, Holwerda, \& Madore}]{Buta2015}
\href{http://arxiv.org/abs/1501.00454}{{Buta, R., Sheth, K., Athanassoula, E.,
  {et~al.}}} 2015, ApJSS, 217, 32

\bibitem[{Capetti {et~al.}(1996)Capetti, Axon, Macchetto, Sparks, \&
  Boksenberg}]{Capetti1996}
\href{http://adsabs.harvard.edu/doi/10.1086/177501}{{Capetti, A., Axon, D.~J.,
  Macchetto, F., Sparks, W.~B., \& Boksenberg, A.}} 1996, ApJ, 466, 169

\bibitem[{Cappellari \& Copin(2003)}]{Cappellari2003}
Cappellari, M., \& Copin, Y. 2003, MNRAS, 342, 345

\bibitem[{Cardelli {et~al.}(1989)Cardelli, Clayton, \& Mathis}]{Cardelli1989}
Cardelli, J.~A., Clayton, G.~C., \& Mathis, J.~S. 1989, ApJ, 345, 245

\bibitem[{Catinella {et~al.}(2005)Catinella, Haynes, \&
  Giovanelli}]{Catinella2005}
\href{http://adsabs.harvard.edu/cgi-bin/bib{\_}query?2005AJ....130.1037C
  http://vizier.u-strasbg.fr/viz-bin/VizieR-3?-source=J/AJ/130/1037{\&}-out.max=50{\&}-out.form=HTML
  Table{\&}-out.add={\_}r{\&}-out.add={\_}RAJ,{\_}DEJ{\&}-sort={\_}r{\&}-oc.form=sexa}{{Catinella,
  B., Haynes, M.~P., \& Giovanelli, R.}} 2005, AJ, 130, 1037

\bibitem[{Colina {et~al.}(2015)Colina, {Piqueras L{\'{o}}pez}, Arribas, Riffel,
  Riffel, Rodriguez-Ardila, Pastoriza, Storchi-Bergmann, Alonso-Herrero, \&
  Sales}]{Colina2015}
Colina, L., {Piqueras L{\'{o}}pez}, J., Arribas, S., {et~al.} 2015, A{\&}A,
  578, 48

\bibitem[{Combes \& Leon(2002)}]{Combes2002}
\href{http://arxiv.org/abs/astro-ph/0209267}{{Combes, F., \& Leon, S.}} 2002,
  in SF2A-2002, ed. F.~Combes \& D.~Barret (Paris: EDP-Sciences), 1--2

\bibitem[{Condon(1992)}]{Condon1992}
Condon, J.~J. 1992, ARA{\&}A, 30, 575

\bibitem[{Davies {et~al.}(2003)Davies, Sternberg, Lehnert, \&
  Tacconi-Garman}]{Davies2003}
\href{http://adsabs.harvard.edu/cgi-bin/nph-data{\_}query?bibcode=2003ApJ...597..907D{\&}link{\_}type=ABSTRACT{\%}5Cnpapers2://publication/doi/10.1086/378634}{{Davies,
  R.~I., Sternberg, a., Lehnert, M., \& Tacconi-Garman, L.~E.}} 2003, ApJ, 597,
  907

\bibitem[{Davies {et~al.}(2005)Davies, Sternberg, Lehnert, \&
  Tacconi-Garman}]{Davies2005}
Davies, R.~I., Sternberg, A., Lehnert, M.~D., \& Tacconi-Garman, L.~E. 2005,
  ApJ, 633, 105

\bibitem[{Davies {et~al.}(2015)Davies, Burtscher, Rosario, Storchi-Bergmann,
  Contursi, Genzel, Carpio, Hicks, Janssen, Koss, Lin, Lutz, Maciejewski,
  S{\'{a}}nchez, de~Xivry, Ricci, Riffel, Riffel, Schartmann, M{\"{u}}ller,
  Sternberg, Sturm, Tacconi, \& Veilleux}]{Davies2015}
\href{http://stacks.iop.org/0004-637X/806/i=1/a=127?key=crossref.3005b9026bba182382c39c5b13098e2a}{{Davies,
  R.~I., Burtscher, L., Rosario, D., {et~al.}}} 2015, ApJ, 806, 127

\bibitem[{Davies {et~al.}(2014{\natexlab{a}})Davies, Kewley, Ho, \&
  Dopita}]{Davies2014b}
Davies, R.~L., Kewley, L.~J., Ho, I.~T., \& Dopita, M.~A. 2014{\natexlab{a}},
  MNRAS, 444, 3961

\bibitem[{Davies {et~al.}(2014{\natexlab{b}})Davies, Rich, Kewley, \&
  Dopita}]{Davies2014a}
Davies, R.~L., Rich, J.~A., Kewley, L.~J., \& Dopita, M.~A. 2014{\natexlab{b}},
  MNRAS, 439, 3835

\bibitem[{Davies {et~al.}(2016)Davies, Groves, Kewley, Dopita, Hampton,
  Shastri, Scharwachter, Sutherland, Kharb, Bhatt, Jin, Banfield, Zaw, James,
  Juneau, \& Srivastava}]{Davies2016}
Davies, R.~L., Groves, B., Kewley, L.~J., {et~al.} 2016, MNRAS, 462, 1616

\bibitem[{Davies {et~al.}(2017)Davies, Groves, Kewley, Medling, Shastri,
  Maithil, Kharb, Banfield, Longbottom, Dopita, Hampton, Scharw{\"{a}}chter,
  Sutherland, Jin, Zaw, James, \& Juneau}]{Davies2017a}
\href{https://arxiv.org/pdf/1707.03404.pdf
  http://academic.oup.com/mnras/article/470/4/4974/3896160/Dissecting-galaxies-separating-star-formation}{{Davies,
  R.~L., Groves, B., Kewley, L.~J., {et~al.}}} 2017, MNRAS, 470, 4974

\bibitem[{de~Vaucouleurs {et~al.}(1991)de~Vaucouleurs, de~Vaucouleurs, {Corwin
  H. G.}, Buta, Paturel, \& Fouqu{\'{e}}}]{DeVaucouleurs1991}
de~Vaucouleurs, G., de~Vaucouleurs, A., {Corwin H. G.}, J., {et~al.} 1991,
  {Third Reference Catalogue of Bright Galaxies.} (New York: Springer-Verlag)

\bibitem[{Dopita {et~al.}(2015)Dopita, Shastri, Davies, Kewley, Hampton,
  Scharw‰chter, Sutherland, Kharb, Jose, Bhatt, Ramya, Jin, Banfield, Zaw,
  Juneau, James, \& Srivastava}]{Dopita2015a}
Dopita, M.~A., Shastri, P., Davies, R., {et~al.} 2015, ApJSS, 217, 12

\bibitem[{Durr{\'{e}} \& Mould(2014)}]{Durre2014}
\href{http://adsabs.harvard.edu/abs/2014ApJ...784...79D}{{Durr{\'{e}}, M., \&
  Mould, J.}} 2014, ApJ, 784, 79

\bibitem[{Durr{\'{e}} {et~al.}(2017)Durr{\'{e}}, Mould, Schartmann, Uddin, \&
  Cotter}]{Durre2017}
\href{http://stacks.iop.org/0004-637X/838/i=2/a=102?key=crossref.4652be8fbbb86c6af289f4b20597b29d
  http://adsabs.harvard.edu/abs/2017ApJ...838..102D}{{Durr{\'{e}}, M., Mould,
  J., Schartmann, M., Uddin, S.~A., \& Cotter, G.}} 2017, ApJ, 838, 102

\bibitem[{Elitzur(2008)}]{Elitzur2008}
Elitzur, M. 2008, New A Rev., 52, 274

\bibitem[{ESO(2012)}]{ESO2012}
ESO. 2012, {GASGANO: Data File Organizer}, Astrophysics Source Code
  Library:1210.020

\bibitem[{Evans {et~al.}(2010)Evans, Primini, Glotfelty, Anderson, Bonaventura,
  Chen, Davis, Doe, Evans, Fabbiano, Galle, Gibbs, Grier, Hain, Hall, Harbo,
  He, Houck, Karovska, Kashyap, Lauer, McCollough, McDowell, Miller, Mitschang,
  Morgan, Mossman, Nichols, Nowak, Plummer, Refsdal, Rots, Siemiginowska,
  Sundheim, Tibbetts, {Van Stone}, Winkelman, \& Zografou}]{Evans2010a}
\href{http://stacks.iop.org/0067-0049/189/i=1/a=37?key=crossref.42a9a7044ff77f3df9aa9f5aa3c7d167}{{Evans,
  I.~N., Primini, F.~A., Glotfelty, K.~J., {et~al.}}} 2010, ApJSS, 189, 37

\bibitem[{George {et~al.}(2016)George, Gr{\"{a}}ff, Feuchtgruber, Hartl,
  Eisenhauer, Buron, Davies, Genzel, Huber, Rau, Plattner, Wiezorrek, Weisz,
  Amico, Glindeman, Hau, Kuntschner, \& Modigliani}]{George2016}
\href{http://arxiv.org/abs/1608.02457{\%}0Ahttp://dx.doi.org/10.1117/12.2231285}{{George,
  E.~M., Gr{\"{a}}ff, D., Feuchtgruber, H., {et~al.}}} 2016, Proc. SPIE, 9908,
  1

\bibitem[{Gorkom(1982)}]{Gorkom1982}
Gorkom, J. H.~V. 1982, The Messenger, 33, 45

\bibitem[{Ho {et~al.}(2011)Ho, Li, Barth, Seigar, \& Peng}]{Ho2011}
\href{http://stacks.iop.org/0067-0049/197/i=2/a=21?key=crossref.4a8109a6abfa22b7017456ddb56fa0cd}{{Ho,
  L.~C., Li, Z.-Y., Barth, A.~J., Seigar, M.~S., \& Peng, C.~Y.}} 2011, ApJSS,
  197, 21

\bibitem[{Hollenbach \& McKee(1989)}]{Hollenbach1989}
\href{http://adsabs.harvard.edu/doi/10.1086/167595}{{Hollenbach, D., \& McKee,
  C.~F.}} 1989, ApJ, 342, 306

\bibitem[{Hopkins \& Hernquist(2006)}]{Hopkins2006}
Hopkins, P.~F., \& Hernquist, L. 2006, ApJSS, 166, 1

\bibitem[{Hummer \& Storey(1987)}]{Hummer1987}
Hummer, D.~G., \& Storey, P.~J. 1987, MNRAS, 224, 801

\bibitem[{Kennicutt {et~al.}(2009)Kennicutt, Hao, Calzetti, Moustakas, Dale,
  Bendo, Engelbracht, Johnson, \& Lee}]{Kennicutt2009}
Kennicutt, R.~C., Hao, C.-N., Calzetti, D., {et~al.} 2009, ApJ, 703, 1672

\bibitem[{Kewley {et~al.}(2006)Kewley, Groves, Kauffmann, \&
  Heckman}]{Kewley2006}
\href{http://arxiv.org/abs/astro-ph/0605681}{{Kewley, L.~J., Groves, B.,
  Kauffmann, G., \& Heckman, T.}} 2006, MNRAS, 372, 961

\bibitem[{Krabbe {et~al.}(2000)Krabbe, {Sams III}, Genzel, Thatte, \&
  Prada}]{Krabbe2000}
\href{http://ads.inasan.ru/abs/2000A{\&}A...354..439K}{{Krabbe, A., {Sams III},
  B., Genzel, R., Thatte, N., \& Prada, F.}} 2000, A{\&}A, 354, 439

\bibitem[{Krist {et~al.}(2011)Krist, Hook, \& Stoehr}]{Krist2011}
\href{http://proceedings.spiedigitallibrary.org/proceeding.aspx?doi=10.1117/12.892762}{{Krist,
  J.~E., Hook, R.~N., \& Stoehr, F.}} 2011, Proc. SPIE, 8127, 81270J

\bibitem[{Kwok(2007)}]{Kwok2007}
\href{http://adsabs.harvard.edu/abs/2007oepn.book.....K}{{Kwok, S.}} 2007, {The
  Origin and Evolution of Planetary Nebulae} (Cambridge, UK: Cambridge
  University Press)

\bibitem[{Larkin {et~al.}(1998)Larkin, Armus, Knop, Soifer, \&
  Matthews}]{Larkin1998}
Larkin, J.~E., Armus, L., Knop, R.~A., Soifer, B.~T., \& Matthews, K. 1998,
  ApJSS, 114, 59

\bibitem[{Leitherer {et~al.}(1999)Leitherer, Schaerer, Goldader, Delgado,
  Robert, Kune, de~Mello, Devost, \& Heckman}]{Leitherer1999}
\href{http://arxiv.org/abs/astro-ph/9902334{\%}0A
  http://dx.doi.org/10.1086/313233}{{Leitherer, C., Schaerer, D., Goldader,
  J.~D., {et~al.}}} 1999, ApJSS, 123, 3

\bibitem[{Lena {et~al.}(2015)Lena, Robinson, Storchi-Bergman, Schnorr-M¸ller,
  Seelig, Riffel, Nagar, Couto, \& Shadler}]{Lena2015}
Lena, D., Robinson, A., Storchi-Bergman, T., {et~al.} 2015, ApJ, 806, 84

\bibitem[{Maloney {et~al.}(1996)Maloney, Hollenbach, \& Tielens}]{Maloney1996}
\href{http://adsabs.harvard.edu/doi/10.1086/177532}{{Maloney, P.~R.,
  Hollenbach, D.~J., \& Tielens, A. G. G.~M.}} 1996, ApJ, 466, 561

\bibitem[{Mazzalay {et~al.}(2013{\natexlab{a}})Mazzalay,
  Rodr{\'{i}}guez-Ardila, Komossa, \& McGregor}]{Mazzalay2013}
Mazzalay, X., Rodr{\'{i}}guez-Ardila, A., Komossa, S., \& McGregor, P.~J.
  2013{\natexlab{a}}, MNRAS, 430, 2411

\bibitem[{Mazzalay {et~al.}(2013{\natexlab{b}})Mazzalay, Saglia, Erwin,
  Fabricius, Rusli, Thomas, Bender, Opitsch, Nowak, \&
  Williams}]{Mazzalay2013a}
Mazzalay, X., Saglia, R.~P., Erwin, P., {et~al.} 2013{\natexlab{b}}, MNRAS,
  428, 2389

\bibitem[{Menezes {et~al.}(2015)Menezes, da~Silva, Ricci, Steiner, May, \&
  Borges}]{Menezes2015a}
Menezes, R.~B., da~Silva, P., Ricci, T.~V., {et~al.} 2015, MNRAS, 450, 369

\bibitem[{Menezes {et~al.}(2014)Menezes, Steiner, \& Ricci}]{Menezes2014}
Menezes, R.~B., Steiner, J.~E., \& Ricci, T.~V. 2014, MNRAS, 438, 2597

\bibitem[{Mezcua {et~al.}(2015)Mezcua, Prieto, Fern{\'{a}}ndez-Ontiveros,
  Tristram, Neumayer, \& Kotilainen}]{Mezcua2015}
Mezcua, M., Prieto, M.~A., Fern{\'{a}}ndez-Ontiveros, J.~A., {et~al.} 2015,
  MNRAS, 452, 4128

\bibitem[{Mould {et~al.}(2000)Mould, Huchra, Freedman, {Kennicutt, Jr.},
  Ferrarese, Ford, Gibson, Graham, Hughes, Illingworth, Kelson, Macri, Madore,
  Sakai, Sebo, Silbermann, \& Stetson}]{Mould2000a}
\href{http://arxiv.org/abs/astro-ph/9909260{\%}5Cnhttp://stacks.iop.org/0004-637X/529/i=2/a=786}{{Mould,
  J.~R., Huchra, J.~P., Freedman, W.~L., {et~al.}}} 2000, ApJ, 529, 786

\bibitem[{Mouri(1994)}]{Mouri1994}
Mouri, H. 1994, ApJ, 427, 777

\bibitem[{Mouri {et~al.}(2000)Mouri, Kawara, \& Taniguchi}]{Mouri2000}
Mouri, H., Kawara, K., \& Taniguchi, Y. 2000, ApJ, 528, 186

\bibitem[{{M{\"{u}}ller S{\'{a}}nchez} {et~al.}(2006){M{\"{u}}ller
  S{\'{a}}nchez}, Davies, Eisenhauer, Tacconi, Genzel, \&
  Sternberg}]{MullerSanchez2006}
{M{\"{u}}ller S{\'{a}}nchez}, F., Davies, R.~I., Eisenhauer, F., {et~al.} 2006,
  A{\&}A, 454, 481

\bibitem[{M{\"{u}}ller-S{\'{a}}nchez {et~al.}(2011)M{\"{u}}ller-S{\'{a}}nchez,
  Prieto, Hicks, Vives-Arias, Davies, Malkan, Tacconi, \&
  Genzel}]{Muller-Sanchez2011}
M{\"{u}}ller-S{\'{a}}nchez, F., Prieto, M.~A., Hicks, E. K.~S., {et~al.} 2011,
  ApJ, 739, 69

\bibitem[{Murayama \& Taniguchi(1998)}]{Murayama1998}
Murayama, T., \& Taniguchi, Y. 1998, ApJL, 497, L9

\bibitem[{Muxlow {et~al.}(1994)Muxlow, Pedlar, Wilkinson, Axon, Sanders, \&
  de~Bruyn}]{Muxlow1994}
\href{https://academic.oup.com/mnras/article-lookup/doi/10.1093/mnras/266.2.455}{{Muxlow,
  T. W.~B., Pedlar, A., Wilkinson, P.~N., {et~al.}}} 1994, MNRAS, 266, 455

\bibitem[{Netzer(2015)}]{Netzer2015}
\href{http://arxiv.org/abs/1505.00811{\%}0Ahttp://dx.doi.org/10.1146/annurev-astro-082214-122302}{{Netzer,
  H.}} 2015, ARA{\&}A, 53, 365

\bibitem[{Oliva {et~al.}(2001)Oliva, Marconi, Maiolino, Testi, Mannucci,
  Ghinassi, Licandro, Origlia, Baffa, Checcucci, Comoretto, Gavryussev,
  Gennari, Giani, Hunt, Lisi, Lorenzetti, Marcucci, Miglietta, Sozzi,
  Stefanini, \& Vitali}]{Oliva2001}
\href{http://arxiv.org/abs/astro-ph/0102159}{{Oliva, E., Marconi, A., Maiolino,
  R., {et~al.}}} 2001, A{\&}A, 369, L5

\bibitem[{Osterbrock \& Ferland(2006)}]{Osterbrock2006}
Osterbrock, D.~E., \& Ferland, G.~J. 2006, {Astrophysics of gaseous nebulae and
  active galactic nuclei}, 2nd edn. (Sausalito, CA: University Science Books)

\bibitem[{Ott(2012)}]{Ott2012}
\href{https://ascl.net/1210.019}{{Ott, T.}} 2012, {QFitsView: FITS file
  viewer}, Astrophysics Source Code Library:1210.019

\bibitem[{Pan {et~al.}(2013)Pan, Lim, Matsushita, Wong, \& Ryder}]{Pan2013}
Pan, H.-A., Lim, J., Matsushita, S., Wong, T., \& Ryder, S. 2013, ApJ, 768, 57

\bibitem[{Pogge \& Martini(2002)}]{Pogge2002}
Pogge, R.~W., \& Martini, P. 2002, ApJ, 569, 624

\bibitem[{Prada \& Guti{\'{e}}rrez(1999)}]{Prada1999}
Prada, F., \& Guti{\'{e}}rrez, C.~M. 1999, ApJ, 20, 123

\bibitem[{Predehl \& Schmitt(1995)}]{Predehl1995}
Predehl, P., \& Schmitt, J. H. M.~M. 1995, A{\&}A, 293, 889

\bibitem[{Riffel {et~al.}(2009)Riffel, Pastoriza, Rodriguez-Ardila, \&
  Bonatto}]{Riffel2009b}
\href{http://arxiv.org/abs/0907.4144}{{Riffel, R., Pastoriza, M.~G.,
  Rodriguez-Ardila, A., \& Bonatto, C.}} 2009, MNRAS, 400, 273

\bibitem[{Riffel {et~al.}(2013{\natexlab{a}})Riffel, Rodr{\'{i}}guez-Ardila,
  Aleman, Brotherton, Pastoriza, Bonatto, \& Dors}]{Riffel2013a}
Riffel, R., Rodr{\'{i}}guez-Ardila, A., Aleman, I., {et~al.}
  2013{\natexlab{a}}, MNRAS, 430, 2002

\bibitem[{Riffel {et~al.}(2006)Riffel, Rodr{\'{i}}guez-Ardila, \&
  Pastoriza}]{Riffel2006}
Riffel, R., Rodr{\'{i}}guez-Ardila, A., \& Pastoriza, M.~G. 2006, A{\&}A, 457,
  61

\bibitem[{Riffel \& Storchi-Bergmann(2011)}]{Riffel2011}
Riffel, R.~A., \& Storchi-Bergmann, T. 2011, MNRAS, 417, 2752

\bibitem[{Riffel {et~al.}(2010)Riffel, Storchi-Bergmann, \&
  Nagar}]{Riffel2010a}
Riffel, R.~A., Storchi-Bergmann, T., \& Nagar, N.~M. 2010, MNRAS, 404, 166

\bibitem[{Riffel {et~al.}(2015)Riffel, Storchi-Bergmann, \&
  Riffel}]{Riffel2015}
Riffel, R.~A., Storchi-Bergmann, T., \& Riffel, R. 2015, MNRAS, 451, 3587

\bibitem[{Riffel {et~al.}(2013{\natexlab{b}})Riffel, Storchi-Bergmann, \&
  Winge}]{Riffel2013b}
Riffel, R.~A., Storchi-Bergmann, T., \& Winge, C. 2013{\natexlab{b}}, MNRAS,
  430, 2249

\bibitem[{Riffel {et~al.}(2014)Riffel, Vale, Storchi-Bergmann, \&
  McGregor}]{Riffel2014a}
Riffel, R.~A., Vale, T.~B., Storchi-Bergmann, T., \& McGregor, P.~J. 2014,
  MNRAS, 442, 656

\bibitem[{Rodr{\'{i}}guez-Ardila
  {et~al.}(2004{\natexlab{a}})Rodr{\'{i}}guez-Ardila, Pastoriza, Viegas, Sigut,
  \& Pradhan}]{Rodriguez-Ardila2004}
Rodr{\'{i}}guez-Ardila, A., Pastoriza, M.~G., Viegas, S., Sigut, T. A.~A., \&
  Pradhan, A.~K. 2004{\natexlab{a}}, A{\&}A, 425, 457

\bibitem[{Rodr{\'{i}}guez-Ardila
  {et~al.}(2004{\natexlab{b}})Rodr{\'{i}}guez-Ardila, Prieto, Viegas, Ho, \&
  Schmitt}]{Rodriguez-Ardila2004a}
Rodr{\'{i}}guez-Ardila, A., Prieto, A., Viegas, S.~M., Ho, L.~C., \& Schmitt,
  H.~R. 2004{\natexlab{b}}, in Interplay Among Black Holes, Stars ISM Galact.
  Nucl., ed. T.~Storchi-Bergmann, Vol. 222, 283--286

\bibitem[{Rodr{\'{i}}guez-Ardila {et~al.}(2017)Rodr{\'{i}}guez-Ardila, Prieto,
  Mazzalay, Fern{\'{a}}ndez-Ontiveros, Luque, \&
  M{\"{u}}ller-S{\'{a}}nchez}]{Rodriguez-Ardila2017}
Rodr{\'{i}}guez-Ardila, A., Prieto, M.~A., Mazzalay, X., {et~al.} 2017, MNRAS,
  470, 2845

\bibitem[{Rodr{\'{i}}guez-Ardila {et~al.}(2011)Rodr{\'{i}}guez-Ardila, Prieto,
  Portilla, \& Tejeiro}]{Rodriguez-Ardila2011}
Rodr{\'{i}}guez-Ardila, A., Prieto, M.~A., Portilla, J.~G., \& Tejeiro, J.~M.
  2011, ApJ, 743, 100

\bibitem[{Rodr{\'{i}}guez-Ardila {et~al.}(2005)Rodr{\'{i}}guez-Ardila, Riffel,
  \& Pastoriza}]{Rodriguez-Ardila2005}
Rodr{\'{i}}guez-Ardila, A., Riffel, R., \& Pastoriza, M.~G. 2005, MNRAS, 364,
  1041

\bibitem[{Rousselot {et~al.}(2000)Rousselot, Lidman, Cuby, Moreels, \&
  Monnet}]{Rousselot2000}
Rousselot, P., Lidman, C., Cuby, J.-G., Moreels, G., \& Monnet, G. 2000,
  A{\&}A, 354, 1134

\bibitem[{Rubin(1980)}]{Rubin1980}
\href{http://adsabs.harvard.edu/doi/10.1086/158041}{{Rubin, V.~C.}} 1980, ApJ,
  238, 808

\bibitem[{Ryter(1996)}]{Ryter1996}
Ryter, C.~E. 1996, Ap{\&}SS, 236, 285

\bibitem[{Schlafly \& Finkbeiner(2011)}]{Schlafly2011}
Schlafly, E.~F., \& Finkbeiner, D.~P. 2011, ApJ, 737, 103

\bibitem[{Schmitt \& Kinney(1996)}]{Schmitt1996}
\href{http://adsabs.harvard.edu/doi/10.1086/177264}{{Schmitt, H.~R., \& Kinney,
  A.~L.}} 1996, ApJ, 463, 498

\bibitem[{Schommer {et~al.}(1988)Schommer, Caldwell, Wilson, Baldwin, Phillips,
  Williams, \& Turtle}]{Schommer1988}
\href{http://adsabs.harvard.edu/doi/10.1086/165887}{{Schommer, R.~A., Caldwell,
  N., Wilson, A.~S., {et~al.}}} 1988, ApJ, 324, 154

\bibitem[{Sch{\"{o}}nell {et~al.}(2017)Sch{\"{o}}nell, Storchi-Bergmann,
  Riffel, \& Riffel}]{Schonell2017}
Sch{\"{o}}nell, A.~J., Storchi-Bergmann, T., Riffel, R.~A., \& Riffel, R. 2017,
  MNRAS, 464, 1771

\bibitem[{Shimizu {et~al.}(2018)Shimizu, Davies, Koss, Ricci, Lamperti, Oh,
  Schawinski, Trakhtenbrot, Burtscher, Genzel, Lin, Lutz, Rosario, Sturm, \&
  Tacconi}]{Shimizu2018}
\href{http://arxiv.org/abs/1710.09117{\%}0Ahttp://dx.doi.org/10.3847/1538-4357/aab09e}{{Shimizu,
  T.~T., Davies, R.~I., Koss, M., {et~al.}}} 2018, ApJ, 856, 154

\bibitem[{{Sim{\~{o}}es Lopes} {et~al.}(2007){Sim{\~{o}}es Lopes},
  Storchi-Bergmann, {de F{\'{a}}tima Saraiva}, \& Martini}]{SimoesLopes2007}
\href{http://adsabs.harvard.edu/abs/2007ApJ...655..718S}{{{Sim{\~{o}}es Lopes},
  R.~D., Storchi-Bergmann, T., {de F{\'{a}}tima Saraiva}, M., \& Martini, P.}}
  2007, ApJ, 655, 718

\bibitem[{Skrutskie {et~al.}(2006)Skrutskie, Cutri, Stiening, Weinberg,
  Schneider, Carpenter, Beichman, Capps, Chester, Elias, Huchra, Liebert,
  Lonsdale, Monet, Price, Seitzer, Jarrett, Kirkpatrick, Gizis, Howard, Evans,
  Fowler, Fullmer, Hurt, Light, Kopan, Marsh, McCallon, Tam, {Van Dyk}, \&
  Wheelock}]{Skrutskie2006}
\href{http://stacks.iop.org/1538-3881/131/i=2/a=1163}{{Skrutskie, M.~F., Cutri,
  R.~M., Stiening, R., {et~al.}}} 2006, AJ, 131, 1163

\bibitem[{Storchi-Bergmann {et~al.}(2009)Storchi-Bergmann, McGregor, Riffel,
  {Sim{\~{o}}es Lopes}, Beck, \& Dopita}]{Storchi-Bergmann2009}
Storchi-Bergmann, T., McGregor, P.~J., Riffel, R.~A., {et~al.} 2009, MNRAS,
  394, 1148

\bibitem[{Urry \& Padovani(1995)}]{Urry1995}
Urry, C.~M., \& Padovani, P. 1995, PASP, 107, 803

\bibitem[{V{\'{e}}ron-Cetty \& V{\'{e}}ron(2006)}]{Veron-Cetty2006}
V{\'{e}}ron-Cetty, M.-P., \& V{\'{e}}ron, P. 2006, A{\&}A, 455, 773

\bibitem[{Wilman {et~al.}(2005)Wilman, Edge, \& Johnstone}]{Wilman2005}
Wilman, R.~J., Edge, A.~C., \& Johnstone, R.~M. 2005, MNRAS, 359, 755

\bibitem[{Wilson {et~al.}(1993)Wilson, Braatz, Heckman, Krolik, \&
  Miley}]{Wilson1993}
\href{http://adsabs.harvard.edu/doi/10.1086/187137}{{Wilson, A.~S., Braatz,
  J.~A., Heckman, T.~M., Krolik, J.~H., \& Miley, G.~K.}} 1993, ApJ, 419, L61

\bibitem[{Wright \& Eastman(2014)}]{Wright2014}
\href{http://arxiv.org/abs/1409.4774}{{Wright, J.~T., \& Eastman, J.~D.}} 2014,
  Publ. Astron. Soc. Pacific, 126, 838

\bibitem[{Zhu {et~al.}(2017)Zhu, Tian, Li, \& Zhang}]{Zhu2017}
\href{http://arxiv.org/abs/1706.07109{\%}0Ahttp://dx.doi.org/10.1093/mnras/stx1580}{{Zhu,
  H., Tian, W., Li, A., \& Zhang, M.}} 2017, MNRAS, 3528, 3494

\end{thebibliography}
\end{document}